\documentclass[12pt]{article}
\usepackage{amsmath}
\usepackage{graphicx,psfrag,epsf}
\usepackage{enumerate}
\usepackage{booktabs}
\usepackage{siunitx}
\usepackage{multirow}
\usepackage{makecell}
\usepackage{placeins}
\usepackage{url} 
\usepackage{setspace, graphicx, caption, subcaption, float, relsize, rotating, color, verbatim, multirow}
\usepackage[round, comma]{natbib}
\usepackage[colorlinks=true, urlcolor=blue,linkcolor=blue, citecolor=magenta]{hyperref}

\newcommand{\blind}{0}

\usepackage{empheq}

\usepackage{enumerate}

\usepackage{dsfont}
\usepackage{mathtools}
\usepackage{bm}

\usepackage{stmaryrd}
\usepackage{amssymb} 
\usepackage{amsmath}
\usepackage{amsfonts}
\usepackage{amsthm} 
\usepackage{mathrsfs}
\usepackage{bigints} 
\usepackage{enumerate} 

\usepackage{color} 

\usepackage{bbm}

\usepackage{extarrows}

\usepackage{amsfonts}
\usepackage{graphicx}

\usepackage{times}
\usepackage{bm}

\usepackage[plain,noend]{algorithm2e}

\newtheorem{thm}{Theorem}

\newtheorem{rk}{Remark}

\newtheorem{assumption}[]{Assumption}
\newtheorem{lemma}{Lemma}

\counterwithin{table}{section}

\newcommand{\nc}{\newcommand}

\nc{\dps}{\displaystyle}
\nc{\tr}{\text{tr}}

\def\bbE{\mathbb{E}}
\def\Var{{\rm Var \hskip-0.01cm }}

\def\boxit#1{\vbox{\hrule\hbox{\vrule\kern6pt
			\vbox{\kern6pt#1\kern6pt}\kern6pt\vrule}\hrule}}

\numberwithin{equation}{section}

\allowdisplaybreaks
\begin{document}

	\def\spacingset#1{\renewcommand{\baselinestretch}%
		{#1}\small\normalsize} \spacingset{1}

	
	\if0\blind
	{
		\title{\bf Non-parametric Quantile Regression and Uniform Inference with Unknown Error Distribution \thanks{
				The authors are alphabetically ordered. Haoze Hou is a PhD student. The present work constitutes part of his research study leading to
				his doctoral thesis at the Renmin University of China.  Zheng Zhang is the corresponding author. E-mail: zhengzhang@ruc.edu.cn.}}
		\author{Haoze Hou\\
			Institute of Statistics \& Big Data, Renmin
			University of China\\and\\Wei Huang\\School of Mathematics and Statistics, University of Melbourne\\and\\Zheng Zhang\\
	Center for Applied Statistics, Institute of Statistics \& Big Data\\ Renmin University of China}
		\maketitle
	} \fi
	
	\if1\blind
	{
		\bigskip
		\bigskip
		\bigskip
		\begin{center}
			{\LARGE\bf Non-parametric Quantile Regression and Uniform Inference with Unknown Error Distribution}
		\end{center}
		\medskip
	} \fi  
	
	\bigskip
	\begin{abstract}
		This paper studies the non-parametric estimation and uniform inference for the conditional quantile regression function (CQRF) with covariates exposed to measurement errors.  We consider the case that the distribution of the measurement error is unknown and allowed to be either ordinary or super smooth.  We estimate the density of the measurement error by the repeated measurements and propose the deconvolution kernel estimator for the CQRF. We derive the uniform Bahadur representation of the proposed estimator and construct the uniform confidence bands for the CQRF, uniformly in the sense for all covariates and a set of quantile indices, and establish the theoretical validity of the proposed inference. A data-driven approach for selecting the tuning parameter is also included. Monte Carlo simulations and a real data application demonstrate the usefulness of the proposed method.
	\end{abstract}
	
	\vfill
	
	\newpage
	\spacingset{1.8} 
	\section{Introduction}\label{sec:introduction}
	
	The quantile regression model provides a valuable tool for analyzing the heterogeneous effects of covariates on the response outcome, which has been widely exploited by statisticians, economists, and machine learning researchers. Given fully observed data, there is extensive literature studying quantile regression methods with various applications, see \cite{koenker1978regression,fan1994robust,takeuchi2006nonparametric,firpo2007efficient,dette2008non,wang2012quantile,donald2014estimation,zheng2015globally,chiang2019robust,tan2022high,he2023smoothed,zhang2023bootstrap,lin2024quantile} among others.
 
 Measurement errors arise very often in practice. There, the underlying true covariate is not observable. Instead, we only observe the sum of the true covariate and a random error. {In particular, we focus on a common scenario known as the \textit{classical measurement error}, where the error is assumed to be independent of the underlying true covariate.} For example, in the National Health and Nutrition Examination Survey (NHANES), since observing the long-term diet directly is expensive, the long-term nutrition data are computed based on the participants' recall of their diet in the previous 24 hours, which is known to subject to classical measurement errors \citep{carroll2006measurement,camirand2022semiparametric}. It is verified that over 75\% of the variance {is attributed to} measurement errors. The literature on classical measurement errors is extensive, including works by \cite{fan1991asymptotic,fan1993nonparametric,schennach2004estimation,delaigle2008deconvolution,delaigle2016methodology,kato2018uniform,kato2019uniform}, among others. For a comprehensive review of classical and other measurement errors, we refer readers to the recent books by \cite{SCHENNACH2020487} and \cite{grace2021handbook}. As pointed out by \cite{chesher2017understanding}, ignoring measurement errors in the quantile regression model may produce severe estimation bias and erroneous conclusions. 

 In this paper, we aim to construct uniform confidence bands for a nonparametric quantile regression function when the covariate is contaminated by classical measurement error. Here, we allow for uniformity over a compact set of both the covariate and quantile levels. To relax the common assumption of knowing the distribution of the measurement error, we follow the approach of \cite{diggle1993fourier,neumann1997effect,kato2018uniform,kato2019uniform}, assuming that observations from the distribution of the measurement error are available. This assumption is often satisfied in real-world scenarios, such as (i) when additional validation data, possibly from a different source, is available {\citep[see, e.g.][]{siddique2019measurement,luijken2019impact}}, or (ii) when repeated measurements of the covariate are available  {\citep[see, e.g.][]{carroll2006measurement,delaigle2008deconvolution}}.

 The existing literature on quantile regression with measurement errors primarily addresses the identification and estimation of the quantile regression function. \cite{he2000quantile} first proposed a consistent estimator for the quantile regression function under a linear model. \cite{wei2009quantile} relaxed the assumptions on the symmetry of measurement error distribution in \cite{he2000quantile} and provided an estimator implemented via the EM algorithm. \cite{firpo2017measurement} considered the estimation and inference for linear quantile regression models with repeated mismeasured covariates.
Recognizing the risk of misspecification in parametric models, \cite{hu2008instrumental} considered the nonparametric identification of models with nonclassical measurement errors and the availability of instrumental variables. \cite{schennach2008quantile} also studied the nonparametric identification and estimation of the conditional quantile regression function, assuming the presence of instrumental variables. However, the asymptotic distribution of the estimator and corresponding inference methods were not established. Our work contributes to the existing literature by proposing a theoretically guaranteed uniform methodology for the nonparametric quantile regression function in the presence of an unknown measurement error distribution. Unlike previous approaches, our method does not require instrumental variables but instead relies on access to samples from the measurement error distribution. See Section D of the supplemental material for more details.
 
   While statistical inferences with classical measurement error exist in the literature, they focus exclusively on the density, distribution function of the true covariate and the regression function with error-contaminated covariates. \cite{delaigle2015confidence} constructed the pointwise confidence bands in non-parametric error-in-variables regression. \cite{bissantz2007non} proposed a uniform confidence band for 
 density function, later extended by \cite{kato2018uniform} to the unknown measurement error case. Subsequently, \cite{kato2021robust} proposed a uniform band for density function based on  Kotlarski’s identity under the repeated measurement case, imposing relatively weak assumptions on the distribution of measurement error. \cite{kato2019uniform,adusumilli2020inference} proposed uniform inference methods for the regression and distribution functions, respectively. To our best knowledge, there has been no attempt to develop a uniform confidence band for the quantile regression function in the context of measurement error. Unlike density, distribution function, or mean regression, the quantile regression estimators lack explicit expressions and their estimation criterion is non-smooth, which imposes new challenges for uniform inference.

 Our construction of the uniform confidence band builds upon the {deconvolution kernel estimator and} the multiplier bootstrap technique. The deconvolution kernel is a widely applied nonparametric method in the context of classical measurement error, initially proposed by \cite{stefanski1990deconvolving}. Regarding the multiplier bootstrap technique, it offers a superior convergence rate compared to limiting-distribution-based confidence bands, as demonstrated by \cite{hall1991convergence}. However, the validity of multiplier bootstrap uniform confidence band based on nonparametric function estimation has been, until recently, unclear. \cite{chernozhukov2014anti} first showed a set of high-level conditions when the multiplier bootstrap uniform confidence band for the non-Donsker function class, a broader class containing nonparametric function estimators, is valid. They showed that the validity of the multiplier bootstrap confidence band only requires the Gaussian approximation of the suprema of the empirical process, instead of the Gaussian approximation of the whole empirical process (Komlós–Major–Tusnády approximation, KMT), which results in weaker conditions. We present the low-level conditions under which the high-level conditions in \cite{chernozhukov2014anti} are satisfied for the nonparametric quantile regression, and show the validity of our uniform confidence bands.
 
Our work also sheds light on the estimation of nonparametric quantile regression function in the presence of measurement error. We contribute by deriving the uniform Bahadur representation of our nonparametric quantile regression estimator, which is a fundamental step in showing the validity of our uniform confidence bands. Our technique builds upon the M-estimator theory \citep[see e.g.][]{chen2003estimation,van1996weak} and the local maximal inequality in  \cite{chernozhukov2014gaussian}. The uniform Bahadur representation itself is also of independent interest, for example, it is useful for constructing other semiparametric parameters, as discussed in \cite{kong2010uniform}.

We also provide a practical data-driven selection of the undersmoothing bandwidth for implementing the uniform bands, extending the method of \cite{bissantz2007non} to our framework. The numerical simulation shows our method performs stable and well. A real data application of NHANES data is also included. 
 		
	The rest of the paper is organized as follows. Section~\ref{Section:Setup} introduces the basic framework, notations, and the goal of the paper. Section~\ref{sec:estandinf} introduces our estimator and uniform inference method. Section~\ref{Section:Asymp} presents the uniform Bahadur representation of our estimators and the asymptotic validity of the uniform band. We present our simulation results in Section \ref{Section:Simulation}, followed by a real data application in Section~\ref{sec:realdata}.

	\section{Basic Framework} \label{Section:Setup}
	We consider the nonparametric conditional quantile regression models:
	\begin{align}\label{mode:CQR}
		Y=\theta_{\tau}(X)+\epsilon(\tau),
	\end{align}
	where $\tau\in (0,1)$, $Y\in \mathbb{R}$ is the univariate response variable, $X\in \mathbb{R}^d$ is the  $d$-dimensional covariate for some $d\in\mathbb{N}$, and $\epsilon(\tau)$ is the error term whose $\tau^{th}$ quantile is zero conditional on $X$. For simplicity of notation, we present the result for $d=1$. The conditional quantile regression function (CQRF), $\theta_{\tau}(\cdot)$, can be written as the solution of the following conditional moment equation:
	\begin{align}\label{id:beta*_potential}
		\mathbb{E}[\psi_{\tau}\left\{Y-\theta_{\tau}(X)\right\}|X]=0\,,
	\end{align}
	where $\psi_{\tau}(u) := \tau - \mathbbm{1}\{u<0\}$, and $\mathbbm{1}\{\cdot\}$ denotes the indicator function.
	
	Suppose that the true covariate $X$ is unobservable due to the classical measurement errors, that is, we observe
	\begin{align}\label{model:ME}
		W = X+U,
	\end{align}
	where $U$ is independent of $(X,Y)$. The observed sample $\{W_i,Y_i\}_{i=1}^n$ are independent and identically distributed ($i.i.d.$) as that of $(W,Y)$. In the literature {on} classical measurement error, a {commonly} used identification assumption is {that the measurement error distribution is known}.  However, this assumption is restrictive in practice. To relax {it, we assume the availability of} auxiliary samples $\{U_j\}_{j=1}^m$ that are independent and identically distributed ($i.i.d.$) as  $U$. Such an assumption is widely accepted in the literature, see e.g., \cite{diggle1993fourier,neumann1997effect,kato2019uniform}.\footnote{Another possible framework for identifying the CQRF from measurement error data is to leverage an instrumental variable \cite[see, e.g.,][]{schennach2008quantile}. However, establishing the uniform asymptotic behavior and inference method for this approach is fundamentally different from our setting and presents significant challenges, which are beyond the scope of our paper.} 

We {provide} two real-world scenarios {where the auxiliary} samples are available: (i) {When repeated measurements of $W$ are available, as discussed in \cite{carroll2006measurement, kato2018uniform}. Specifically, denote the repeated measurements of $W$ as $W^{(1)}$ and $W^{(2)}$, such that $W^{(1)}=X+U^{(1)}$ and $W^{(2)}=X+U^{(2)}$, where $U^{(1)}$ and $U^{(2)}$ are identically distributed and the conditional distribution $U^{(2)}|U^{(1)}$ is symmetric. Then $\overline{U}:=(W^{(1)}-W^{(2)})/2 \overset{d}{=} \widetilde{U}:= (U^{(1)}+U^{(2)})/2$. Defining $\widetilde{W}=(W^{(1)}+W^{(2)})/2$, in this case, we observe the primary sample $(Y_i, \widetilde{W}_i)_{i=1}^n$ and the auxiliary sample $\{\overline{U}_j\}_{j=1}^n$; (ii) When we have access to a validation study where an auxiliary sample of $U$ is available \citep[see, e.g.][]{siddique2019measurement,luijken2019impact}.}
 
 The goal of this paper is to estimate the CQRF $\theta_{\tau}(x)$ and construct its uniform confidence bands over $(x,\tau)\in \mathcal{X}\times\mathcal{T}$ when we have {a} primary sample $\{Y_i, W_i\}_{i=1}^n$ and {an} auxilary sample $\{U_i\}_{i=1}^{m}$, where $\mathcal{X}\subset \mathbb{R}$ is a compact subset of the support\footnote{In practice, the support of $X$ {can be determined} through prior information or {by using the} data-driven boundary estimation method {developed} in \cite{delaigle2006data,delaigle2006estimation}.} of $X$ and $\mathcal{T}$ is a singleton or a compact set of quantile indices such that $\mathcal{T}\subset (0,1)$.

	\section{Estimation \& Inference Procedure}\label{sec:estandinf}
	\subsection{Estimation Procedure}
	We start with a nonparametric consistent estimator of $\theta_{\tau}(x)$ by exploiting the deconvolution kernel (DK) method  \citep{stefanski1990deconvolving}. Let $K(\cdot): \mathbb{R} \rightarrow (0, \infty)$ be a univariate kernel function whose characteristic function is denoted by $\phi_K(t):=\int_{-\infty}^{+\infty} \exp(it\cdot v)K(v)dv$. Let $h$ be a bandwidth and
	$K_{h}(x):= K(x/h)/h$ be the normalized kernel function. If $X$ was observable, $\theta_{\tau}(x)$ defined by \eqref{id:beta*_potential} can be identified by
	\begin{align}\label{eq:identify}
		\theta_{\tau}(x):=\arg\min_{\theta\in\Theta}\lim_{h\rightarrow 0 }\left|\mathbb{E}\left\{\psi_{\tau}(Y-\theta)K_{h}(x-X)\right\}\right|
	\end{align}
	in the sense that $\lim_{h\rightarrow 0 }\left|\mathbb{E}\left\{\psi_{\tau}(Y-\theta)K_{h}(x-X)\right\}\right| =\left| \mathbb{E}\left\{\psi_{\tau}(Y-\theta)|X=x\right\}f_X(x)\right|.$ This motivates the local constant quantile regression estimator in \cite{hardle2010confidence,kong2010uniform}.
	
	In our case, $X$ is not observable.  Let $\phi_X$, $\phi_U$ and $\phi_W$ be the characteristic functions of $X$, $U$ and $W$, respectively. In light of  \eqref{model:ME} and the Fourier inversion theorem, the probability density function $f_X(x)$ can be represented by
	\begin{align*}
		f_X(x)=\frac{1}{2\pi}\int_{-\infty}^{+\infty} \frac{\phi_W(t)}{\phi_U(t)}\exp \left(-ix\cdot  t\right)dt.
	\end{align*}
	Then $\phi_{{W}}({t}) = \int_{-\infty}^{+\infty} \exp(it\cdot w)f_{{W}}({w})\,d{w}$ can be estimated by the kernel estimator $\widehat{\phi}_{{W}}({t})=n^{-1}\sum^n_{i=1}\int_{-\infty}^{+\infty} \exp(it\cdot  w)K_{{h}}({w}-{W}_i)\,d{w}$. If $\phi_U(t)$ is known, then the deconvolution kernel density estimator of $f_{{X}}({x})$ is given by
	\begin{equation}\label{eq:fXhat}
		\widehat{f}_{{X}}(x):=\frac{1}{2\pi}\int_{-\infty}^{+\infty} \frac{\widehat{\phi}_W(t)}{{\phi}_U(t)}\exp \left(-ix\cdot  t\right)dt=\frac{1}{n}\sum^n_{i=1}K_{{U},{h}}({x}-{W}_i)\ ,,
	\end{equation}
	where 
	$$
	K_{U,h}(x):= \frac{1}{2\pi h} \int_{-\infty}^{+\infty} \exp(-itx/h) \frac{\phi_K(t)}{\phi_{U}(t/h)} dt
	$$
	is called the \emph{deconvolution kernel}. Note that $K_{U,h}$ satisfies $\mathbb{E}\left\{K_{U,h}(x-W)|X\right\} = K_{h}(x-X)$ .  Based on \eqref{model:ME}, \eqref{eq:identify} and this property, for every $x\in\mathcal{X}$, the CQRF $\theta_{\tau}(x)$ can be written as 
	\begin{align}\label{id:theta_tau}
		\theta_{\tau}(x)= \arg\min_{\theta\in\Theta} \lim_{h\to 0} \left|\mathbb{E}\left[ \psi_{\tau}(Y-\theta)K_{U,h}(x-W)\right]\right|.
	\end{align}
	Equation \eqref{id:theta_tau} provides the pointwise identification of the CQRF $\theta_{\tau}(x)$, from the observations $(Y,W)$.
	
	Note that $\phi_U(t)$ in the definition of the the deconvolution kernel $K_{U,h}(x)$ is unknown in our context, we first estimate $\phi_U(t)$ and $K_{U,h}(x)$ respectively by 
	\begin{align}\label{def:repeateddecon}
		\widehat{\phi}_U(t):=\frac{1}{m}\sum_{j=1}^m \exp(itU_j) \ \text{and} \	\widehat{K}_{U,h}(x):= \frac{1}{2\pi h} \int_{-\infty}^{+\infty} \exp(-itx/h) \frac{\phi_K(t)}{\widehat{\phi}_{U}(t/h)} dt.
	\end{align}
	Then, using \eqref{id:theta_tau}, we propose the DK estimator of $\theta_{\tau}(x)$  by
	\begin{align}\label{def:repeatedestimator}
		\widehat{\theta}_{\tau}(x):=\arg\min_{\theta\in\Theta}|\widehat{M}_n(\theta;x,\tau)|:=\arg\min_{\theta\in\Theta}\left|\frac{1}{n}\sum_{i=1}^n\psi_{\tau}(Y_i-\theta)\widehat{K}_{U,h}(x-W_i)\right|.
	\end{align}
 
 To numerically solve \eqref{def:repeatedestimator}, we employ the following algorithm:
\begin{enumerate}
    \item Sort the samples $\{Y_i\}_{i=1}^n$ in ascending order: $Y_{(1)} \leq Y_{(2)} \leq \dots \leq Y_{(n)}$.
    \item Calculate the value of $|\widehat{M}_n(\theta;x,\tau)|$ for $\theta = \frac{Y_{(k)} + Y_{(k+1)}}{2}$, where $k = 1, 2, \dots, n$ and $Y_{(n+1)} = Y_{(n)}$.
    \item Select the $\theta$ that yields the lowest $|\widehat{M}_n(\theta;x,\tau)|$ as our solution to \eqref{def:repeatedestimator}.
\end{enumerate}
This approach ensures that we find a minimizer of the objective function. 

	\subsection{Bootstrap Uniform Band}
	In this section, we conduct the uniform confidence band for the parameter $\theta_{\tau}(x)$ over $(x,\tau)\in\mathcal{X}\times \mathcal{T}$. To introduce our bootstrap procedure, we introduce an estimator of the variance of $\widehat{\theta}_{\tau}(x)$, $ \widehat{\sigma}^2_n(x,\tau)$. We shall show rigorously in Section \ref{Section:Asymp} that the asymptotic variance of $\widehat{\theta}_{\tau}(x)$ is given by $$\sigma_n^2(x,\tau)=\frac{1}{nf^2_{X,Y}\{x,\theta_\tau(x)\}}\mathbb{E}\left[K_{U,h}^2(x-W_i)\psi_{\tau}^2\{Y_i-\theta_{\tau}(x)\}\right]\,,$$ where $f_{X,Y}(x,y)$ is the joint density function of $X$ and $Y$ for $(x,y)\in\mathcal{X}\times\mathbb{R}$. We thus propose to estimate $\sigma_n^2(x,\tau)$ by \begin{align*}
		\widehat{\sigma}_n^2(x,\tau) := \frac{1}{\left[n \widehat{f}_{X,Y}\left\{x,\widehat{\theta}_{\tau}(x)\right\}\right]^2} \sum_{i=1}^n \widehat{K}_{U,h}^2(x-W_i)\psi_{\tau}^2\left\{Y_i-\widehat{\theta}_{\tau}(x)\right\}\,,
	\end{align*} where $\widehat{K}_{U,h}$ is the estimated devoconlution kernel with bandwidth $h$ used for estimating $\widehat{\theta}_\tau$ in \eqref{def:repeatedestimator}, and $$\widehat{f}_{X,Y}(x,y):= \frac{1}{n}\sum_{i=1}^n K_{h_Y}\left(y-Y_i\right)\widehat{K}_{U,h_W}\left(x-W_i\right)$$ is a bivariate kernel-type estimator of the density $f_{X,Y}$ with bandwidth $h_Y$ and $h_W$. Note that both $\widehat{K}_{U,h}$ and $\widehat{f}_{X,Y}(x,y)$ always take real values. The proof is provided in Section~\ref{appendix:realvalue} of the supplementary material. The selection of $h$ and $(h_Y,h_W)$ in practice can be found in Section \ref{Section:bandchoose} and Section~\ref{sec:hwhy} in the supplementary material, respectively. 
	
	\noindent
	\textbf{Uniform Confidence Bands by Multiplier Bootstrap Algorithm:}
	\begin{itemize}
		\item[1.] Let $\chi$ be a random variable that is independent of $(W,Y)$ with $\mathbb{E}[\chi]=1$ and $\operatorname{Var}(\chi)<\infty$, and its distribution has sub-Gaussian tails. Draw an i.i.d sample $\{\chi_i\}_{i=1}^n$ from the distribution $\chi$.
		\item[2.] Repeat step 1 for B times and denote each sample as $\{\chi_i^{(b)}\}_{i=1}^n$, for $b=1,2,\ldots, B$. For each iteration, compute the bootstrap-version estimator on $(x,\tau)\in \mathcal{X}\times\mathcal{T}$,
		\begin{align*}
			\widehat{\theta}_{\tau}^{(b)}(x) = \arg\min_{\theta\in\Theta}\left|\sum_{i=1}^n\chi_i^{(b)} {\widehat{K}_{{U},h}({x}-{W}_i)}\psi_{\tau}(Y_i-\theta)\right|.
		\end{align*}
		\item[3.] For $b=1,2,\dots,B$, compute
		$$
		M_b^{1-\text {sided}}=\sup _{(x,\tau) \in \mathcal{X}\times\mathcal{T}} \frac{\widehat{\theta}_{\tau}^{(b)}\left(x\right)-\widehat{\theta}_{\tau}\left(x\right)}{\widehat{\sigma}_n\left(x,\tau\right)}, \quad M_b^{2-\text {sided}}=\sup _{(x,\tau) \in \mathcal{X}\times\mathcal{T}} \frac{\left|\widehat{\theta}_{\tau}^{(b)}\left(x\right)-\widehat{\theta}_{\tau}\left(x\right)\right|}{\widehat{\sigma}_n\left(x,\tau\right)},
		$$
		where the supremum is approximated by the maximum over a chosen grid of $\mathcal{X}\times\mathcal{T}$.
		
		\item[4.] Given a confidence level $1-\alpha$, find the empirical $(1-\alpha)$ quantile of the sets of numbers $\left\{M_b^{1 \text {-sided }}: b=1, \ldots, B\right\}$ and $\left\{M_b^{2 \text {-sided }}: b=1, \ldots, B\right\}$. Denote these quantiles as $\widehat{C}_\alpha^{1-\text {sided}}$ and $\widehat{C}_\alpha^{2-\text {sided}}$, respectively.
		\item[5.] The left and right one-sided and two-sided uniform confidence bands are respectively constructed as
		$$
		\begin{aligned}
			& I_L=\left\{\left(\widehat{\theta}_{\tau}\left(x\right)-\widehat{C}_\alpha^{1-\text {sided}} \widehat{\sigma}_n\left(x,\tau\right), \infty\right): (x,\tau) \in \mathcal{X}\times\mathcal{T}\right\}, \\
			& I_R=\left\{\left(-\infty, \widehat{\theta}_{\tau}\left(x\right)+\widehat{C}_\alpha^{1-\text { sided }}\widehat{\sigma}_n\left(x,\tau\right)\right): (x,\tau) \in \mathcal{X}\times\mathcal{T}\right\}, \\
			& I_2 = \left\{\left(\widehat{\theta}_{\tau}\left(x\right)-\widehat{C}_\alpha^{2-\text {sided}} \widehat{\sigma}_n\left(x,\tau\right),\widehat{\theta}_{\tau}\left(x\right)+\widehat{C}_\alpha^{2-\text {sided}} \widehat{\sigma}_n\left(x,\tau\right) \right): (x,\tau) \in \mathcal{X}\times\mathcal{T}\right\}.
		\end{aligned}
		$$
	\end{itemize}
		
	\begin{rk} As discussed in Section~\ref{sec:introduction}, under the errors-in-variables setting, i.e. $W=X+U$ in \eqref{model:ME}, existing literature concerning nonparametric statistical estimation and inference mainly focus on the probability density function, the conditional mean regression function, and the cumulative distribution function. 
		
		On the contrary, in the literature, the conditional quantile regression (CQRF) considered in this paper does not have a closed expression, and the check loss function used in quantile regression is non-smooth. These significant differences bring non-trivial challenges to the statistical analysis. We contribute to the existing literature by exploiting the DK method to estimate the CQRF function nonparametrically, developing the uniform asymptotic distribution and constructing the uniform confidence bands, uniformly in both the covariates $x\in\mathcal{X}$ and the quantile index $\tau\in\mathcal{T}$. 
	\end{rk}

	\section{Large Sample Properties}\label{Section:Asymp}
	This section gives the asymptotic validity of the uniform bands. Towards the asymptotic validity of our uniform band, a key intermediate result is the uniform Bahadur expansion of the estimators $\widehat{\theta}_{\tau}(x)$ and $\widehat{\theta}_{\tau}^{(b)}$, which is presented in Section \ref{sec:osbahadur} and \ref{sec:ssbahadur}. Based on the asymptotic uniform properties of $\widehat{\theta}_{\tau}(x)$ and $\widehat{\theta}_{\tau}^{(b)}$, we establish the asymptotic validity of the uniform band in Section \ref{Section:Band}.
	
	The asymptotic properties of the proposed estimators and the uniform confidence band depend on the distribution of the measurement error $U$. Here, we introduce the ordinary smooth and the supersmooth errors, which are commonly assumed in the literature of errors-in-variables problems (see e.g. \citealp{stefanski1990deconvolving,fan1993nonparametric,delaigle2009design,Meister2009,huang2023nonparametric}). The distribution of $U\in\mathbb{R}$ is called \emph{ordinary smooth} of order $\beta>0$ if
	\begin{equation}\label{def:os}
		\lim_{t\rightarrow \infty} t^\beta \phi_{U}(t) = c \text{\ \ \ and\ \ \ } t^{\beta+1}\phi_{U}'(t)=-c\beta \text{\ \ as\ \ } t\rightarrow \infty,
	\end{equation}
	for some constant $c>0$. The distribution of $U$ is called \emph{supersmooth} of order $(\beta,\beta_0)$ if
	\begin{equation}\label{def:ss}
		d_0|t|^{\beta_0}\exp(-|t|^\beta/\gamma) \le |\phi_{U}(t)| \le d_1|t|^{\beta_0}\exp(-|t|^\beta/\gamma) \text{\ \ as\ \ } t\rightarrow \infty,
	\end{equation}
	for some positive constants $d_0,d_1,\gamma,\beta$ and some constant $\beta_0$. Examples of ordinary smooth distributions include Laplace and Gamma together with their convolutions. Examples of super smooth distributions include Cauchy and Gaussian together with their convolutions. 
	
	\subsection{Asymptotics under the Ordinary Smooth Errors}\label{sec:osbahadur}
	This section establishes the uniform asymptotic results for  $\widehat{\theta}_{\tau}(x)-\theta_{\tau}(x)$ over $(x,\tau)\in\mathcal{X}\times\mathcal{T}$ under the ordinary smooth error condition \eqref{def:os}. Denote $\varepsilon_{\tau} (x;\theta) := \mathbb{E}\{\psi_{\tau}(Y-\theta)|X=x\}$ and $M(\theta;x,\tau) = \varepsilon_{\tau} (x;\theta)f_X(x)$.
	\begin{assumption}\label{Assumption:boundedsupport}  Let  $\mathcal{X}$ be a compact subset of the support of $X$ and $\mathcal{T}\subset  (0,1)$ is a compact set of quantile indices.
		\begin{itemize}
			\item [(i)] There is a compact set $\Theta\subset \mathbb{R}$ such that  the conditional quantile function $\theta_{\tau}(x)\in\Theta$ for all $(x,\tau)\in \mathcal{X}\times \mathcal{T}$;
			\item  [(ii)] For $(x,y)\in \mathcal{X}\times \Theta$, the conditional density $f_{Y|X}(y|x)$ exists and satisfies $\inf_{(x,y)\in \mathcal{X}\times \Theta} f_{Y|X}(y|x)\geq c_1$ and $\inf_{x\in \mathcal{X}} f_{X}(x)\geq c_2 $ for some  constants $c_1, c_2>0$;
			\item [(iii)] The characteristic function of $X$, $\phi_X(t)$ satisfies $\int|\phi_X(t)|dt<\infty.$
		\end{itemize}
	\end{assumption}
	\begin{assumption}\label{Assumption:Continuity}
		Let $s\in\mathbb{N}$ be a nonnegative integer, we assume
		\begin{itemize}
			\item 	[(i)] $\varepsilon_{\tau}(x;\theta)$ and $f_{X}(x)$ are $(s+1)$-times continuously differentiable in $x$ for any $\theta\in\Theta$;
			\item [(ii)] $\partial_x^{j}M(\theta;x,\tau)$ is uniformly bounded over $(x,\tau,\theta)\in \mathbb{R}\times\mathcal{T}\times\Theta$ for $j=s,s+1$;
			\item [(iii)] $\partial_x^{s} M(\theta;x,\tau)$ is continuous in $\theta\in\Theta$.
		\end{itemize}
	\end{assumption}

	\begin{assumption}\label{Assumption:Kernel}
		$K(\cdot)$ is a symmetric univariate kernel function of order $s$. The Fourier transform of $K(\cdot)$, denoted by $\phi_K$, has bounded support on $[-1,1]$. Furthermore, $\kappa_{j1} := \int x^j K(x)dx = 0$ for $j=0,\dots,s-1$ and  $\kappa_{j1}<\infty$ for $j = s,\ s+1$.
	\end{assumption}
	
	\begin{assumption}\label{Assumption:ME}
		$\phi_U(t)\neq 0$ for  all $t\in\mathbb{R}$.
	\end{assumption}
	
	\begin{assumption}\label{Assumption:PartialBoundAway} The conditional density 
		$f_{Y|X}(y|x)$ is differentiable in $y$. Both $f_{Y|X}(y|x)$ and $\partial_yf_{Y|X}(y|x)$  are continuous with respect to $y$ and bounded over $(x,y)\in\mathcal{X}\times{\Theta}$. $f_Y(y)$ is uniformly continuous over $y\in\Theta$, {and $f_X(x)$ is bounded over $x\in \mathcal{X}$.}
	\end{assumption}
	
	\noindent
	{{\bf Assumption OS.}} (i) $h\rightarrow 0$ as $n\to \infty$; (ii) $nh^{2\beta+1}/\log(1/h)\rightarrow \infty$. 
	
	Assumption \ref{Assumption:boundedsupport} imposes compactness and boundedness conditions that are required for establishing the uniform asymptotic results over $\mathcal{X}\times \mathcal{T}$. In addition, Assumption \ref{Assumption:boundedsupport}(ii) guarantees that (2.2) admits a unique solution almost surely. Assumption \ref{Assumption:boundedsupport}(iii) restricts the characteristic function $\phi_X$ belongs to $L_1$ space, which implies the density function $f_X(x)$ is continuous over $x\in\mathbb{R}$, which is also imposed in \cite{kato2019uniform}. Assumptions \ref{Assumption:Continuity} and  \ref{Assumption:Kernel} impose smoothness and regularity conditions on nuisance functions and the smoothing kernel that are needed for controlling the approximation bias. Following the results and recommendation of \cite{delaigle2006optimal}, in our numerical implementation, we use  $\phi_K(t)=(1-t^2)^3\mathbbm{1}_{[-1,1]}(t)$ for computing the deconvolution kernel estimator $\widehat{\theta}_{\tau}(x)$ in \eqref{def:repeatedestimator}, which is a second order kernel function.  Assumption \ref{Assumption:ME} ensures the non-degeneracy of the deconvolution kernel which is standard in the literature \citep[e.g.][]{fan1993nonparametric,kato2018uniform,kato2019uniform}. {For example, commonly used measurement error distributions, such as Gaussian and Laplace, satisfy Assumption \ref{Assumption:ME}. In cases where Assumption \ref{Assumption:ME} is violated, such as with a uniform measurement error distribution, the deconvolution kernel is not applicable, and alternative techniques are required \citep[see e.g.,][]{hall2007ridge,delaigle2011nonparametric}, which are beyond the scope of this work.} Assumption \ref{Assumption:PartialBoundAway} imposes the continuity and boundedness on the distributions $f_{Y|X}$,  {$f_Y$, and $f_X$}. The restriction on $f_{Y|X}$ is standard in the literature of quantile regression {without measurement error} \citep[see e.g.][]{hardle2010confidence}. 
 Assumption OS {is} required for our estimator to be consistent under the ordinary smooth error. 
		
  Let $\mathcal{B}_n(\theta;x,\tau) := -\kappa_{s1}\cdot h^s\cdot \partial_x^s (\varepsilon_\tau f_X)(x;\theta)\big/ f_{X,Y}\{x,\theta_\tau(x)\}$, $$\mathcal{V}_n(\theta;x,\tau):=-\frac{1}{n\cdot f_{X,Y}\{x,\theta_\tau(x)\}}\sum_{i=1}^n\left[ K_{U,h}(x-W_i)\psi_{\tau}(Y_i-\theta) -\mathbb{E}\left\{K_{U,h}(x-W_i)\psi_{\tau}(Y-\theta)\right\}\right]\,,$$
$$\mathcal{V}_n^{\chi}(\theta;x,\tau)=-\frac{1}{n\cdot f_{X,Y}\{x,\theta_\tau(x)\}}\sum_{i=1}^n\left\{\chi_i^{(b)} -1\right\}K_{U,h}(x-W_i)\psi_{\tau}(Y_i-\theta)\,,  $$ and
$\sigma_n^2(x,\tau):=\Var \left\{\mathcal{V}_n(\theta;x,\tau)\right\} = \Var \left\{\mathcal{V}^{\chi}_n(\theta;x,\tau)\right\}$.
The first theorem establishes the uniform Bahadur representation for $\widehat{\theta}_{\tau}(x)$ and $\widehat{\theta}_{\tau}^{(b)}$, over $(x,\tau)\in\mathcal{X}\times\mathcal{T}$ under the ordinary smooth error. The proof of Theorem \ref{thm:criteron_consistency} is relegated to the Appendix.
	\begin{thm}\label{thm:criteron_consistency}
		Denote $R_{U,OS}=n^{-1/2}m^{-1/2}h^{-2\beta-1} + m^{-1/2}h^{-\beta}$. Let $U$ be the ordinary smooth error satisfying \eqref{def:os} of order $\beta$, 	under Assumptions \ref{Assumption:boundedsupport}-\ref{Assumption:PartialBoundAway} and OS,  we have
		\begin{align*}
			\widehat{\theta}_{\tau}(x)-\theta_{\tau}(x) = &\mathcal{B}_n\{\theta_{\tau}(x);x,\tau\} + \mathcal{V}_n\{\theta_{\tau}(x);x,\tau\}\\
			& + o_P\left(h^s\right) + O_P\left(\left\{ \frac{\log(1/h)}{{nh^{2\beta+1}}}\right\}^{3/4} \right) + O_P\left(R_{U,OS}\right),
		\end{align*}
		and
		\begin{align*}
			\widehat{\theta}^{(b)}_{\tau}(x)-\widehat{\theta}_{\tau}(x) = \mathcal{V}_n^{\chi}\{\theta_{\tau}(x);x,\tau\}
			 + o_P\left(h^s\right) + O_P\left(\left\{ \frac{\log(1/h)}{{nh^{2\beta+1}}}\right\}^{3/4} \right) + O_P\left(R_{U,OS}\right),
		\end{align*}
		where $o_P(\cdot)$ and $O_P(\cdot)$ hold uniformly over $(x,\tau)\in \mathcal{X}\times\mathcal{T}$, {$\mathcal{V}_n\{\theta_{\tau}(x);x,\tau\}$ and $\mathcal{V}_n^{\chi}\{\theta_{\tau}(x);x,\tau\}$ are mean zero with variance satisfying $ \inf_{x\in\mathcal{X},\tau\in\mathcal{T}} \sigma^2_n(x,\tau) \asymp \sup_{x\in\mathcal{X},\tau\in\mathcal{T}} \sigma^2_n(x,\tau) \asymp n^{-1}h^{-2\beta-1}$. Furthermore,} 
  $$\sup_{x\in\mathcal{X},\tau\in\mathcal{T}} |\mathcal{V}_n\{\theta_\tau(x); x,\tau\}| \quad \text{and} \quad \sup_{x\in\mathcal{X},\tau\in\mathcal{T}} |\mathcal{V}^{\chi}_n\{\theta_\tau(x); x,\tau\}| \lesssim\left\{ \frac{\log(1/h)}{{nh^{2\beta+1}}}\right\}^{1/2}\,.
$$
  Consequently, 
        $$\sup_{x\in\mathcal{X},\tau\in\mathcal{T}}\left|\widehat{\theta}_{\tau}(x)-\theta_{\tau}(x)\right| = O_P(h^s) + O_P\left(\left\{ \frac{\log(1/h)}{{nh^{2\beta+1}}}\right\}^{1/2} \right) + O_P(R_{U,OS})\,.$$\end{thm}
        The term $R_{U,OS}$ in Theorem \ref{thm:criteron_consistency} {captures} the effect of estimating $\phi_U$. If the density of $U$ is known, one can replace the $\widehat{K}_{U,h}$ in \eqref{def:repeatedestimator} with $K_{U,h}$, then the term $O_P(R_{U,OS})$ will vanish. {In that case, with} $h\asymp n^{-c}$ for some constant $c>0$, abusing the terminology, the ``variance part'' of the residual term has the convergence rate $\{\log n/((nh^{2\beta+1})\}^{3/4}$ uniformly in $(x,\tau)\in\mathcal{X}\times\mathcal{T}$. This rate has an additional $h^{-2\beta}$ factor to the best uniform rate for the remainder of the Bahadur representation of a local constant quantile regressor in the error-free case established in \cite{kong2010uniform}, which is $\{\log n/(nh)\}^{\lambda(\omega)}$ with $\lambda(\omega)=\min \left\{\frac{1}{\omega+1},\frac{3+2\omega}{4\omega+4}\right\}\leq \lambda(0) = 3/4$, for some $ \omega \geq 0$. They claimed that this result can not be further improved. In errors-in-variables regression {with a known measurement error distribution}, \cite{fan1993nonparametric} showed that the order of the variance part is $(nh^{2\beta+1})^{-1}$, also containing an extra factor $h^{-2\beta}$ compared to the error-free case. Based on these findings, we believe our result of the order of the remainder term cannot be further improved either. 

If we take $h\sim (n/\log n)^{-1/(2\beta+2s+1)}$ and suppose $m$ increases sufficiently fast such that $R_{U,OS}$ vanishes, then the $L^{\infty}$-convergence rate of $\widehat{\theta}_{\tau}(x)$ is $(n/\log n)^{-s/(2\beta+2s+1)}$. Compared to the pointwise minimax rate of the mean regression with known measurement error distribution in \cite{fan1993nonparametric}, which is $n^{-s/(2\beta+2s+1)}$, this uniform convergence rate has an additional factor of up to $\log n$. This uniform convergence rate is also new to the literature.

	\begin{rk}
		In the case of multiple covariates, i.e. $\mathbf{X}=(X_1,X_2,\ldots,X_d)^{\top}\in\mathbb{R}^d$ for $d\geq 2$, the vector of measurement error  $\mathbf{U}=(U_1,U_2,\ldots,U_d)^{\top}\in\mathbb{R}^d$ satisfies that $U_j$ is an ordinary smooth error of order $\beta$ for $j=1,\ldots,d$, then it can be similarly shown that optimal uniform convergence rate of our estimator $\widehat{\theta}_{\tau}(x)$ is of $(n/\log n)^{-2s/\{2s+(2\beta+1)d\}}$ given that the density function of every component of $\mathbf{X}$ is  $s$-smooth in H\"older sense.
	\end{rk}

\begin{rk}
While we assume the order of the kernel function and the smoothness of $\varepsilon_{\tau}$ and
$f_X$ are all $s$, our method remains valid for any order of the kernel function, $\alpha_K$, and any smoothness of $\varepsilon_{\tau}(x;\theta)$, $\alpha_\varepsilon$, and $f_X$, $\alpha_f$. In this more general setting, the estimation bias of our estimator will be of $O(h^
{\min\{\alpha_K,\alpha_\varepsilon,\alpha_f\}})$. For representation simplicity, we maintain the assumption that  $\alpha_K=\alpha_\varepsilon=\alpha_f\equiv s$, which is a common practice in the kernel regression literature (see e.g. \citealp{kong2010uniform,abrevaya2015estimating}). 
\end{rk}
    
	\subsection{Asymptotics under the Supersmooth Errors}\label{sec:ssbahadur}
	This section establishes the large sample properties for  $\widehat{\theta}_{\tau}(x)$ and $\widehat{\theta}^{(b)}_{\tau}(x)$ under supersmooth errors. Let $e_L(h):=h^{(\ell_K+1)\beta+\beta_0-0.5}\exp(h^{-\beta} /\gamma)$ and $e_U(h) := h^{-1/2}\{\log(1/h)\}^{\ell_K }e_L(h)$, where $\ell_k$ satisfies the following conditions.

\noindent{{\bf Assumption SS.}}\   (i) There exists positive constants $\delta_K$, $c_K$, $d_K$ $\ell_K$ such that $d_K\cdot(1-y)^{\ell_K}\le|\phi_K(y)|\le c_K\cdot(1-y)^{\ell_K}$, for $y\in(1-\delta_K,1)$; (ii) The real part $R_U(y)$ and imaginary part $I_U(y)$ of $\phi_U(y)$ satisfy $R_U(y) =o\{I_U(y)\}$ or $I_U(y)=o\{R_U(y)\}$ as $y\rightarrow \infty$. (iii) $h\rightarrow 0$, $n^{-1/2}e_U(h)\log (1/h)\rightarrow 0$; 
	
	Assumptions SS (i) and (ii) enable us to apply the upper and lower bound of the deconvolution kernel for supersmooth measurement error derived by \cite{fan1992multivariate}, which are used for the proof of the uniform Bahadur representation. These assumptions are satisfied by the commonly used kernel discussed below Assumption OS and the error distribution such as Gaussian, Cauchy, and Gaussian mixture. \citet[Lemma 3.2]{fan1992multivariate}  has showed for any $L_2$ integrable function $g$, $e_L^2(h)\lesssim\mathbb{E}\left\{K_{U,h}^2(x-W)g^2(Y_i)\right\}\lesssim e_U^2(h)$ when $U$ is a supersmooth measurement error of type \eqref{def:ss}. Note that $e_U(h)/e_L(h) =  h^{-1/2}\{\log(1/h)\}^{\ell_K }\rightarrow \infty$ for $h\rightarrow 0$, which indicates an unmatched upper and lower bound. One possible way to obtain an exact order is to employ the techniques in \cite{Supervar}, which proves the asymptotic normality of the deconvolution density estimator for supersmooth error. However, their assumption excludes the Cauchy error. To make our result suitable for general supersmooth measurement error, we apply the bounds in \cite{fan1992multivariate}. Assumption SS(iii) is required for our estimator to be consistent under the super smooth error.
	\begin{thm}\label{thm:criteron_consistency_supersmooth}
		Denote $R_{U,SS}=m^{-1/2}n^{-1/2}h^{2\beta_0-1}\exp(2h^{-\beta}/\gamma) + m^{-1/2}h^{\beta_0}\exp(h^{-\beta}/\gamma)$. Under Assumptions \ref{Assumption:boundedsupport}-\ref{Assumption:PartialBoundAway} and SS, $U$ is a supersmooth error satisfying \eqref{def:ss} of order $(\beta,\beta_0)$, then we have
		\begin{align*}
			\widehat{\theta}_{\tau}(x)-\theta_{\tau}(x) = &\mathcal{B}_n\{\theta_{\tau}(x);x,\tau\} + \mathcal{V}_n\{\theta_{\tau}(x);x,\tau\}\\
			&  + o_P\left(h^s\right)+ O_P\left(\left\{ \frac{e_U^2(h)\log(1/h)}{{n}}\right\}^{3/4}\right) + O_P\left(R_{U,SS}\right)
		\end{align*}
		and
		\begin{align*}
			\widehat{\theta}^{(b)}_{\tau}(x)-\widehat{\theta}_{\tau}(x) = \mathcal{V}_n^{\chi}\{\theta_{\tau}(x);x,\tau\}+ o_P\left(h^s\right)+ O_P\left(\left\{ \frac{e_U^2(h)\log(1/h)}{{n}}\right\}^{3/4}\right) + O_P\left(R_{U,SS}\right),
		\end{align*}
		uniformly over $(x,\tau)\in \mathcal{X}\times\mathcal{T}${, $\mathcal{V}_n\{\theta_{\tau}(x);x,\tau\}$ and $\mathcal{V}^{\chi}_n\{\theta_{\tau}(x);x,\tau\}$ are mean zero with variance satisfying
  $
  n^{-1} e_L^2(h) \lesssim \inf_{x\in\mathcal{X},\tau\in\mathcal{T}}\sigma_n^2(x,\tau) \lesssim \sup_{x\in\mathcal{X},\tau\in\mathcal{T}}\sigma_n^2(x,\tau) \lesssim n^{-1}e_U^2(h)\,.$
  } Furthermore,
  $$\sup_{x\in\mathcal{X},\tau\in\mathcal{T}} |\mathcal{V}_n\{\theta_\tau(x); x,\tau\}|\quad \text{and} \quad \sup_{x\in\mathcal{X},\tau\in\mathcal{T}} |\mathcal{V}^{\chi}_n\{\theta_\tau(x); x,\tau\}| \lesssim \left\{ \frac{e_U^2(h)\log(1/h)}{{n}}\right\}^{1/2} \,.
  $$
  Consequently,
  $$\sup_{x\in\mathcal{X},\tau\in\mathcal{T}}\left|\widehat{\theta}_{\tau}(x)-\theta_{\tau}(x)\right| = O_P(h^s) + O_P\left(\left\{ \frac{e_U^2(h)\log(1/h)}{{n}}\right\}^{1/2}\right) + O_P(R_{U,SS})\,.$$
	\end{thm}
	As in Theorem \ref{thm:criteron_consistency}, the residual term, $R_{U,SS}$ arises from estimating $\phi_U(t)$ by $\widehat{\phi}_U(t)$ for supersmooth errors. When $m$ is large enough such that $R_{U,SS}$ converges faster than $(\log n)^{-s/\beta}$, then the optimal uniform convergence rate of $\widehat{\theta}_{\tau}(x)$ is $(\log n)^{-s/\beta}$ when we take $h=C(\log n)^{-1/\beta}$ for some constant $C>(2/\gamma)^{1/\beta}$. This coincides with the minimax optimal uniform convergence rate derived in \cite{fan1993nonparametric} for nonparametric mean regression with errors following certain supersmooth distributions such as normal, Cauchy and their mixtures. 
    
 \begin{rk}
			In the case of multiple covariates, i.e. $\mathbf{X}=(X_1,X_2,\ldots,X_d)^{\top}\in\mathbb{R}^d$ for $d\geq 2$, the vector of measurement error $\mathbf{U}=(U_1,U_2,\ldots,U_d)^{\top}\in\mathbb{R}^d$ satisfies that $U_j$ is a supersmooth error with order $(\beta_{j},\beta_{0j})$ for $j=1,\ldots,d$, then it can be similarly shown that optimal uniform convergence rate of our estimator $\widehat{\theta}_{\tau}(x)$ is of $(\log n)^{-2 s/(\max_j \beta_j) }$ given that the density function of every component of $\mathbf{X}$ is $s$-smooth in H\"older sense.
	\end{rk}
	\subsection{Asymptotic Validity}\label{Section:Band}
	
	In this section, we establish the asymptotic validity of the uniform bands we proposed, which requires the following additional assumptions.
	
\begin{assumption}\label{Assumption:lowerbound}
$f_W(x)>0$ for all $x\in\mathcal{X}$.
	\end{assumption}\begin{assumption}\label{Assumption:VarianceContinuity} The characteristic function of $Y$, satisfies $\int |\phi_Y(t)|dt < \infty$;
	\end{assumption}
	
	\noindent{{\bf Assumption OS$'$.}}
	\begin{itemize}
		\item [(i)] $nh^{2s+2\beta+1}\log(1/h)\rightarrow 0$;
		\item [(ii)] $\{\log(1/h)\}^5/(nh^{2\beta+1})\rightarrow 0$;
		\item [(iii)]  $mh^{2\beta+2}\{\log(1/h)\}^{-2}\rightarrow \infty$ {and $nh\log(1/h)/m\rightarrow 0$};
		\item [(iv)] $n^{-1/2}h_Y^{-1/2}h_W^{-\beta-1/2}(\log\{1/(h_Y\land h_W)\})^{1/2} + m^{-1/2}n^{-1/2}h_W^{-2\beta-1}h_Y^{-1}+m^{-1/2}h_W^{-\beta} + h_Y^s + h_{W}^s  = o\left(\{\log(1/h)\}^{-1}\right)$
	\end{itemize}
	
	\noindent{{\bf Assumption SS$'$.}}
	\begin{itemize}
		\item [(i)] $nh^{2s}\log(1/h)e_L^{-2}(h)\rightarrow 0$;
		\item [(ii)] $e_U^2(h)\{\log(1/h)\}^{5+4\ell_k}/(nh^2)\rightarrow 0;$
		\item [(iii)] $mh^{2(\ell_k+1)\beta-2\beta_0+2}\{\log(1/h)\}^{-2\ell_k-2}\exp(-2h^{-\beta}/\gamma)\rightarrow \infty$ {and $n\log(1/h)/\left[mh^{2(\ell_k+1)\beta-1}\right]\rightarrow 0$;}
		\item [(iv)] $m^{-1/2}n^{-1/2}h^{2\beta_0-1}\exp(2h^{-\beta}/\gamma)+m^{-1/2}h^{\beta_0}\exp(h^{-\beta}/\gamma) = o\left(h\{\log(1/h)\}^{-2\ell_k-1}\right)$ and $$\sqrt{\frac{e_U^2(h_W)\log\{1/(h_Y\land h_W)\}}{nh_Y}} +h_Y^s + h_{W}^s=o\left(\frac{h}{\{\log(1/h)\}^{2\ell_k+1}}\right).$$
	\end{itemize}
	
	Assumption \ref{Assumption:lowerbound} {ensures that} the asymptotic variance of $\widehat{\theta}_{\tau}(x)$, {$\sigma^2_n(x,\tau)$, is uniformly bounded away from 0 over $(x,\tau)\in\mathcal{X}\times \mathcal{T}$}, with the formal proof {provided} in Lemma \ref{lemma:variancelowerboundos} in the supplementary material. {A sufficient condition of Assumption \ref{Assumption:lowerbound} is $f_U(u)>0$ for any $u\in\mathbb{R}$.}  Assumption \ref{Assumption:VarianceContinuity} {helps control the error rate from} estimating {$\sigma_n^2$ by} $\widehat{\sigma}_n^2$ for inference, which is analogous to Assumption \ref{Assumption:boundedsupport}(iii). {Under Assumptions~\ref{Assumption:lowerbound} and \ref{Assumption:VarianceContinuity}, we show the consistency of  $\widehat{\sigma}_n^2(x,\tau)$ and its convergence rate uniformly in $(x,\tau)\in\mathcal{X}\times \mathcal{T}$ in Lemma~\ref{thm:varianceconvergence} in the supplementary material.} Assumption OS$'$(i) is the undersmooth condition on the bandwidth $h$, for ordinary smooth, which is commonly used in the nonparametric inference literature to the bias induced by nonparametric estimation \citep[see e.g.][]{kato2018uniform,kato2019uniform,adusumilli2020inference}. It restricts the bandwidth to be not too large so that the bias converges faster to zero than the variance. Assumptions OS$'$(ii) is a technical condition imposed on the bandwidth for the validity of the uniform bands. It is slightly more restrictive than Assumptions OS. Assumption OS$'$(iii) restricts the rate of $m$ {relative} to $n$. Assumption OS$'$(iv) is a condition on $m,h_Y$ and $h_W$. It is imposed such that the variance estimator, $\widehat{\sigma}_n^2$, has a desired convergence rate for the validity of the uniform inference (see details in Lemma \ref{thm:varianceconvergence} in the supplementary material). Under some conditions on the bandwidth $h$ which we will discuss in the paragraph after next, Assumption OS$'$ does not exclude that case that $m =  n$. This makes the method applicable to a wide range of auxiliary sample scenarios, including the repeated measurement case discussed in Section 2.
 
 Assumptions SS$'$(i) to (iv) play the same role as Assumption OS$'$ (i) to (iv), respectively, for the supersmooth errors. {However}, Assumption SS$'$(iii) restricts $m$ to be at least of order $n(\log n)^{2(\ell_k+1)-{1}/\beta}$. This implies that Assumption SS$'$(iii) excludes the case {$m= n$, when $(\ell_k+1)\beta>{1/2}$, which, unfortunately, applies to the commonly assumed Gaussian measurement error scenario.} This unsatisfactory restriction is due to {a technical issue that has been recognized and remains an open problem in the literature \citep[see, e.g.,][]{kato2018uniform}. While this assumption may not always be realistic in practice, we believe our theoretical insights into uniform bands for supersmooth measurement errors remain valuable. Addressing this limitation is an important direction for future research.} 
 

In the following, we give a set of possible ($h$, $h_W$, $h_Y$, $m$) under Assumption OS and OS$'$. Suppose $U$ is an ordinary smooth error of order $\beta$ and $h_W$ and $h_Y$, {are chosen optimally such that} $h_W\asymp h_Y \asymp n^{-1/(2\beta+2s+2)}$.  Then one can choose $h\asymp c_n \cdot n^{-1/(2\beta+2s+1)}$ with $c_n\rightarrow 0$ {as the undersmoothing parameter}. The restriction on $m$ should then satisfy $m\gtrsim \left(c_n\cdot n^{1-\frac{1}{2\beta+2s+1}}\log n \right)\vee \left( c_n^{-2\beta-2} n^{\frac{2\beta+2}{2\beta+2s+1}}\log^2 n\right) \vee \left(n^{\frac{\beta}{\beta+s+1}\vee \frac{s}{\beta+s+1}} \log^2 n\right)$.  Accordingly, if $c_n > n^{\frac{1-2s}{(2\beta+2)(2\beta+2s+1)}}$, the sample size of auxiliary data $m\asymp n$ will satisfy these conditions. For the supersmooth case, due to the technical reason discussed above, we require $m\gtrsim n(\log n)^{r}$ with some $r>0$  to satisfy Assumption SS$'$(iii). In practice, the bandwidth selection $(h,h_W,h_Y)$, {especially the choice of $h$,} is crucial for the performance of uniform confidence bands. We discuss how to select $h$ and $(h_W,h_Y)$ in Section~\ref{Section:bandchoose} and Section~\ref{sec:hwhy} in the supplementary material, respectively. {Our numerical studies indicate that uniform inference with our chosen tuning parameters performs well for both ordinary smooth and supersmooth measurement errors when the auxiliary sample size is $m=n$.}
	\begin{thm}\label{thm:validity} Under Assumptions \ref{Assumption:boundedsupport}-\ref{Assumption:VarianceContinuity}, if (i) $U$ is an ordinary smooth error satisfying \eqref{def:os} of order $\beta$ and Assumptions OS, OS$'$ hold or (ii) $U$ is a supersmooth error satisfying \eqref{def:ss} of order $(\beta,\beta_0)$ and Assumptions SS, SS$'$ hold, we have
		\begin{align*}
			&\lim _{N \rightarrow \infty} \mathbb{P}\left(\theta_\tau(x) \in I_L,  \forall(x,\tau)\in\mathcal{X}\times\mathcal{T}\right)=1-\alpha,\\
			&\lim _{N \rightarrow \infty} \mathbb{P}\left(\theta_\tau(x) \in I_R, \ \forall(x,\tau)\in\mathcal{X}\times\mathcal{T}\right)=1-\alpha,\\
			&\lim _{N \rightarrow \infty} \mathbb{P}\left(\theta_\tau(x) \in I_2, \ \forall(x,\tau)\in\mathcal{X}\times\mathcal{T}\right)=1-\alpha.
		\end{align*}
	\end{thm}
	
	Our asymptotic results cover Cauchy error and other supersmooth measurement errors that \cite{bissantz2007non,kato2018uniform,kato2019uniform} do not cover. \cite{kato2018uniform} excludes Cauchy measurement errors from the supersmooth error to derive an exact rate of variance of their estimator. One of our contributions is to show the uniform band is still valid for Cauchy error, as discussed below in Assumption SS.

	\section{Numerical Study}\label{Section:Simulation}
	
	\subsection{Practical Implementation}\label{Section:bandchoose}
	The performance of the confidence band heavily relies on the choice of the undersmooth bandwidth, as discussed below in Assumption SS$'$. We adapt the method of \citep[][Section 5.2]{bissantz2007non} to the context of quantile regression.
	
	The procedure involves two steps: 1) Starting with an optimal bandwidth as the pilot bandwidth {$\widehat{h}_{opt}$}; 2) Finding $c\in(0,1)$ s.t. {$c\cdot \widehat{h}_{opt}$} satisfies the undersmoothing requirements. Then {$c\cdot \widehat{h}_{opt}$} is the desired bandwidth. 
 
 {For the first step, practitioners may use any suitable bandwidth selection method tailored to the measurement error setting, such as those proposed in \cite{dong2023bandwidth,delaigle2008using}. Specifically, we extend the approach of \cite{dong2023bandwidth} to the quantile regression context, which we thoroughly utilise in our simulation and empirical studies. Additionally, an alternative pilot bandwidth based on \cite{delaigle2008using} is detailed in Section~\ref{sec:SIMEX} in the supplementary material.} 
 
 {Let $\tau_0$ denote the midpoint of $\mathcal{T}$. To adapt the cross-validation (CV) method in \cite{dong2023bandwidth}, we replace the least-squares loss function with the quantile loss function and employ $J$-fold CV for efficiency. Splitting the data $(W_i, Y_i)$ into $J$ equal folds, let $I_j\subset\{1,2,\dots,n\}$ represent the index set of the $j$-th fold ($j = 1, \dots, J$), and $\widehat{\theta}_{-I_j,h}(x;\tau)$ denote the estimator in \eqref{def:repeatedestimator} based on the samples $\{W_i,Y_i\}_{i\notin I_j}$. The $J$-fold CV criterion is defined as 
\begin{align*}
 CV(h) =  \frac{1}{n}\sum_{j=1}^J\sum_{i\in I_j}\int_{\mathcal{X}} \mathcal{L}_{\tau_0}\left\{Y_j-\widehat{\theta}_{-I_j,h}(x;\tau_0)\right\} \widehat{K}_{U,b}(x-W_j)dx\,,
 \end{align*}
 where {$\mathcal{L}_{\tau}(x) = (\tau - \mathbbm{1}_{\{x<0\}})\cdot x$ is the check function}, $\widehat{K}_{U,b}$ is an estimated deconvolution kernel defined in \eqref{def:repeateddecon} using a separate bandwidth $b$, and $\mathcal{X}$ is a compact subset of $\mathbb{R}$. Here we use the loss function $\mathcal{L}_{\tau}$ instead of $\psi_\tau$ for improved stability.} 
 
 {In practice, \cite{dong2023bandwidth} use $b$ as proposed by \cite{delaigle2004practical}, who assume $\phi_U$ and $\text{Var}(U)$ are known. In our case, where these quantities are unknown, we substitute the sample characteristic function and sample variance of $\{U_j\}_{j=1}^m$ to calculate $b$. For $\mathcal{X}$, we take $\mathcal{X}=[\min(\{W_i\}_{i=1}^n),\max(\{W_i\}_{i=1}^n)].$}
 
	The second step is to find an undersmooth bandwidth for valid uniform inference. {Following \cite{bissantz2007non}, we start with an oversmooth bandwidth $\widehat{h}=\zeta\cdot \widehat{h}_{opt}$ where $\zeta>1$ is a pre-specified constant and $\widehat{h}_{opt}$ is the bandwidth we get from step 1.} Let $L$ be a specified constant. We calculate $\widehat{\theta}_\ell(\cdot,\tau)$ as $\widehat{\theta}_{\tau}(x)$ in \eqref{def:repeatedestimator} with bandwidth $h_\ell=(\ell/L)\widehat{h}$, for $\ell=1,\dots,L.$ We find the largest $k$ such that $\sup_{x\in\mathcal{X},\tau\in\mathcal{T}}|\widehat{\theta}_k(x,\tau) - \widehat{\theta}_{k-1}(x,\tau)| > \rho\sup_{x\in\mathcal{X},\tau\in\mathcal{T}}|\widehat{\theta}_L(x,\tau) - \widehat{\theta}_{L-1}(x,\tau)|$, where $\rho>1$ is a constant. Then, we set the undersmooth bandwidth $h = \max\{k/L,(\log n)^{-1}\} \widehat{h}.$ Here, we introduce $(\log n)^{-1}$ to avoid excessive undersmoothing. {Throughout our numerical studies, we set {$\zeta=1.3$, $L=20$, $\rho=3$.} The results behave well and stably. We provide further sensitivity analysis on $(L,\rho)$ in Section~\ref{appendix:sensitivity} in the supplementary material.}

    While the bandwidth selection method of \cite{bissantz2007non} has been commonly used in the literature of uniform inference with measurement error \citep[see e.g.][]{kato2018uniform,adusumilli2020inference}, to the best of our knowledge, there has been no any theoretical understanding of this method. It is an interesting future topic but beyond the scope of this paper.
	
	\subsection{Simulation Settings}

        \begin{figure}[t]
        \centering
        \includegraphics[width=\textwidth]{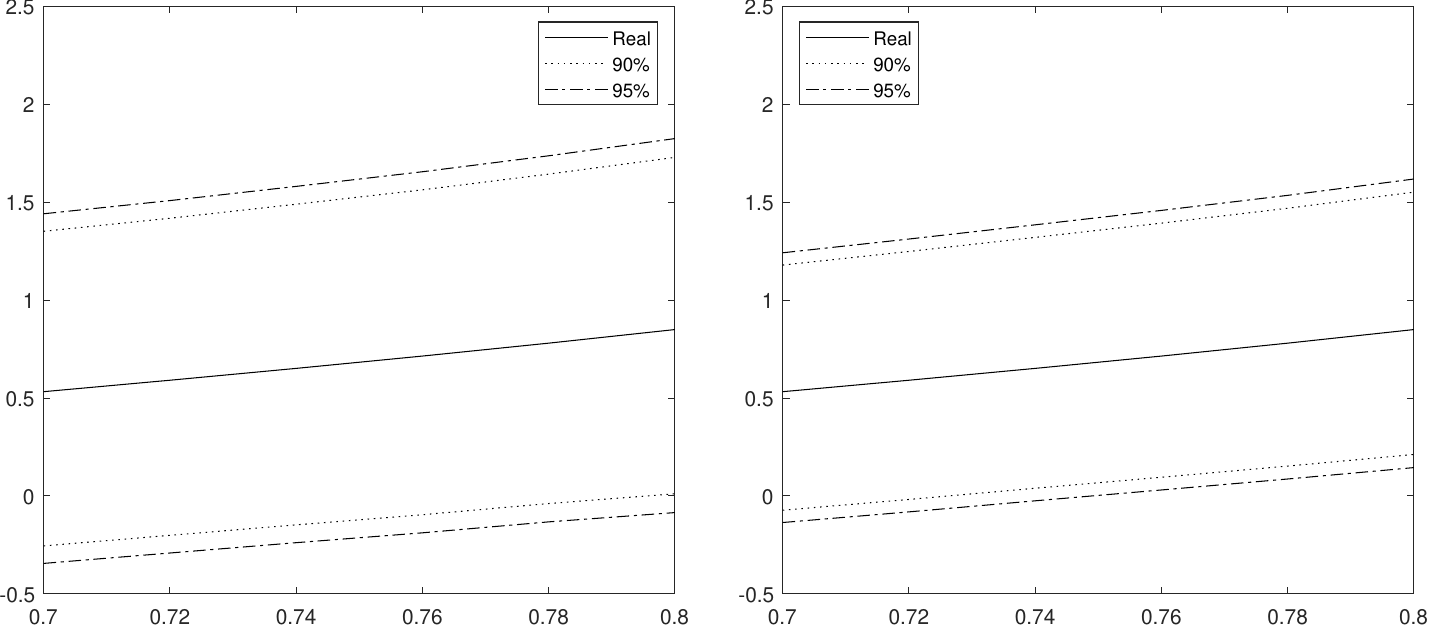}
        \caption{A slice taken at $x=8\times10^{-3}$ from the average of {1000} three-dimensional uniform confidence bands of $\theta_\tau(x)$ for $\tau\in\mathcal{T}=[0.7,0.8]$ and $x\in\mathcal{X} =[-0.8,0.8]$, calculated using {1000} random samples from DGP1 with Laplace error for $n=250$ (left) and $n=500$ (right).}
		\label{DGP1:Slicedtau}
        \end{figure}

        \begin{figure}[t]
        \centering
        \includegraphics[width=\textwidth]{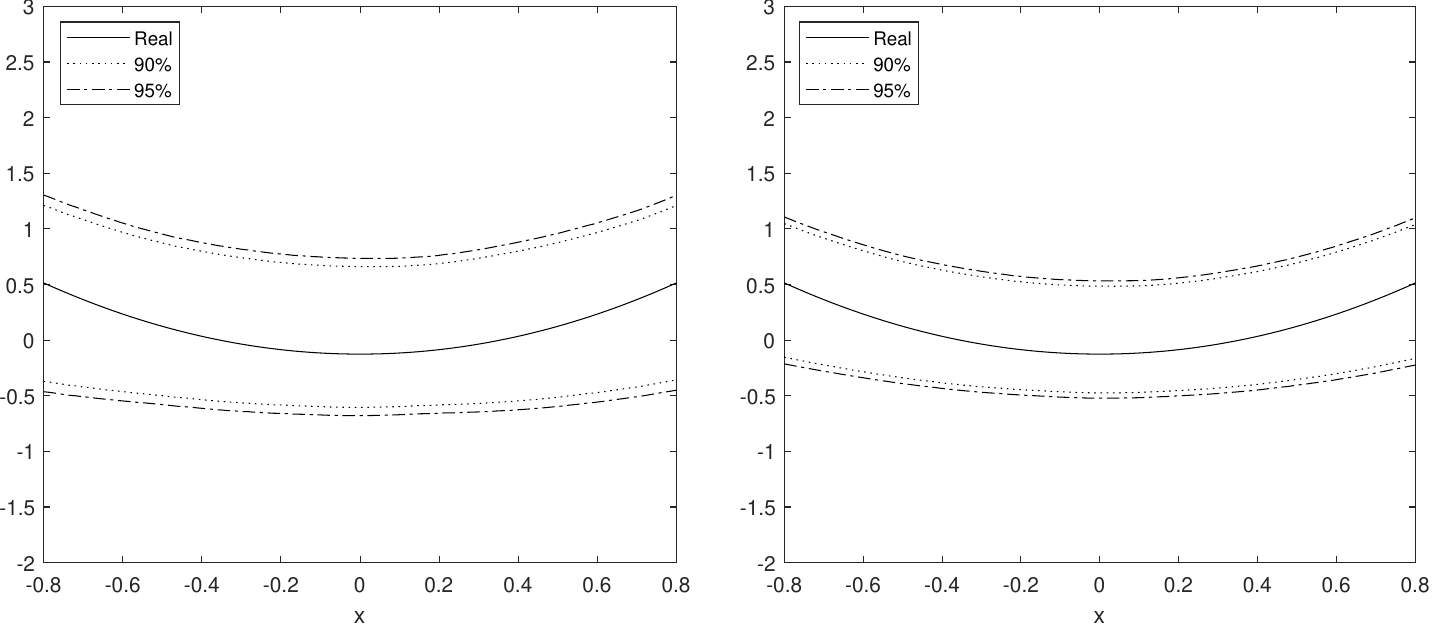}
        \caption{A slice taken at $\tau=0.45$ from the average of {1000} three-dimensional uniform confidence bands of $\theta_\tau(x)$ for $\tau\in\mathcal{T}=[0.45,0.55]$ and $x\in\mathcal{X}=[-0.8,0.8]$, calculated using {1000} random samples from DGP2 with normal error and $n=250$ (left) and $n=500$ (right).}
		\label{DGP2:Slicedx}
        \end{figure}
        
        \begin{figure}[t]
        \centering
        \includegraphics[width=\textwidth]{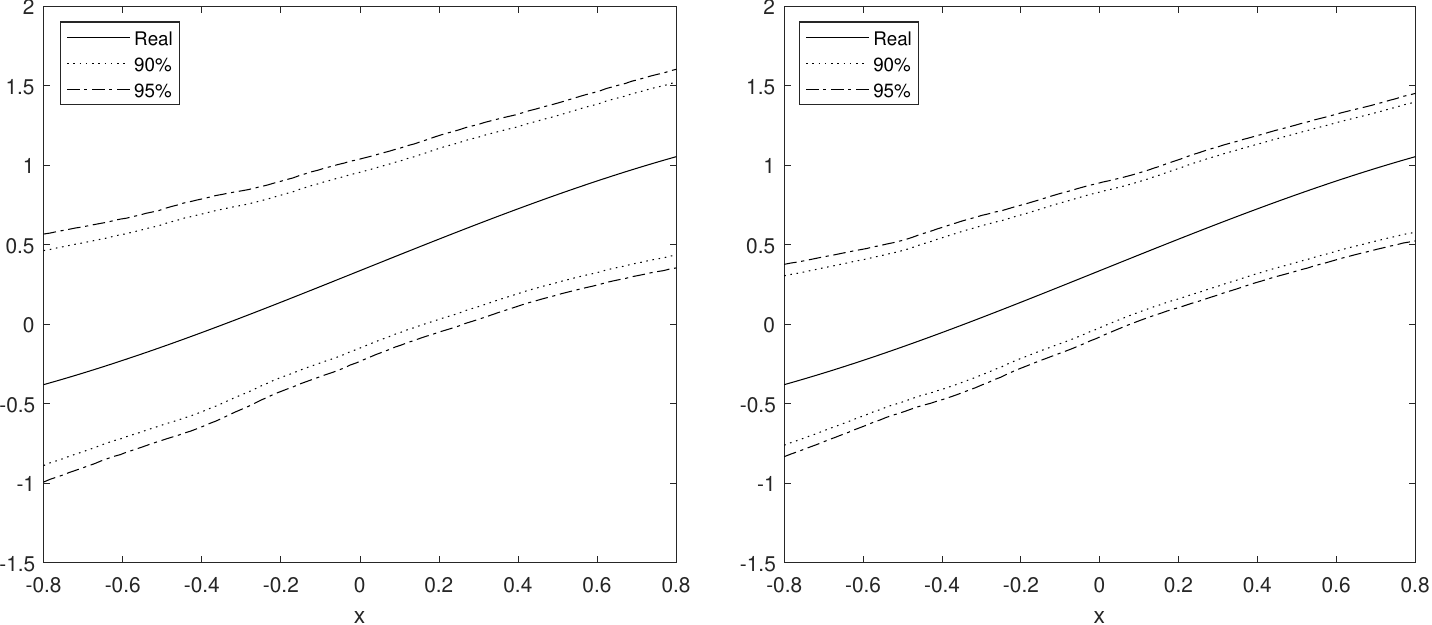}
        \caption{The average of {1000} uniform confidence bands of $\theta_\tau(x)$ for quantile $\tau=0.25$ and $x\in[-0.8,0.8]$, calculated using {1000} random samples from DGP3 with normal Error and sample sizes $n=250$ (left) and $n=500$ (right).}
		\label{fig:DGP3}
        \end{figure}

	Let $X\sim N(0,1)$ and $\epsilon_Y\sim N(0,1)$ be independent variables. We consider three different data generating processes:
	$$\text{DGP1:}\ Y = X + \epsilon_Y;\ \text{DGP2:}\ Y = X^2 + \epsilon_Y;\ \text{DGP3:}\ Y = \sin(X) + 0.5\epsilon_Y.$$
	We observe $Y$ and $W=X+U$ but not $X$. For each data generating process, we consider that $U$ is a Laplace (ordinary smooth) or a centered normal (supersmooth) random variable. For the Laplace error, we set the variance of $U$ by $\text{var}(U)/\text{var}(X)$ = 0.25. For the Normal error, we set the variance of $U$ by $\text{var}(U)/\text{var}(X)$ = 0.2.
	
	In each case, we generate {1000} samples with $n$ = 250 and 500. For each estimator, we construct the $(1-\alpha)$ uniform confidence band for the conditional quantile function $\theta_\tau(x)$ for $x\in\mathcal{X}=[-0.8,0.8]$ and (i) $\tau\in\mathcal{T}=\{0.25\}$, $\{0.5\}$ and $\{0.75\}$ and (ii)  $\tau\in\mathcal{T}=[0.2,0.3]$, $[0.45,0.55]$ and $[0.7,0.8]$. We take $\alpha = 0.1$ and 0.05. To measure the quality of the confidence band, we compute the empirical uniform coverage probability and the average size of the confidence bands. For case (i) where $\mathcal{T}$ is a singleton, the confidence bands are two-dimensional and the size is defined as the area of the band. For case (ii) where the confidence bands are three-dimensional, the size is defined as the volume. Unless otherwise specified, we use the kernel function $K(\cdot)$ for which $\phi_K(t)=(1-t^2)^3\mathbbm{1}_{[-1,1]}(t)$. For comparison, we also compute the pointwise confidence band of $\theta_{\tau}(\cdot)$ constructed from  $(1-\alpha)$  confidence intervals of $\theta_{\tau}(x)$ for each $x\in\mathcal{X}=[-0.8,0.8]$.
	
	\subsection{Simulation Results}
	Tables~\ref{tab:my-table1}, \ref{tab:my-table2} and \ref{tab:my-table3} report the performance of the uniform confidence bands for the three intervals of quantiles in case (ii), respectively. In Appendix \ref{appendix:singleton}, we report the performance of the uniform confidence bands for the three singleton quantiles in case (i). The empirical uniform coverage probability and the average sizes of the confidence bands from {1000} Monte Carlo samples are listed in the tables. We can see from the tables that all the uniform bands behave stable and well. Most empirical uniform coverage probabilities of our method are reasonably close to the corresponding nominal coverage probabilities, with the confidence bands getting narrower as the sample size increases. Another point worth mentioning is that the confidence bands for normal error have lower empirical uniform coverage probabilities in general than the ones for Laplace error, which reveals the difficulty of the inference for supersmooth error to some extent. Not surprisingly, by construction, the pointwise confidence bands are narrower but have a lower uniform coverage probability than our proposed uniform bands.
	
	Figures \ref{DGP1:Slicedtau}, \ref{DGP2:Slicedx} and \ref{fig:DGP3} provide some graphical explanation for the confidence bands in some typical settings. For the convenience of exposition, Figure \ref{DGP1:Slicedtau} shows how the uniform bands vary with $\tau$ for a fixed $x=8\times10^{-3}$, while Figure \ref{DGP2:Slicedx} shows how the curves vary with $x$ at a fixed $\tau=0.45$. {We also provide 3D visualization of our uniform confidence bands in Section~\ref{appendix:3D} of the supplementary material.} As we can see, with the sample sizes increasing, all the confidence bands in Figure \ref{DGP1:Slicedtau}-\ref{fig:DGP3} shrink fast. All the uniform bands can capture the shape of the underlying true conditional quantile function, along both $\tau$ and $x$. Furthermore, our uniform bands behave well for the relatively small sample size $n=250$ and the case of supersmooth Normal error.

\begin{table}
    \centering
    \caption{The empirical uniform coverage probability (ECP) and the average size (Avg. BS) of the uniform and pointwise confidence regions of $\theta_\tau(x)$ for $\tau\in[0.2,0.3]$ and $x\in[-0.8,0.8]$, calculated using {1000} random samples.}
    \label{tab:my-table1}
    \resizebox{.75\textwidth}{!}{
    \begin{tabular}{
        @{}
        l
        c
        c
        S[table-format=1.3]
        S[table-format=1.3]
        S[table-format=1.2]
        S[table-format=1.2]
        S[table-format=1.3]
        S[table-format=1.3]
        S[table-format=1.2]
        S[table-format=1.2]
        @{}
    }
    \toprule
    & & & \multicolumn{4}{c}{Uniform Confidence Band} & \multicolumn{4}{c}{Pointwise Confidence Band} \\
    \cmidrule(lr){4-7} \cmidrule(l){8-11}
    & & & \multicolumn{2}{c}{ECP} & \multicolumn{2}{c}{Avg. BS} & \multicolumn{2}{c}{ECP} & \multicolumn{2}{c}{Avg. BS} \\
    \cmidrule(lr){4-5} \cmidrule(lr){6-7} \cmidrule(lr){8-9} \cmidrule(l){10-11}
    \textbf{DGP} & {1-$\alpha$} & {n} & {Lap} & {Norm} & {Lap} & {Norm} & {Lap} & {Norm} & {Lap} & {Norm} \\
    \midrule

    \multirow{4}{*}{1} & \multirow{2}{*}{0.90} & 250 & 0.976 & 0.955 & 1.804 & 1.686 & 0.177 & 0.221 & 0.981 & 1.040 \\

    & & 500 & 0.963 & 0.945 & 1.322 & 1.217 & 0.164 & 0.194 & 0.780 & 0.810 \\

    \cmidrule{2-11}
    & \multirow{2}{*}{0.95} & 250 & 0.988 & 0.977 & 2.010 & 1.876 & 0.297 & 0.353 & 1.139 & 1.211 \\

    & & 500 & 0.982 & 0.969 & 1.458 & 1.342 & 0.294 & 0.290 & 0.905 & 0.943 \\

    \midrule
    \multirow{4}{*}{2} & \multirow{2}{*}{0.90} & 250 & 0.907 & 0.898 & 1.472 & 1.358 & 0.181 & 0.170 & 0.945 & 0.961 \\

    & & 500 & 0.916 & 0.867 & 1.148 & 1.025 & 0.105 & 0.079 & 0.747 & 0.761 \\

    \cmidrule{2-11}
    & \multirow{2}{*}{0.95} & 250 & 0.952 & 0.932 & 1.634 & 1.503 & 0.356 & 0.332 & 1.100 & 1.128 \\

    & & 500 & 0.954 & 0.919 & 1.261 & 1.129 & 0.227 & 0.171 & 0.868 & 0.895 \\

    \midrule
    \multirow{4}{*}{3} & \multirow{2}{*}{0.90} & 250 & 0.969 & 0.926 & 1.428 & 1.319 & 0.036 & 0.043 & 0.726 & 0.812 \\

    & & 500 & 0.963 & 0.867 & 1.029 & 0.943 & 0.037 & 0.045 & 0.589 & 0.671 \\

    \cmidrule{2-11}
    & \multirow{2}{*}{0.95} & 250 & 0.986 & 0.957 & 1.626 & 1.503 & 0.091 & 0.111 & 0.850 & 0.951 \\

    & & 500 & 0.980 & 0.920 & 1.147 & 1.055 & 0.087 & 0.099 & 0.690 & 0.784 \\

    \bottomrule
    \end{tabular}}
\end{table}

\begin{table}

    \centering
    \caption{The empirical uniform coverage probability (ECP) and the average size (Avg. BS) of the uniform and pointwise confidence regions of $\theta_\tau(x)$ for $\tau\in[0.45,0.55]$ and $x\in[-0.8,0.8]$, calculated using {1000} random samples.}
    \label{tab:my-table2}
    \resizebox{.75\textwidth}{!}{
    \begin{tabular}{%
        @{}
        l
        c
        S[table-format=3.0]
        S[table-format=1.3]
        S[table-format=1.3]
        S[table-format=1.2]
        S[table-format=1.2]
        S[table-format=1.3]
        S[table-format=1.3]
        S[table-format=1.2]
        S[table-format=1.2]
        @{}
    }
    \toprule
    & & & \multicolumn{4}{c}{Uniform Confidence Band} & \multicolumn{4}{c}{Pointwise Confidence Band} \\
    \cmidrule(lr){4-7} \cmidrule(l){8-11}
    & & & \multicolumn{2}{c}{ECP} & \multicolumn{2}{c}{Avg. BS} & \multicolumn{2}{c}{ECP} & \multicolumn{2}{c}{Avg. BS} \\
    \cmidrule(lr){4-5} \cmidrule(lr){6-7} \cmidrule(lr){8-9} \cmidrule(l){10-11}
    \textbf{DGP} & {1-$\alpha$} & {n} & {Lap} & {Norm} & {Lap} & {Norm} & {Lap} & {Norm} & {Lap} & {Norm} \\
    \midrule
    \multirow{4}{*}{1} & \multirow{2}{*}{0.90} & 250 & 0.961 & 0.949 & 1.590 & 1.480 & 0.239 & 0.239 & 0.890 & 0.953 \\
    & & 500 & 0.962 & 0.937 & 1.181 & 1.072 & 0.192 & 0.224 & 0.710 & 0.751 \\
    \cmidrule{2-11}
    & \multirow{2}{*}{0.95} & 250 & 0.978 & 0.970 & 1.776 & 1.655 & 0.374 & 0.383 & 1.041 & 1.119 \\
    & & 500 & 0.979 & 0.964 & 1.304 & 1.185 & 0.323 & 0.356 & 0.830 & 0.883 \\
    \midrule
    \multirow{4}{*}{2} & \multirow{2}{*}{0.90} & 250 & 0.950 & 0.924 & 1.525 & 1.371 & 0.197 & 0.212 & 0.974 & 1.033 \\
    & & 500 & 0.958 & 0.911 & 1.139 & 1.040 & 0.175 & 0.188 & 0.790 & 0.853 \\
    \cmidrule{2-11}
    & \multirow{2}{*}{0.95} & 250 & 0.977 & 0.956 & 1.711 & 1.532 & 0.335 & 0.358 & 1.142 & 1.231 \\
    & & 500 & 0.981 & 0.954 & 1.252 & 1.144 & 0.301 & 0.330 & 0.926 & 1.015 \\
    \midrule
    \multirow{4}{*}{3} & \multirow{2}{*}{0.90} & 250 & 0.975 & 0.940 & 1.223 & 1.153 & 0.063 & 0.076 & 0.645 & 0.738 \\
    & & 500 & 0.963 & 0.882 & 0.841 & 0.791 & 0.064 & 0.086 & 0.516 & 0.611 \\
    \cmidrule{2-11}
    & \multirow{2}{*}{0.95} & 250 & 0.981 & 0.967 & 1.405 & 1.324 & 0.129 & 0.153 & 0.764 & 0.875 \\
    & & 500 & 0.983 & 0.935 & 0.941 & 0.887 & 0.123 & 0.162 & 0.612 & 0.728 \\
    \bottomrule
    \end{tabular}}
\end{table}

\begin{table}
    \centering
    \small
    \caption{The empirical uniform coverage probability (ECP) and the average size (Avg. BS) of the uniform and pointwise confidence regions of $\theta_\tau(x)$ for $\tau\in[0.7,0.8]$ and $x\in[-0.8,0.8]$, calculated using {1000} random samples.}
    \label{tab:my-table3}
    \resizebox{.75\textwidth}{!}{
    \begin{tabular}{
        @{}
        l
        c
        c
        S[table-format=1.3]
        S[table-format=1.3]
        S[table-format=1.2]
        S[table-format=1.2]
        S[table-format=1.3]
        S[table-format=1.3]
        S[table-format=1.2]
        S[table-format=1.2]
        @{}
    }
    \toprule
    & & & \multicolumn{4}{c}{Uniform Confidence Band} & \multicolumn{4}{c}{Pointwise Confidence Band} \\
    \cmidrule(lr){4-7} \cmidrule(l){8-11}
    & & & \multicolumn{2}{c}{ECP} & \multicolumn{2}{c}{Avg. BS} & \multicolumn{2}{c}{ECP} & \multicolumn{2}{c}{Avg. BS} \\
    \cmidrule(lr){4-5} \cmidrule(lr){6-7} \cmidrule(lr){8-9} \cmidrule(l){10-11}
    \textbf{DGP} & {1-$\alpha$} & {n} & {Lap} & {Norm} & {Lap} & {Norm} & {Lap} & {Norm} & {Lap} & {Norm} \\
    \midrule

    \multirow{4}{*}{1} & \multirow{2}{*}{0.90} & 250 & 0.966 & 0.953 & 1.753 & 1.639 & 0.203 & 0.207 & 0.972 & 1.016 \\

    & & 500 & 0.967 & 0.945 & 1.351 & 1.220 & 0.185 & 0.212 & 0.784 & 0.809 \\

    \cmidrule{2-11}
    & \multirow{2}{*}{0.95} & 250 & 0.982 & 0.977 & 1.947 & 1.825 & 0.326 & 0.331 & 1.132 & 1.186 \\

    & & 500 & 0.980 & 0.968 & 1.486 & 1.344 & 0.305 & 0.332 & 0.909 & 0.941 \\

    \midrule
    \multirow{4}{*}{2} & \multirow{2}{*}{0.90} & 250 & 0.977 & 0.972 & 1.982 & 1.876 & 0.179 & 0.224 & 1.226 & 1.355 \\

    & & 500 & 0.971 & 0.952 & 1.428 & 1.309 & 0.127 & 0.179 & 0.978 & 1.081 \\

    \cmidrule{2-11}
    & \multirow{2}{*}{0.95} & 250 & 0.990 & 0.982 & 2.220 & 2.122 & 0.290 & 0.334 & 1.435 & 1.608 \\

    & & 500 & 0.991 & 0.972 & 1.572 & 1.444 & 0.207 & 0.281 & 1.140 & 1.281 \\

    \midrule
    \multirow{4}{*}{3} & \multirow{2}{*}{0.90} & 250 & 0.954 & 0.919 & 1.370 & 1.240 & 0.043 & 0.061 & 0.718 & 0.797 \\

    & & 500 & 0.967 & 0.878 & 1.035 & 0.951 & 0.043 & 0.046 & 0.587 & 0.665 \\

    \cmidrule{2-11}
    & \multirow{2}{*}{0.95} & 250 & 0.982 & 0.954 & 1.550 & 1.406 & 0.098 & 0.114 & 0.840 & 0.932 \\

    & & 500 & 0.985 & 0.918 & 1.159 & 1.067 & 0.076 & 0.095 & 0.687 & 0.780 \\

    \bottomrule
    \end{tabular}}
\end{table}
 \color{black}
	
	\section{Real Data Example}\label{sec:realdata}
	
	We apply our method to the National Health and Nutrition Examination Survey (NHANES) 2017-2018 dataset, which is available on \href{https://www.cdc.gov/nchs/nhanes/index.htm}{https://www.cdc.gov/nchs/nhanes/index.htm}. We propose to examine the relationship between the blood pressure (BP) and the long-term saturated fatty acids intake (SFA). In this dataset, the nutrition intake was measured by a simple 24-hour dietary recall. We consider the individuals aged from 20 to 40 in the dataset, including $n_1=531$ male and $n_2=618$ female. 
 
	In the dataset, we have two repeated measurements on the daily saturated fatty acids intake for each individual, which we denote as $W^{(1)}$ and $W^{(2)}$. {As discussed in the introduction of \cite{camirand2022semiparametric}, the 24-hour dietary recall data, including our data $W^{(1)}$ and $W^{(2)}$, often do not satisfy the classical measurement error assumption; however, it is common to assume that their log transformations do. In particular, the log transformation, $\widetilde{W}^{(j)}=\log(W^{(j)}+5)$, for $j=1,2$, is frequently used to make the observed values approximately normally distributed \citep[see e.g.][Section 4.3]{carroll2006measurement}. Under this transformation, we assume $\widetilde{W}^{(j)} = \widetilde{X} + U^{(j)}$, where $\widetilde{X}=\log(\text{long term saturated fatty acids intake}+5)$ with $U^{(1)}$ and $U^{(2)}$  independent of $\widetilde{X}$. This allows us to adapt the repeated measurement error framework to our framework as described in Section~\ref{Section:Setup}.} We define $\widetilde{W}=(\widetilde{W}^{(1)}+\widetilde{W}^{(2)})/2$ and $U=(\widetilde{W}^{(1)}-\widetilde{W}^{(2)})/2$, then $\widetilde{W} = \widetilde{X} + (U_1+U_2)/2 \stackrel{d}{=}X^{'}+U.$ In this way, we have observations of the distribution $U$, i.e., $\{\widetilde{W}_{1,i}-\widetilde{W}_{2,i}\}_{i=1}^n$.

{We use the method described in Section \ref{Section:bandchoose} to obtain the undersmooth bandwidth. The parameters for choosing bandwidth were set to $J=5,\zeta=1.3, D=30, L=20,\rho=3$, consistent with those used in the simulation.}
	\begin{figure}[t]
		\centering
		\begin{subfigure}[b]{0.32\textwidth}
			\includegraphics[width=\textwidth]{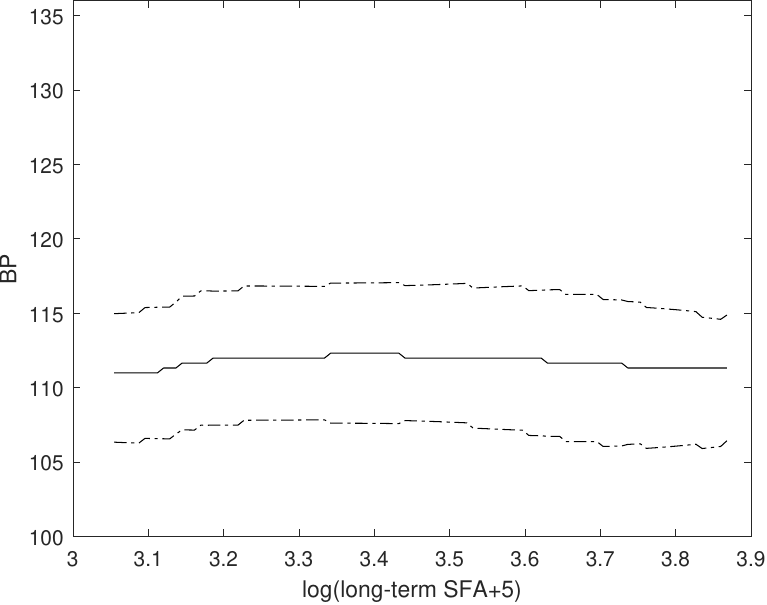}
		\end{subfigure}
		\hfill
		\begin{subfigure}[b]{0.32\textwidth}
			\includegraphics[width=\textwidth]{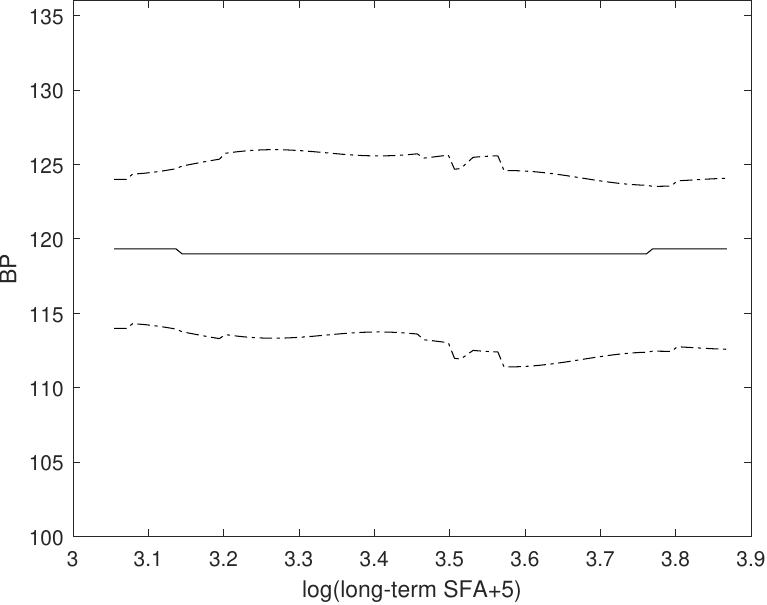}
		\end{subfigure}
		\hfill
		\begin{subfigure}[b]{0.32\textwidth}
			\includegraphics[width=\textwidth]{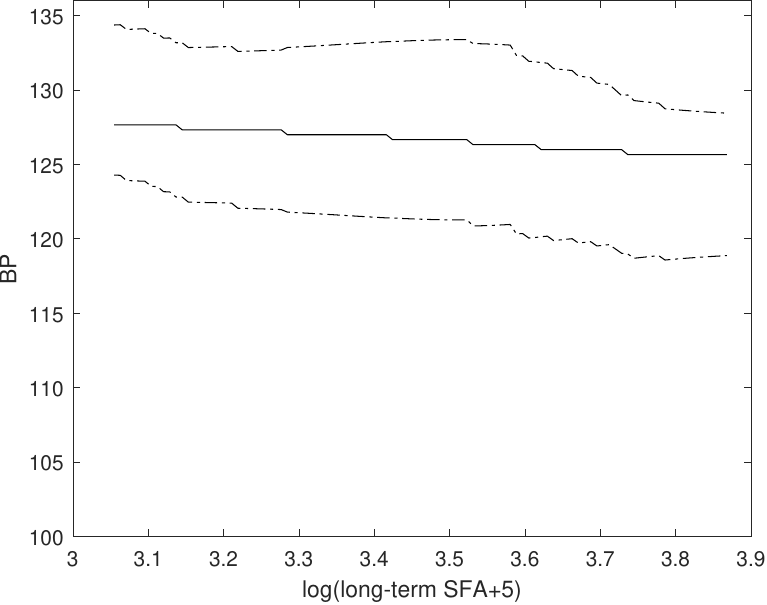}
		\end{subfigure}
		\caption{95\% Uniform Confidence Bands for Quantile $\mathcal{T}=\{0.25\}$ (left), $\{0.5\}$ (center) and $\{0.75\}$ (right) for men aged 20 to 40.}
		\label{figure:realdataman2040}
	\end{figure}
	\begin{figure}[t]
		\centering
		\begin{subfigure}[b]{0.32\textwidth}
			\includegraphics[width=\textwidth]{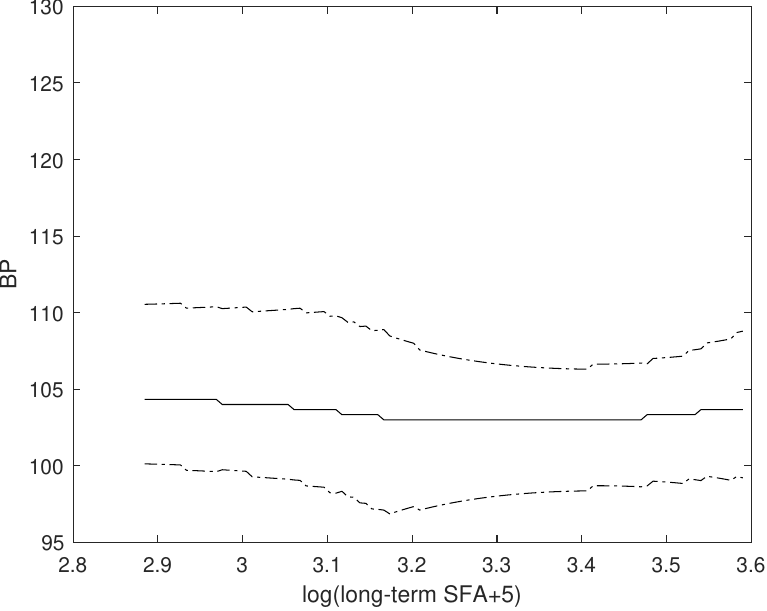}
		\end{subfigure}
		\hfill
		\begin{subfigure}[b]{0.32\textwidth}
			\includegraphics[width=\textwidth]{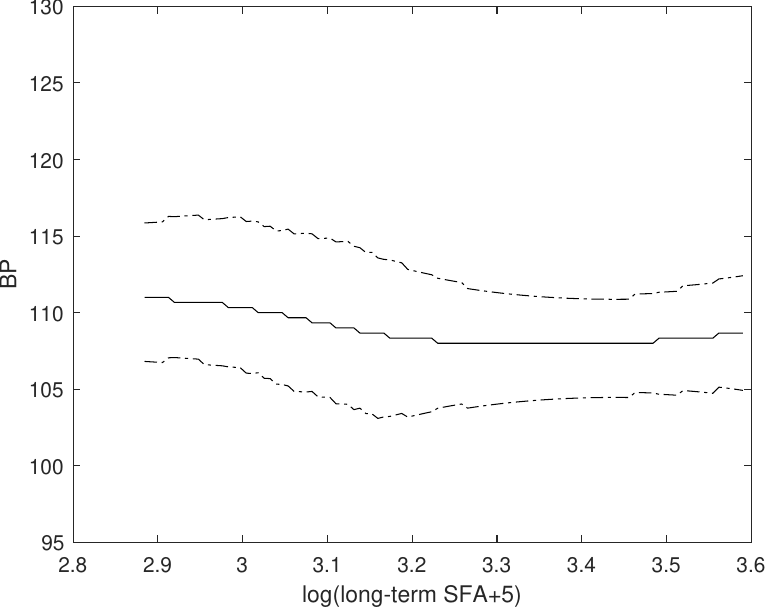}
		\end{subfigure}
		\hfill
		\begin{subfigure}[b]{0.32\textwidth}
			\includegraphics[width=\textwidth]{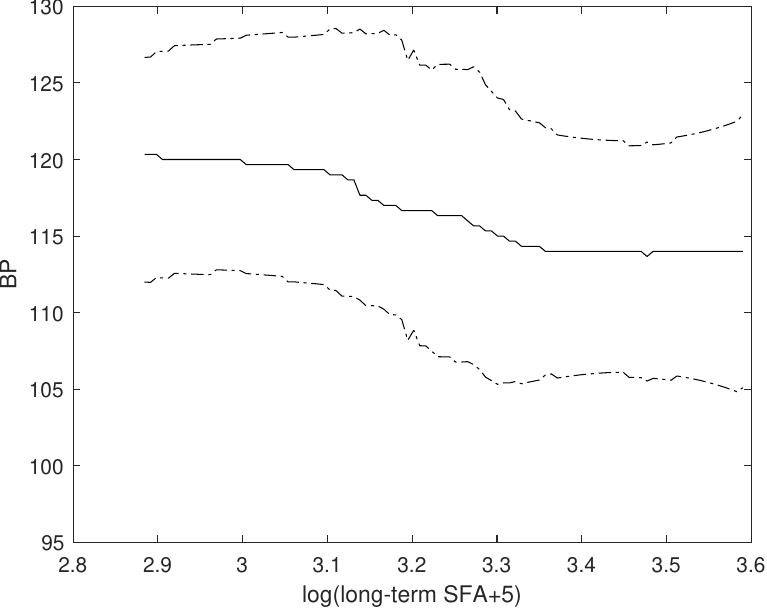}
		\end{subfigure}
		\caption{95\% Uniform Confidence Bands for Quantile $\mathcal{T}=\{0.25\}$ (left), $\{0.5\}$ (center) and $\{0.75\}$ (right) in women age 20 to 40.}
		\label{figure:realdatawoman2040}
	\end{figure}
	
	To eliminate the confounding issue arising from gender, we split the dataset into two groups: men aged from 20 to 40 and women aged from 20 to 40. We consider the uniform bands of conditional quantile curve for BP on long term SFA  intake with $\tau=0.25,0.5$ and $0.75$. The results of the two gender groups are shown in figures \ref{figure:realdataman2040} and \ref{figure:realdatawoman2040}, respectively. Figure \ref{figure:realdataman2040} shows for all the quantiles, there is a decreasing trend and the trend is significant when $\tau=0.25$ and $0.75$. Different from the male group, Figure \ref{figure:realdatawoman2040} shows that the conditional quantile curve for women first decreases and then increases with the SFA intake increasing. These results indicate that (i) the mechanism of how SFA intake affects BP may vary with gender; (ii) For different quantiles of BP, the effect of SFA intake is different, e.g. for 20-40 years old men with high BP (75\% quantile), larger SFA intake leads to more significant BP decreasing.
	
\section*{Acknowledgements}
We thank the editor Ivan Canay, an associate editor, and two referees for their constructive comments. The research of Zheng Zhang is
supported by the funds from the National Key R\&D Program of China [grant number 2022YFA1008300], the Fundamental Research Funds for the Central Universities, and the Research Funds of Renmin University of China [project number 23XNA025]. Haoze Hou acknowledges
the financial support from China Scholarship Council. Wei Huang's research is supported by the Faculty of Science Global Collaboration Awards 2023 at the University of Melbourne. 

\section*{Disclosure Statement}
The authors report there are no competing interests to declare.
 
	\bibliographystyle{chicago}
	
	\bibliography{my}

\begin{thebibliography}{}

\bibitem[\protect\citeauthoryear{Abrevaya, Hsu, and Lieli}{Abrevaya et~al.}{2015}]{abrevaya2015estimating}
Abrevaya, J., Y.-C. Hsu, and R.~P. Lieli (2015).
\newblock Estimating conditional average treatment effects.
\newblock {\em Journal of Business \& Economic Statistics\/}~{\em 33\/}(4), 485--505.

\bibitem[\protect\citeauthoryear{Adusumilli, Kurisu, Otsu, and Whang}{Adusumilli et~al.}{2020}]{adusumilli2020inference}
Adusumilli, K., D.~Kurisu, T.~Otsu, and Y.-J. Whang (2020).
\newblock Inference on distribution functions under measurement error.
\newblock {\em Journal of Econometrics\/}~{\em 215\/}(1), 131--164.

\bibitem[\protect\citeauthoryear{Bissantz, D{\"u}mbgen, Holzmann, and Munk}{Bissantz et~al.}{2007}]{bissantz2007non}
Bissantz, N., L.~D{\"u}mbgen, H.~Holzmann, and A.~Munk (2007).
\newblock Non-parametric confidence bands in deconvolution density estimation.
\newblock {\em Journal of the Royal Statistical Society Series B: Statistical Methodology\/}~{\em 69\/}(3), 483--506.

\bibitem[\protect\citeauthoryear{Camirand~Lemyre, Carroll, and Delaigle}{Camirand~Lemyre et~al.}{2022}]{camirand2022semiparametric}
Camirand~Lemyre, F., R.~J. Carroll, and A.~Delaigle (2022).
\newblock Semiparametric estimation of the distribution of episodically consumed foods measured with error.
\newblock {\em Journal of the American Statistical Association\/}~{\em 117\/}(537), 469--481.

\bibitem[\protect\citeauthoryear{Carroll, Ruppert, Stefanski, and Crainiceanu}{Carroll et~al.}{2006}]{carroll2006measurement}
Carroll, R.~J., D.~Ruppert, L.~A. Stefanski, and C.~M. Crainiceanu (2006).
\newblock {\em Measurement error in nonlinear models: a modern perspective}.
\newblock Chapman and Hall/CRC.

\bibitem[\protect\citeauthoryear{Chen, Linton, and Van~Keilegom}{Chen et~al.}{2003}]{chen2003estimation}
Chen, X., O.~Linton, and I.~Van~Keilegom (2003).
\newblock Estimation of semiparametric models when the criterion function is not smooth.
\newblock {\em Econometrica\/}~{\em 71\/}(5), 1591--1608.

\bibitem[\protect\citeauthoryear{Chernozhukov, Chetverikov, and Kato}{Chernozhukov et~al.}{2014a}]{chernozhukov2014anti}
Chernozhukov, V., D.~Chetverikov, and K.~Kato (2014a).
\newblock {Anti-concentration and honest, adaptive confidence bands}.
\newblock {\em The Annals of Statistics\/}~{\em 42\/}(5), 1787 -- 1818.

\bibitem[\protect\citeauthoryear{Chernozhukov, Chetverikov, and Kato}{Chernozhukov et~al.}{2014b}]{chernozhukov2014gaussian}
Chernozhukov, V., D.~Chetverikov, and K.~Kato (2014b).
\newblock Gaussian approximation of suprema of empirical processes.
\newblock {\em The Annals of Statistics\/}~{\em 42\/}(4), 1564--1597.

\bibitem[\protect\citeauthoryear{Chernozhukov, Chetverikov, and Kato}{Chernozhukov et~al.}{2016}]{chernozhukov2016empirical}
Chernozhukov, V., D.~Chetverikov, and K.~Kato (2016).
\newblock Empirical and multiplier bootstraps for suprema of empirical processes of increasing complexity, and related gaussian couplings.
\newblock {\em Stochastic Processes and their Applications\/}~{\em 126\/}(12), 3632--3651.

\bibitem[\protect\citeauthoryear{Chesher}{Chesher}{2017}]{chesher2017understanding}
Chesher, A. (2017).
\newblock Understanding the effect of measurement error on quantile regressions.
\newblock {\em Journal of Econometrics\/}~{\em 200\/}(2), 223--237.

\bibitem[\protect\citeauthoryear{Chiang, Hsu, and Sasaki}{Chiang et~al.}{2019}]{chiang2019robust}
Chiang, H.~D., Y.-C. Hsu, and Y.~Sasaki (2019).
\newblock Robust uniform inference for quantile treatment effects in regression discontinuity designs.
\newblock {\em Journal of Econometrics\/}~{\em 211\/}(2), 589--618.

\bibitem[\protect\citeauthoryear{Delaigle, Fan, and Carroll}{Delaigle et~al.}{2009}]{delaigle2009design}
Delaigle, A., J.~Fan, and R.~J. Carroll (2009).
\newblock A design-adaptive local polynomial estimator for the errors-in-variables problem.
\newblock {\em Journal of the American Statistical Association\/}~{\em 104\/}(485), 348--359.

\bibitem[\protect\citeauthoryear{Delaigle and Gijbels}{Delaigle and Gijbels}{2004}]{delaigle2004practical}
Delaigle, A. and I.~Gijbels (2004).
\newblock Practical bandwidth selection in deconvolution kernel density estimation.
\newblock {\em Computational statistics \& data analysis\/}~{\em 45\/}(2), 249--267.

\bibitem[\protect\citeauthoryear{Delaigle and Gijbels}{Delaigle and Gijbels}{2006a}]{delaigle2006data}
Delaigle, A. and I.~Gijbels (2006a).
\newblock Data-driven boundary estimation in deconvolution problems.
\newblock {\em Computational statistics \& data analysis\/}~{\em 50\/}(8), 1965--1994.

\bibitem[\protect\citeauthoryear{Delaigle and Gijbels}{Delaigle and Gijbels}{2006b}]{delaigle2006estimation}
Delaigle, A. and I.~Gijbels (2006b).
\newblock Estimation of boundary and discontinuity points in deconvolution problems.
\newblock {\em Statistica Sinica\/}~{\em 16\/}(3), 773--788.

\bibitem[\protect\citeauthoryear{Delaigle and Hall}{Delaigle and Hall}{2006}]{delaigle2006optimal}
Delaigle, A. and P.~Hall (2006).
\newblock On optimal kernel choice for deconvolution.
\newblock {\em Statistics \& Probability Letters\/}~{\em 76\/}(15), 1594--1602.

\bibitem[\protect\citeauthoryear{Delaigle and Hall}{Delaigle and Hall}{2008}]{delaigle2008using}
Delaigle, A. and P.~Hall (2008).
\newblock Using simex for smoothing-parameter choice in errors-in-variables problems.
\newblock {\em Journal of the American Statistical Association\/}~{\em 103\/}(481), 280--287.

\bibitem[\protect\citeauthoryear{Delaigle and Hall}{Delaigle and Hall}{2016}]{delaigle2016methodology}
Delaigle, A. and P.~Hall (2016).
\newblock Methodology for non-parametric deconvolution when the error distribution is unknown.
\newblock {\em Journal of the Royal Statistical Society: Series B (Statistical Methodology)\/}~{\em 78\/}(1), 231--252.

\bibitem[\protect\citeauthoryear{Delaigle, Hall, and Jamshidi}{Delaigle et~al.}{2015}]{delaigle2015confidence}
Delaigle, A., P.~Hall, and F.~Jamshidi (2015).
\newblock Confidence bands in non-parametric errors-in-variables regression.
\newblock {\em Journal of the Royal Statistical Society: Series B (Statistical Methodology)\/}~{\em 77\/}(1), 149--169.

\bibitem[\protect\citeauthoryear{Delaigle, Hall, and Meister}{Delaigle et~al.}{2008}]{delaigle2008deconvolution}
Delaigle, A., P.~Hall, and A.~Meister (2008).
\newblock On deconvolution with repeated measurements.
\newblock {\em The Annals of Statistics\/}~{\em 36\/}(2), 665--685.

\bibitem[\protect\citeauthoryear{Delaigle and Meister}{Delaigle and Meister}{2011}]{delaigle2011nonparametric}
Delaigle, A. and A.~Meister (2011).
\newblock Nonparametric function estimation under fourier-oscillating noise.
\newblock {\em Statistica Sinica\/}~{\em 21\/}(3), 1065--1092.

\bibitem[\protect\citeauthoryear{Dette and Volgushev}{Dette and Volgushev}{2008}]{dette2008non}
Dette, H. and S.~Volgushev (2008).
\newblock Non-crossing non-parametric estimates of quantile curves.
\newblock {\em Journal of the Royal Statistical Society Series B: Statistical Methodology\/}~{\em 70\/}(3), 609--627.

\bibitem[\protect\citeauthoryear{Diggle and Hall}{Diggle and Hall}{1993}]{diggle1993fourier}
Diggle, P.~J. and P.~Hall (1993).
\newblock A fourier approach to nonparametric deconvolution of a density estimate.
\newblock {\em Journal of the Royal Statistical Society Series B: Statistical Methodology\/}~{\em 55\/}(2), 523--531.

\bibitem[\protect\citeauthoryear{Donald and Hsu}{Donald and Hsu}{2014}]{donald2014estimation}
Donald, S.~G. and Y.-C. Hsu (2014).
\newblock Estimation and inference for distribution functions and quantile functions in treatment effect models.
\newblock {\em Journal of Econometrics\/}~{\em 178}, 383--397.

\bibitem[\protect\citeauthoryear{Dong, Otsu, and Taylor}{Dong et~al.}{2023}]{dong2023bandwidth}
Dong, H., T.~Otsu, and L.~Taylor (2023).
\newblock Bandwidth selection for nonparametric regression with errors-in-variables.
\newblock {\em Econometric Reviews\/}~{\em 42\/}(4), 393--419.

\bibitem[\protect\citeauthoryear{Es and Uh}{Es and Uh}{2005}]{Supervar}
Es, B.~V. and H.-W. Uh (2005).
\newblock Asymptotic normality of kernel-type deconvolution estimators.
\newblock {\em Scandinavian Journal of Statistics\/}~{\em 32\/}(3), 467--483.

\bibitem[\protect\citeauthoryear{Fan}{Fan}{1991}]{fan1991asymptotic}
Fan, J. (1991).
\newblock Asymptotic normality for deconvolution kernel density estimators.
\newblock {\em Sankhy{\=a}: The Indian Journal of Statistics, Series A\/}~{\em 53\/}(1), 97--110.

\bibitem[\protect\citeauthoryear{Fan, Hu, and Truong}{Fan et~al.}{1994}]{fan1994robust}
Fan, J., T.-C. Hu, and Y.~K. Truong (1994).
\newblock Robust non-parametric function estimation.
\newblock {\em Scandinavian Journal of Statistics\/}~{\em 21\/}(4), 433--446.

\bibitem[\protect\citeauthoryear{Fan and Masry}{Fan and Masry}{1992}]{fan1992multivariate}
Fan, J. and E.~Masry (1992).
\newblock Multivariate regression estimation with errors-in-variables: asymptotic normality for mixing processes.
\newblock {\em Journal of Multivariate Analysis\/}~{\em 43\/}(2), 237--271.

\bibitem[\protect\citeauthoryear{Fan and Truong}{Fan and Truong}{1993}]{fan1993nonparametric}
Fan, J. and Y.~K. Truong (1993).
\newblock Nonparametric regression with errors in variables.
\newblock {\em The Annals of Statistics\/}~{\em 21\/}(4), 1900--1925.

\bibitem[\protect\citeauthoryear{Firpo}{Firpo}{2007}]{firpo2007efficient}
Firpo, S. (2007).
\newblock Efficient semiparametric estimation of quantile treatment effects.
\newblock {\em Econometrica\/}~{\em 75\/}(1), 259--276.

\bibitem[\protect\citeauthoryear{Firpo, Galvao, and Song}{Firpo et~al.}{2017}]{firpo2017measurement}
Firpo, S., A.~F. Galvao, and S.~Song (2017).
\newblock Measurement errors in quantile regression models.
\newblock {\em Journal of Econometrics\/}~{\em 198\/}(1), 146--164.

\bibitem[\protect\citeauthoryear{Grace, Delaigle, and Gustafson}{Grace et~al.}{2021}]{grace2021handbook}
Grace, Y.~Y., A.~Delaigle, and P.~Gustafson (2021).
\newblock {\em Handbook of measurement error models}.
\newblock CRC Press.

\bibitem[\protect\citeauthoryear{Hall}{Hall}{1991}]{hall1991convergence}
Hall, P. (1991).
\newblock On convergence rates of suprema.
\newblock {\em Probability Theory and Related Fields\/}~{\em 89\/}(4), 447--455.

\bibitem[\protect\citeauthoryear{Hall and Meister}{Hall and Meister}{2007}]{hall2007ridge}
Hall, P. and A.~Meister (2007).
\newblock A ridge-parameter approach to deconvolution.
\newblock {\em Annals of Statistics\/}~{\em 35}, 1535--1558.

\bibitem[\protect\citeauthoryear{H{\"a}rdle and Song}{H{\"a}rdle and Song}{2010}]{hardle2010confidence}
H{\"a}rdle, W.~K. and S.~Song (2010).
\newblock Confidence bands in quantile regression.
\newblock {\em Econometric Theory\/}~{\em 26\/}(4), 1180--1200.

\bibitem[\protect\citeauthoryear{He and Liang}{He and Liang}{2000}]{he2000quantile}
He, X. and H.~Liang (2000).
\newblock Quantile regression estimates for a class of linear and partially linear errors-in-variables models.
\newblock {\em Statistica Sinica\/}~{\em 10\/}(1), 129--140.

\bibitem[\protect\citeauthoryear{He, Pan, Tan, and Zhou}{He et~al.}{2023}]{he2023smoothed}
He, X., X.~Pan, K.~M. Tan, and W.-X. Zhou (2023).
\newblock Smoothed quantile regression with large-scale inference.
\newblock {\em Journal of Econometrics\/}~{\em 232\/}(2), 367--388.

\bibitem[\protect\citeauthoryear{Hu and Schennach}{Hu and Schennach}{2008}]{hu2008instrumental}
Hu, Y. and S.~M. Schennach (2008).
\newblock Instrumental variable treatment of nonclassical measurement error models.
\newblock {\em Econometrica\/}~{\em 76\/}(1), 195--216.

\bibitem[\protect\citeauthoryear{Huang and Zhang}{Huang and Zhang}{2023}]{huang2023nonparametric}
Huang, W. and Z.~Zhang (2023).
\newblock Nonparametric estimation of the continuous treatment effect with measurement error.
\newblock {\em Journal of the Royal Statistical Society Series B: Statistical Methodology\/}~{\em 85\/}(2), 474--496.

\bibitem[\protect\citeauthoryear{Kato and Sasaki}{Kato and Sasaki}{2018}]{kato2018uniform}
Kato, K. and Y.~Sasaki (2018).
\newblock Uniform confidence bands in deconvolution with unknown error distribution.
\newblock {\em Journal of Econometrics\/}~{\em 207\/}(1), 129--161.

\bibitem[\protect\citeauthoryear{Kato and Sasaki}{Kato and Sasaki}{2019}]{kato2019uniform}
Kato, K. and Y.~Sasaki (2019).
\newblock Uniform confidence bands for nonparametric errors-in-variables regression.
\newblock {\em Journal of Econometrics\/}~{\em 213\/}(2), 516--555.

\bibitem[\protect\citeauthoryear{Kato, Sasaki, and Ura}{Kato et~al.}{2021}]{kato2021robust}
Kato, K., Y.~Sasaki, and T.~Ura (2021).
\newblock Robust inference in deconvolution.
\newblock {\em Quantitative Economics\/}~{\em 12\/}(1), 109--142.

\bibitem[\protect\citeauthoryear{Koenker and Bassett~Jr}{Koenker and Bassett~Jr}{1978}]{koenker1978regression}
Koenker, R. and G.~Bassett~Jr (1978).
\newblock Regression quantiles.
\newblock {\em Econometrica\/}~{\em 46\/}(1), 33--50.

\bibitem[\protect\citeauthoryear{Kong, Linton, and Xia}{Kong et~al.}{2010}]{kong2010uniform}
Kong, E., O.~Linton, and Y.~Xia (2010).
\newblock Uniform bahadur representation for local polynomial estimates of m-regression and its application to the additive model.
\newblock {\em Econometric Theory\/}~{\em 26\/}(5), 1529--1564.

\bibitem[\protect\citeauthoryear{Lin, Hsiao, and Hsu}{Lin et~al.}{2024}]{lin2024quantile}
Lin, H.-Y., Y.-H. Hsiao, and Y.-C. Hsu (2024).
\newblock Quantile policy effects: An application to us macroprudential policy.
\newblock {\em Journal of Business \& Economic Statistics\/}~{\em 0\/}(0), 1--17.

\bibitem[\protect\citeauthoryear{Luijken, Groenwold, Van~Calster, Steyerberg, and van Smeden}{Luijken et~al.}{2019}]{luijken2019impact}
Luijken, K., R.~H. Groenwold, B.~Van~Calster, E.~W. Steyerberg, and M.~van Smeden (2019).
\newblock Impact of predictor measurement heterogeneity across settings on the performance of prediction models: A measurement error perspective.
\newblock {\em Statistics in medicine\/}~{\em 38\/}(18), 3444--3459.

\bibitem[\protect\citeauthoryear{Meister}{Meister}{2009}]{Meister2009}
Meister, A. (2009).
\newblock {\em Deconvolution Problems in Nonparametric Statistics}.
\newblock Springer-Verlag Berlin Heidelberg.

\bibitem[\protect\citeauthoryear{Neumann and H{\"o}ssjer}{Neumann and H{\"o}ssjer}{1997}]{neumann1997effect}
Neumann, M.~H. and O.~H{\"o}ssjer (1997).
\newblock On the effect of estimating the error density in nonparametric deconvolution.
\newblock {\em Journal of Nonparametric Statistics\/}~{\em 7\/}(4), 307--330.

\bibitem[\protect\citeauthoryear{Schennach}{Schennach}{2004}]{schennach2004estimation}
Schennach, S.~M. (2004).
\newblock Estimation of nonlinear models with measurement error.
\newblock {\em Econometrica\/}~{\em 72\/}(1), 33--75.

\bibitem[\protect\citeauthoryear{Schennach}{Schennach}{2008}]{schennach2008quantile}
Schennach, S.~M. (2008).
\newblock Quantile regression with mismeasured covariates.
\newblock {\em Econometric Theory\/}~{\em 24\/}(4), 1010--1043.

\bibitem[\protect\citeauthoryear{Schennach}{Schennach}{2020}]{SCHENNACH2020487}
Schennach, S.~M. (2020).
\newblock Chapter 6 - mismeasured and unobserved variables.
\newblock In S.~N. Durlauf, L.~P. Hansen, J.~J. Heckman, and R.~L. Matzkin (Eds.), {\em Handbook of Econometrics, Volume 7A}, Volume~7 of {\em Handbook of Econometrics}, pp.\  487--565. Elsevier.

\bibitem[\protect\citeauthoryear{Scott}{Scott}{2015}]{scott2015multivariate}
Scott, D.~W. (2015).
\newblock {\em Multivariate density estimation: theory, practice, and visualization}.
\newblock John Wiley \& Sons.

\bibitem[\protect\citeauthoryear{Siddique, Daniels, Carroll, Raghunathan, Stuart, and Freedman}{Siddique et~al.}{2019}]{siddique2019measurement}
Siddique, J., M.~J. Daniels, R.~J. Carroll, T.~E. Raghunathan, E.~A. Stuart, and L.~S. Freedman (2019).
\newblock Measurement error correction and sensitivity analysis in longitudinal dietary intervention studies using an external validation study.
\newblock {\em Biometrics\/}~{\em 75\/}(3), 927--937.

\bibitem[\protect\citeauthoryear{Stefanski and Carroll}{Stefanski and Carroll}{1990}]{stefanski1990deconvolving}
Stefanski, L.~A. and R.~J. Carroll (1990).
\newblock Deconvolving kernel density estimators.
\newblock {\em Statistics\/}~{\em 21\/}(2), 169--184.

\bibitem[\protect\citeauthoryear{Takeuchi, Le, Sears, and Smola}{Takeuchi et~al.}{2006}]{takeuchi2006nonparametric}
Takeuchi, I., Q.~Le, T.~Sears, and A.~Smola (2006).
\newblock Nonparametric quantile estimation.
\newblock {\em Journal of Machine Learning Research\/}~{\em 7}, 1231–1264.

\bibitem[\protect\citeauthoryear{Tan, Wang, and Zhou}{Tan et~al.}{2022}]{tan2022high}
Tan, K.~M., L.~Wang, and W.-X. Zhou (2022).
\newblock High-dimensional quantile regression: Convolution smoothing and concave regularization.
\newblock {\em Journal of the Royal Statistical Society Series B: Statistical Methodology\/}~{\em 84\/}(1), 205--233.

\bibitem[\protect\citeauthoryear{van~der Vaart and Wellner}{van~der Vaart and Wellner}{1996}]{van1996weak}
van~der Vaart, A. and J.~Wellner (1996).
\newblock {\em Weak Convergence and Empirical Processes: With Applications to Statistics}.
\newblock Springer Series in Statistics. Springer.

\bibitem[\protect\citeauthoryear{Wang, Wu, and Li}{Wang et~al.}{2012}]{wang2012quantile}
Wang, L., Y.~Wu, and R.~Li (2012).
\newblock Quantile regression for analyzing heterogeneity in ultra-high dimension.
\newblock {\em Journal of the American Statistical Association\/}~{\em 107\/}(497), 214--222.

\bibitem[\protect\citeauthoryear{Wei and Carroll}{Wei and Carroll}{2009}]{wei2009quantile}
Wei, Y. and R.~J. Carroll (2009).
\newblock Quantile regression with measurement error.
\newblock {\em Journal of the American Statistical Association\/}~{\em 104\/}(487), 1129--1143.

\bibitem[\protect\citeauthoryear{Zhang, Kato, and Ruppert}{Zhang et~al.}{2023}]{zhang2023bootstrap}
Zhang, T., K.~Kato, and D.~Ruppert (2023).
\newblock Bootstrap inference for quantile-based modal regression.
\newblock {\em Journal of the American Statistical Association\/}~{\em 118\/}(541), 122--134.

\bibitem[\protect\citeauthoryear{Zheng, Peng, and He}{Zheng et~al.}{2015}]{zheng2015globally}
Zheng, Q., L.~Peng, and X.~He (2015).
\newblock Globally adaptive quantile regression with ultra-high dimensional data.
\newblock {\em Annals of Statistics\/}~{\em 43\/}(5), 2225.

\end{thebibliography}
	\clearpage

\newpage
\setcounter{page}{1}

\begin{center}
    {\bf \Large Supplementary Materials for ``Non-parametric Quantile Regression and Uniform Inference with Unknown Error Distribution''}
\end{center} 

\bigskip
	\appendix
 {\section{More Practical Details}}
\subsection{Selection of $h_W$ and $h_Y$}\label{sec:hwhy}
In this section, we discuss the bandwidth choice $(h_W,h_Y)$ aiming at an optimal bias-variance trade-off estimator $\widehat{f}_{X, Y}$ of the bivariate density $f_{X,Y}$.

  We start from the AMISE of $\widetilde{f}_{X,Y}$, defined as
  $$\widetilde{f}_{X,Y}(x,y):= \frac{1}{n}\sum_{i=1}^n K_{h_Y}\left(y-Y_i\right)K_{U,h_W}\left(x-W_i\right).$$
  As suggested by \cite{delaigle2008deconvolution}, one can use the AMISE of $\widetilde{f}_{X,Y}$ to approximate the AMISE of $\widehat{f}_{X,Y}.$ Our Lemma \ref{lemma:effectU} in the Section~\ref{appendix:proof} also shows the validity of this approximation.
  
  We know that
  \begin{align*}
  \int  \mathbb{E}\left\{\widetilde{f}_{X,Y}(x,y) - f_{X,Y}(x,y)\right\}^2 dxdy =&  \int  \mathbb{E}\left\{\widetilde{f}_{X,Y}(x,y) - \mathbb{E}\widetilde{f}_{X,Y}(x,y)\right\}^2 dxdy\\
  +&  \int  \left\{\mathbb{E}\widetilde{f}_{X,Y}(x,y) - f_{X,Y}(x,y)\right\}^2 dxdy\,.
  \end{align*}
  For the variance part,
  \begin{align*}
  &\int  \mathbb{E}\left\{\widetilde{f}_{X,Y}(x,y) - \mathbb{E}\widetilde{f}_{X,Y}(x,y)\right\}^2 dxdy\\
  &= \frac{1}{n}\left[\int  K_{h_Y}^2\left(y-Y_i\right)K_{U,h_W}^2\left(x-W_i\right)dxdy - \int \left\{\mathbb{E}K_{h_Y}\left(y-Y_i\right)K_{U,h_W}\left(x-W_i\right)\right\}^2 dxdy\right]\\
    &=\frac{1}{n}\int  \mathbb{E}K_{h_Y}^2\left(y-Y_i\right)K_{U,h_W}^2\left(x-W_i\right)dxdy + O\left(\frac{1}{n}\right)\\
    &=\frac{1}{n}\mathbb{E} \int K_{h_Y}^2\left(y-Y_i\right)K_{U,h_W}^2\left(x-W_i\right)dxdy + O\left(\frac{1}{n}\right)\\
    &=\frac{1}{nh_Yh_W}\int K^2(y)K_{U}^2(x)dxdy + O\left(\frac{1}{n}\right)\\
    &=\frac{1}{2\pi nh_Yh_W}\int K^2(y) dy\cdot \int \phi_K^2(t)/\phi_U^2(t/h_W)dt  +O\left(\frac{1}{n}\right).
  \end{align*}
  For the bias part,
  \begin{align*}
   &\int  \left\{\mathbb{E}\widetilde{f}_{X,Y}(x,y) - f_{X,Y}(x,y)\right\}^2 dxdy \\
   &= \frac{\kappa_{2,1}^2}{4}\int \left\{h_W^2  f_{11}(x,y) + h_Y^2 f_{22}(x,y)\right\}^2 dxdy\\
   &= \frac{\kappa_{2,1}^2}{4} \left\{ h_W^4 \int f_{11}^2(x,y) dxdy + h_Y^4 \int f_{22}^2(x,y) dxdy + 2h_W^2h_Y^2 \int f_{11}(x,y)f_{22}(x,y)dxdy\right\}.
  \end{align*}
  Using the bivariate rule-of-thumb idea proposed by \citet[][Page 163]{scott2015multivariate}, which approximates the above terms by assuming $(X,Y)$ follows a bivariate normal distribution with mean zero, variance $(\sigma_X^2,\sigma_Y^2)$ and correlation coefficient $\rho$, we have
  \begin{align*}
  &\int f_{11}^2(x,y) dxdy  = 3\left\{16\pi(1-\rho^2)^{5/2}\sigma_X^{5}\sigma_Y\right\}^{-1}\,,\\
  &\int f_{22}^2(x,y) dxdy  = 3\left\{16\pi(1-\rho^2)^{5/2}\sigma_Y^{5}\sigma_X\right\}^{-1}\,,\\
  &\int f_{11}(x,y)f_{22}(x,y)dxdy = \left(1+2\rho^2\right)\left\{16\pi(1-\rho^2)^{5/2}\sigma_X^{3}\sigma_X^{3}\right\}^{-1}.
  \end{align*}
  Combining the results and omitting the remainder, we have
  \begin{align*}
  \text{AMISE}(h_W,h_Y) =& \frac{1}{nh_Yh_W}\int K^2(y) dy\cdot \int \phi_K^2(t)/\phi_U^2(t/h_W)dt \\
  +&\frac{3\kappa_{2,1}^2}{64\pi(1-\rho^2)^{5/2}\sigma_X^{5}\sigma_Y}  h_W^4+\frac{3\kappa_{2,1}^2}{64\pi(1-\rho^2)^{5/2}\sigma_Y^{5}\sigma_X}  h_Y^4\\
  +&\frac{(1+2\rho^2)\kappa_{2,1}^2}{32\pi(1-\rho^2)^{5/2}\sigma_X^3\sigma_Y^{3}}h_W^2h_Y^2.
  \end{align*}
 Denote the sample variance of $W$, $Y$ and $U$ as $\widehat{\sigma}_W^2$, $\widehat{\sigma}_Y^2$ and $\widehat{\sigma}_U^2${, respectively,} and sample covariance of $W$ and $Y$ by $\widehat{\sigma}_{WY}$. We estimate the $\rho$ by $\widehat{\rho} = \widehat{\sigma}_{WY}/\left(\widehat{\sigma}_Y\sqrt{\widehat{\sigma}_{W}^2-\widehat{\sigma}_{U}^2}\right)$ and estimate $\sigma_X^2$ by $\widehat{\sigma}_X^2 = \widehat{\sigma}_W^2 - \widehat{\sigma}_U^2$. Finally, we define our sample version AMISE as
 \begin{align*}
  \widehat{\text{AMISE}}(h_W,h_Y) =& \frac{1}{2\pi nh_Yh_W}\int K^2(y) dy\cdot \int \phi_K^2(t)/\widehat{\phi}_U^2(t/h_W)dt \\
  +&\frac{3\kappa_{2,1}^2}{64\pi(1-\widehat{\rho}^2)^{5/2}\widehat{\sigma}_X^{5}\widehat{\sigma}_Y}  h_W^4+\frac{3\kappa_{2,1}^2}{64\pi(1-\widehat{\rho}^2)^{5/2}\widehat{\sigma}_Y^{5}\widehat{\sigma}_X}  h_Y^4\\
  +&\frac{(1+2\widehat{\rho}^2)\kappa_{2,1}^2}{32\pi(1-\widehat{\rho}^2)^{5/2}\widehat{\sigma}_X^3\widehat{\sigma}_Y^{3}}h_W^2h_Y^2.
  \end{align*}
  In practice, we choose the bandwidths $h_W$ and $h_Y$ to be the minimizer of $\widehat{\text{AMISE}}$.

\subsection{SIMEX bandwidth for quantile regression}\label{sec:SIMEX}
 We extend the SIMEX approach of \cite{delaigle2008using}, originally designed for mean regression, to the quantile regression context. To be specific, we generate two additional sets of samples, ${W}^{*,d}_{i} = {W}_i + {U}^{*,d}_{i}$ and ${W}^{**,d}_{i} = {W}^{*,d}_{i} + {U}^{**,d}_{i}$ where ${U}^{*,d}_{i}$ and ${U}^{**,d}_{i}$ are sampled independently with replacement from $\{U_j\}_{j=1}^m$ for $i = 1,2,\dots,n$ and $d=1,2,\dots,D$ with a sufficiently large constant $D$. For each $d$, we split the data $(W_i,W_i^{*,d},W_i^{**,d},Y_i)$ into $J$ equal size folds and denote $I_j\subset\{1,2,\dots,n\}$ as the index set of the $j$-th fold for $j=1,\dots,J$. Denote $\widehat{\theta}^{*,d}_{-I_j}(x;\tau)$ and $\widehat{\theta}^{**,d}_{-I_j}(x;\tau)$ as the estimator in \eqref{def:repeatedestimator} using the samples $\{W_i^{*,d},Y_i\}_{i\notin I_j}$ and $\{W_i^{**,d},Y_i\}_{i\notin I_j}$, respectively. Let $\tau_0$ be the middle point of $\mathcal{T}$. Then the $J$-fold cross-validations can be defined as
	$$CV^{*,d}(h) = \sum_{j=1}^{J}\sum_{i\in I_j} \mathcal{L}_{\tau_0}\left(Y_i-\widehat{\theta}^{*,d}_{-I_j}(W_i;\tau_0)\right)w(W_i)$$
	and
	$$CV^{**,d}(h) = \sum_{j=1}^{J}\sum_{i\in I_j} \mathcal{L}_{\tau_0}\left(Y_i-\widehat{\theta}^{**,d}_{-I_j}(W_i^{*,d};\tau_0)\right)w(W_i^{*,d}),$$
	where $\mathcal{L}_{\tau}(x) = (\tau - \mathbbm{1}_{\{x<0\}})\cdot x$ and $w(\cdot)$ is a pre-specified function avoiding the unstable behavior near the boundary. For example, $w(\cdot)$ can be the Gaussian kernel estimator of the density function of $W$, with the rule-of-thumb bandwidth $1.06\cdot \text{std}(W)\cdot n^{-1/5}$, where $\text{std}(W)$ is the sample standard deviation of $\{W_i\}_{i=1}^n$.
	Let $\widehat{h}^{*}$ and $\widehat{h}^{**}$ be the bandwidths which minimize $\sum_{d=1}^D CV^{*,d}(h)$ and $\sum_{d=1}^D CV^{**,d}(h)$, respectively. Since the role of ${W}^{**,d}_i$ to ${W}^{*,d}$ is same as that of ${W}^{*,d}$ to ${W}$, intuitively, we choose the bandwidth
	$$\widehat{h} =\left(\widehat{h}^{**}\right)^2/\widehat{h}^{*}$$
	as our pilot bandwidth. As asserted by \cite{delaigle2008using}, such bandwidth is asymptotically optimal in terms of minimizing the MSE of $\widehat{\theta}_{\tau}$. 

 \subsection{Deconvolution estimators take real values}\label{appendix:realvalue}
 Both $\widehat{K}_{U,h}$ and $\widehat{f}_{X, Y}$ are real, for any bandwidth $h$ and $n$. We prove the result for $\widehat{K}_{U,h}$, and a similar argument can be applied to prove that for $\widehat{f}_{X, Y}$.  Since the pre-specified kernel $K$ is symmetric (see Assumption~3), then the characteristic function of $K$, $\phi_K$, is always real, i.e. $\phi_K = \overline{\phi_K}$. The conjugate of $\widehat{K}_{U,h}$ is given by 
\begin{align*}
\overline{\widehat{K}_{U,h}(x)}=&\frac{1}{2\pi h}\overline{ \int_{-\infty}^{+\infty} \exp(-itx/h) \frac{{\phi_K(t)}}{{\widehat{\phi}_{U}(t/h)}} dt}\\
=& \frac{1}{2\pi h} \int_{-\infty}^{+\infty} \exp(itx/h) \frac{\overline{\phi_K(t)}}{\overline{\widehat{\phi}_{U}(t/h)}} dt\\
=&\frac{1}{2\pi h} \int_{-\infty}^{+\infty} \exp(itx/h) \frac{{\phi_K(t)}}{{\widehat{\phi}_{U}(-t/h)}}dt \ \left(\text{By } \phi_K = \overline{\phi_K}\text{ and } \overline{\widehat{\phi}(t)}=\widehat{\phi}(-t)\right)\\
=&-\frac{1}{2\pi h} \int_{+\infty}^{-\infty} \exp(-itx/h) \frac{{\phi_K(-t)}}{{\widehat{\phi}_{U}(t/h)}}dt \ \left(\text{By a change of variable }t=-t\right)\\
=&\frac{1}{2\pi h} \int_{-\infty}^{+\infty} \exp(-itx/h) \frac{{\phi_K(t)}}{{\widehat{\phi}_{U}(t/h)}}dt\  \left(\text{By } \phi_K = \overline{\phi_K}\right)\\
=&\widehat{K}_{U,h}(x),
\end{align*}
which implies $\widehat{K}_{U,h}$ is real.

 \color{black}

	\newpage
	\section{More Simulation Results}\label{appendix:moresimulation}

\subsection{3D visualization of uniform confidence bands}\label{appendix:3D}

In this subsection, we provide 3D visualization of uniform confidence bands for the simulation studies.

\begin{figure}[htbp]
    \centering
    \begin{minipage}{0.48\textwidth}
        \centering
        \includegraphics[width=\linewidth]{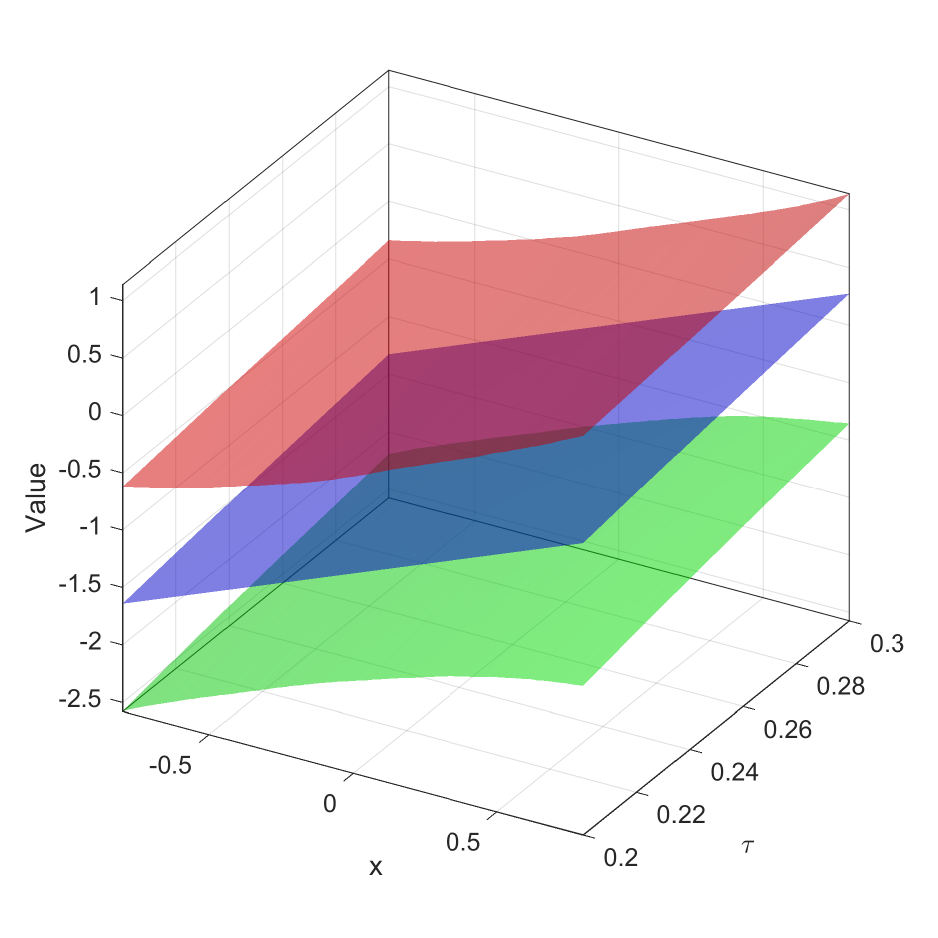}

    \end{minipage}%
    \hspace{0.1cm}
    \begin{minipage}{0.48\textwidth}
        \centering
        \includegraphics[width=\linewidth]{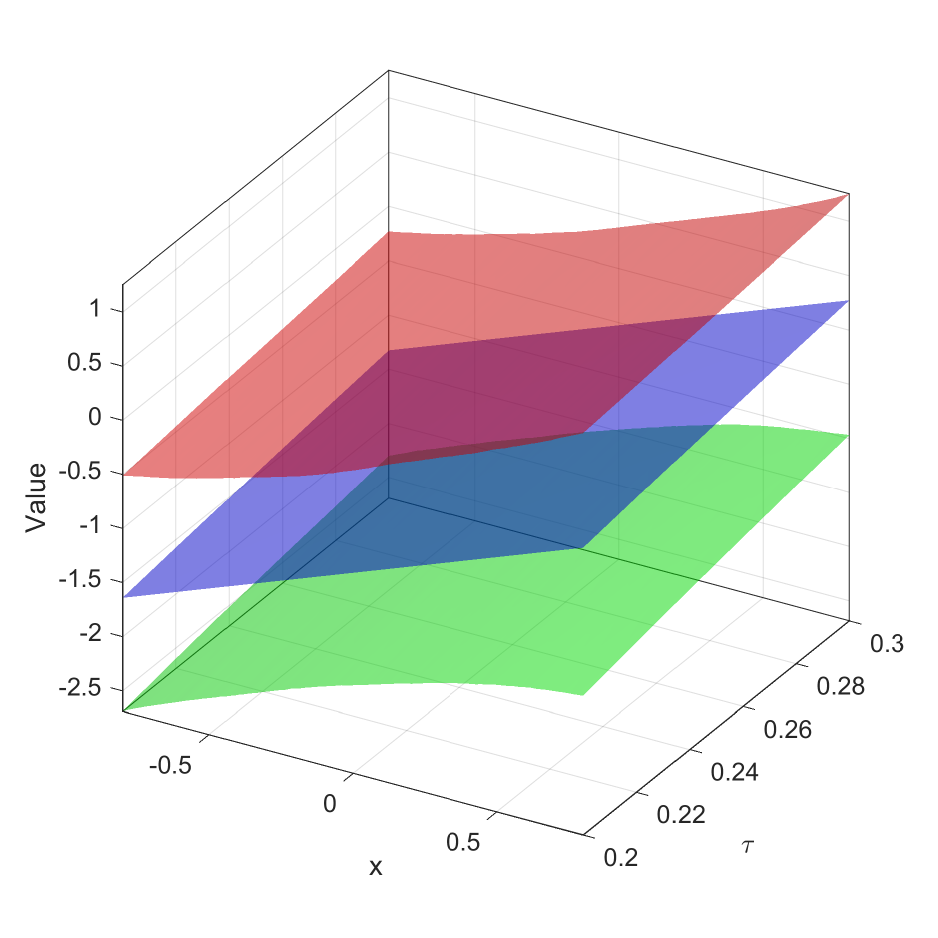}

    \end{minipage}
    
    \vspace{0.2cm}
    
    \begin{minipage}{0.48\textwidth}
        \centering
        \includegraphics[width=\linewidth]{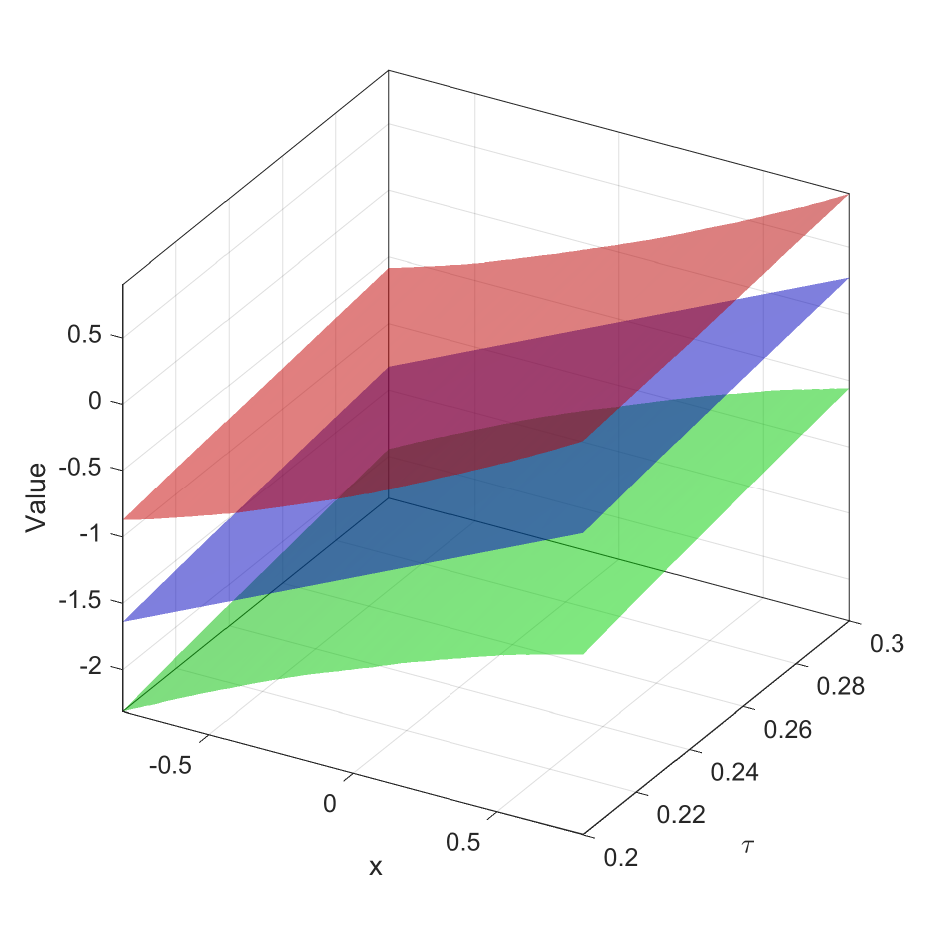}

    \end{minipage}%
    \hspace{0.1cm}
    \begin{minipage}{0.48\textwidth}
        \centering
        \includegraphics[width=\linewidth]{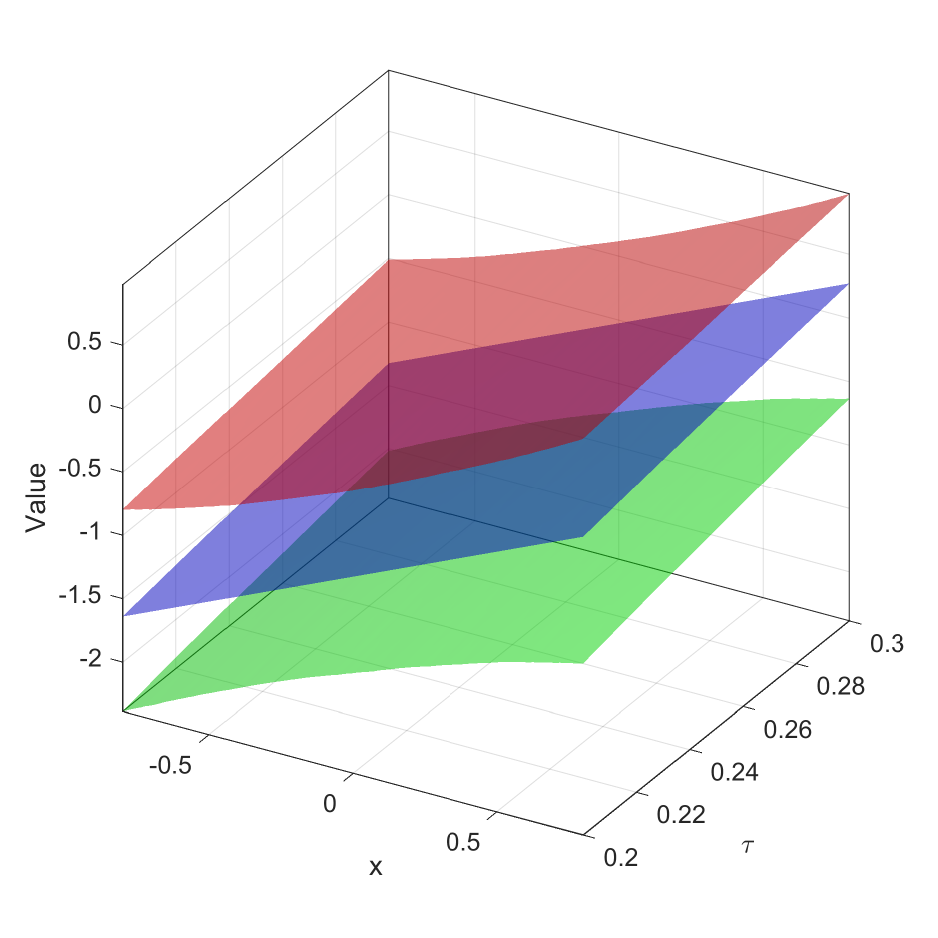}
    \end{minipage}
    
    \caption{The average of 1000 $(1-\alpha)$ uniform confidence bands for DGP1 with Laplace error for $\tau\in[0.2,0.3]$ and $x\in[-0.8,0.8]$, for $\alpha = 10\%$ (left) and $\alpha = 5\%$ (right), with $n=250$ (Row 1) and $n=500$ (Row 2). The red and green surfaces represent the upper and lower bound of the  confidence band, and the blue surface is the real  quantile regression function.}
\end{figure}
\begin{figure}[htbp]
    \centering
    \begin{minipage}{0.48\textwidth}
        \centering
        \includegraphics[width=\linewidth]{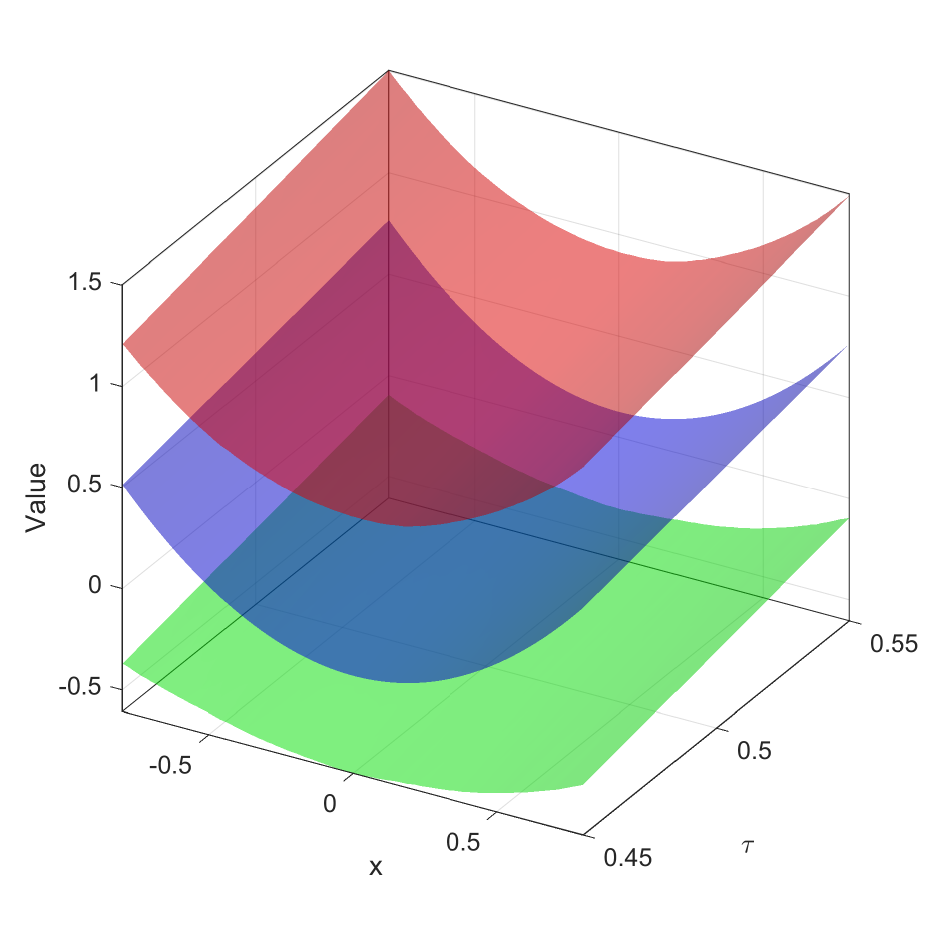}

    \end{minipage}%
    \hspace{0.1cm}
    \begin{minipage}{0.48\textwidth}
        \centering
        \includegraphics[width=\linewidth]{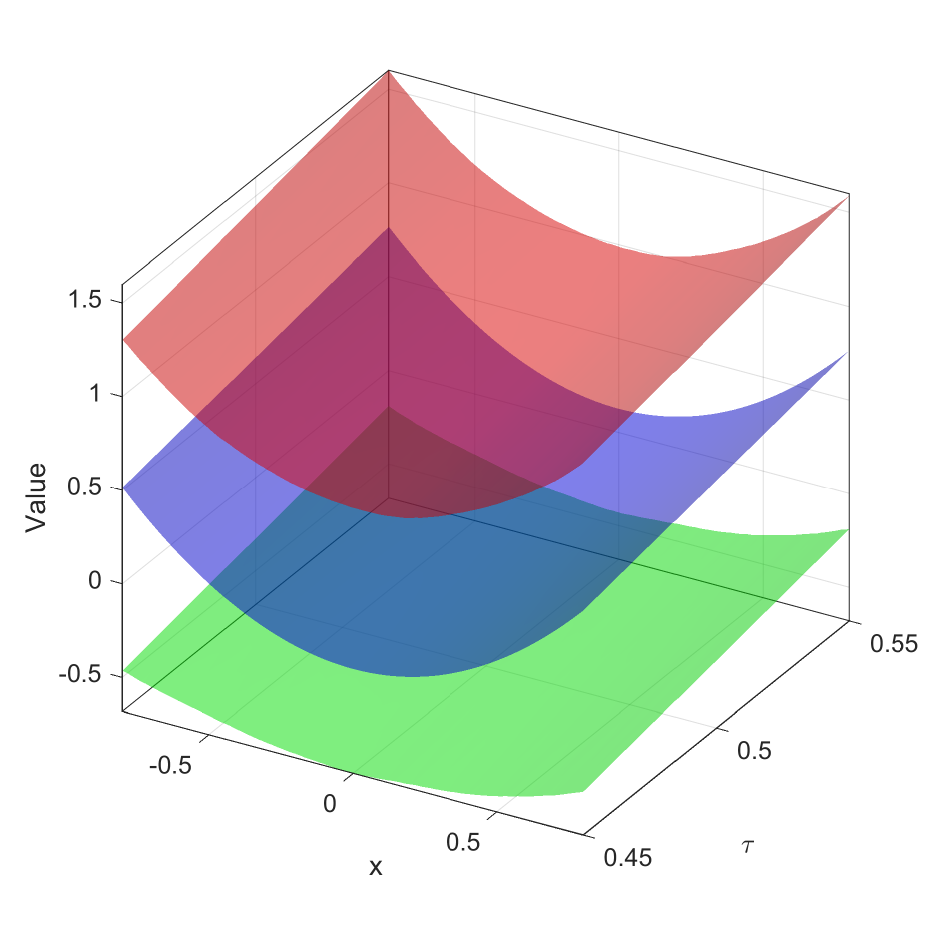}

    \end{minipage}
    
    \vspace{0.2cm}
    
    \begin{minipage}{0.48\textwidth}
        \centering
        \includegraphics[width=\linewidth]{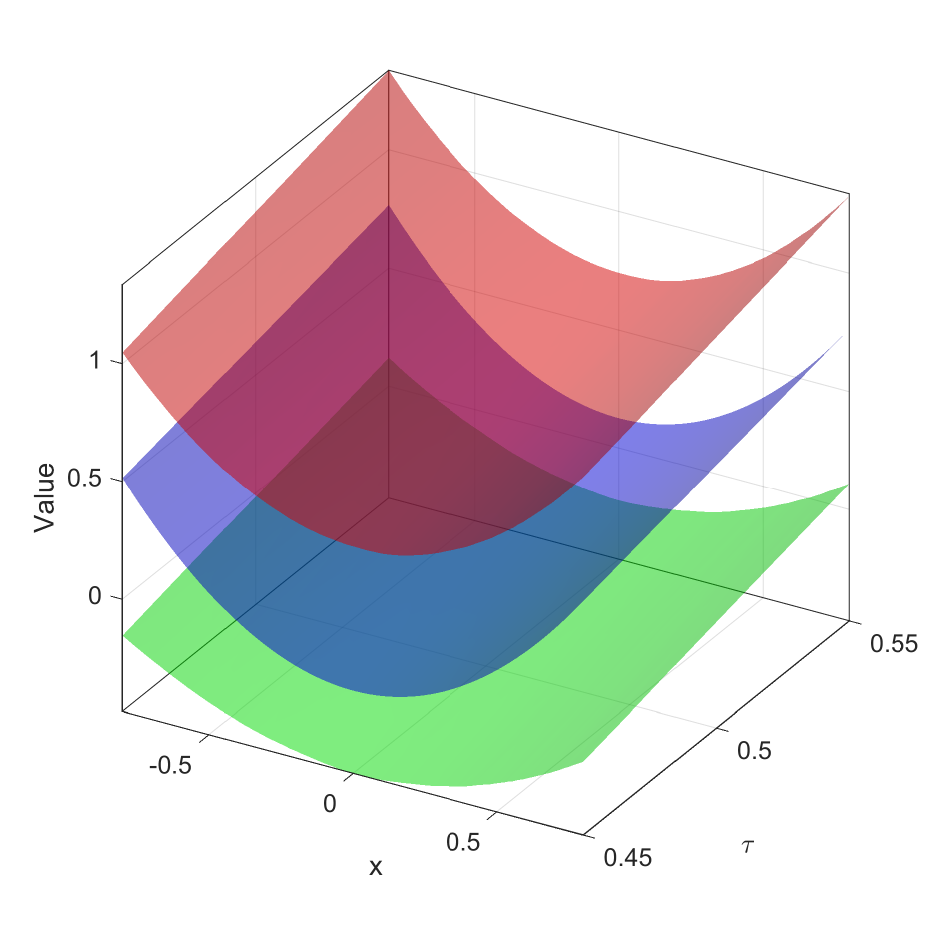}

    \end{minipage}%
    \hspace{0.1cm}
    \begin{minipage}{0.48\textwidth}
        \centering
        \includegraphics[width=\linewidth]{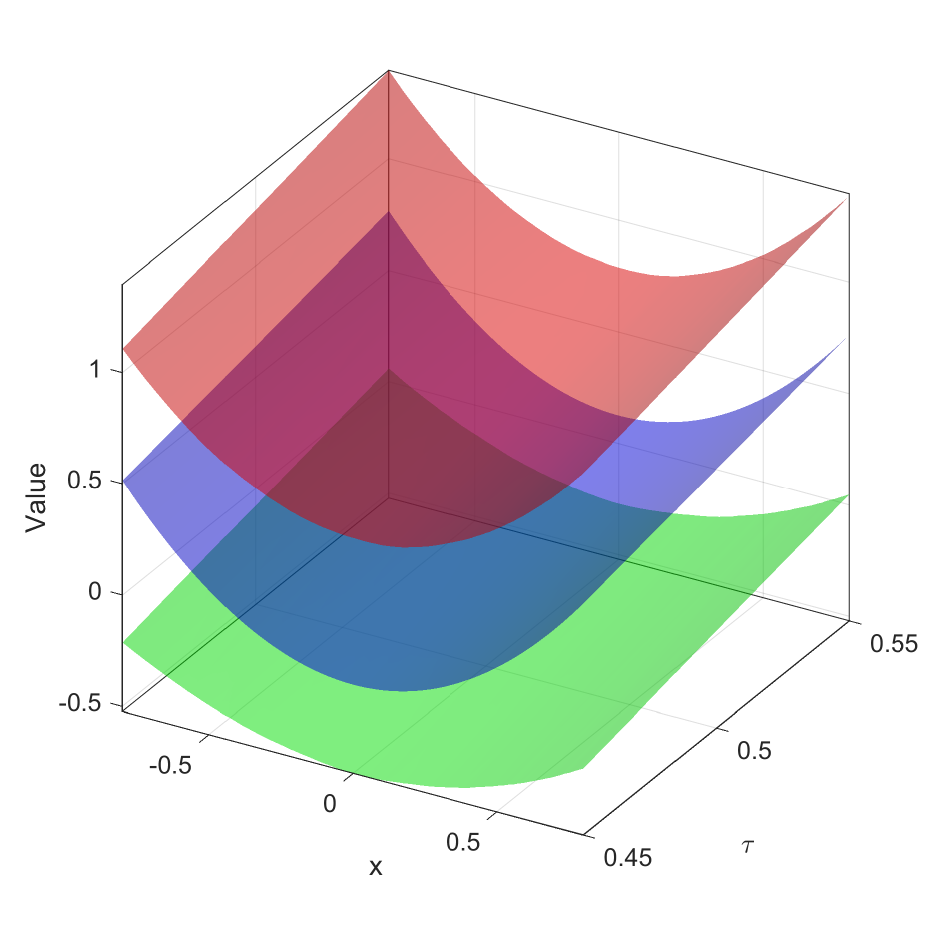}
    \end{minipage}
    
    \caption{The average of 1000 $(1-\alpha)$ uniform confidence bands for DGP2 with Gaussian error for $\tau\in[0.45,0.55]$ and $x\in[-0.8,0.8]$, for $\alpha = 10\%$ (left) and $\alpha = 5\%$ (right), with $n=250$ (Row 1) and $n=500$ (Row 2). The red and green surfaces represent the upper and lower bound of the  confidence band, and the blue surface is the real  quantile regression function.}
\end{figure}
\begin{figure}[htbp]
    \centering
    \begin{minipage}{0.48\textwidth}
        \centering
        \includegraphics[width=\linewidth]{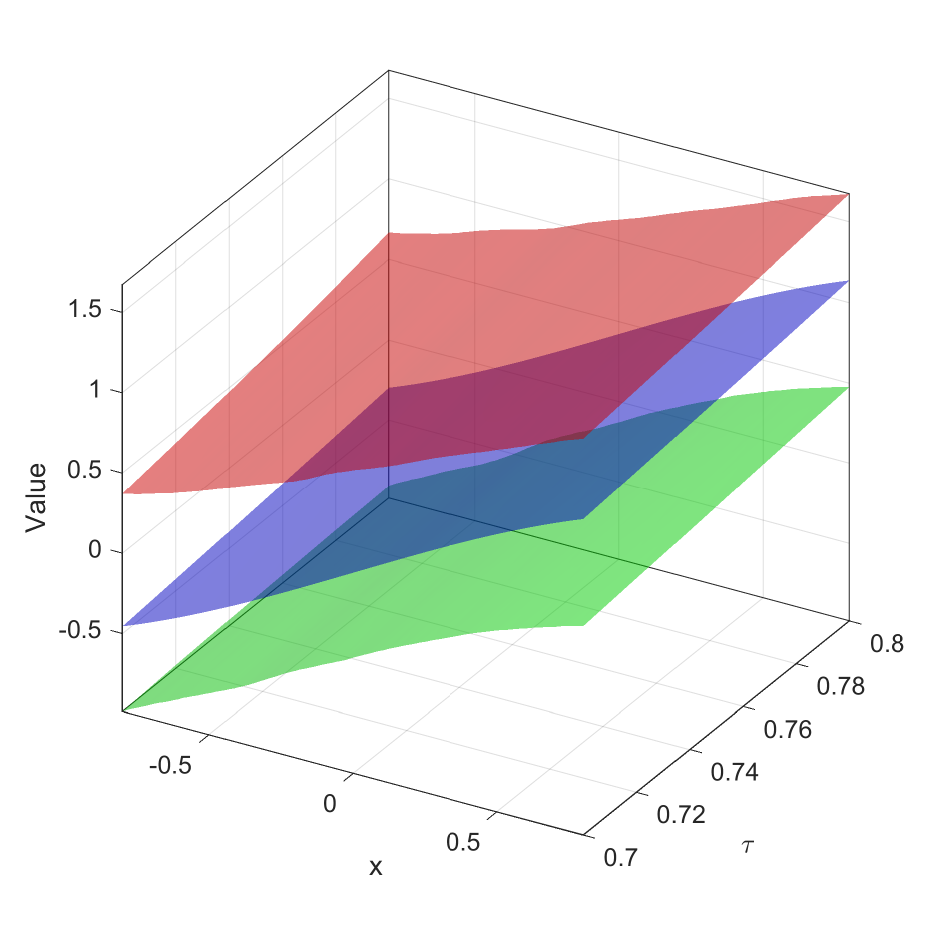}

    \end{minipage}%
    \hspace{0.1cm}
    \begin{minipage}{0.48\textwidth}
        \centering
        \includegraphics[width=\linewidth]{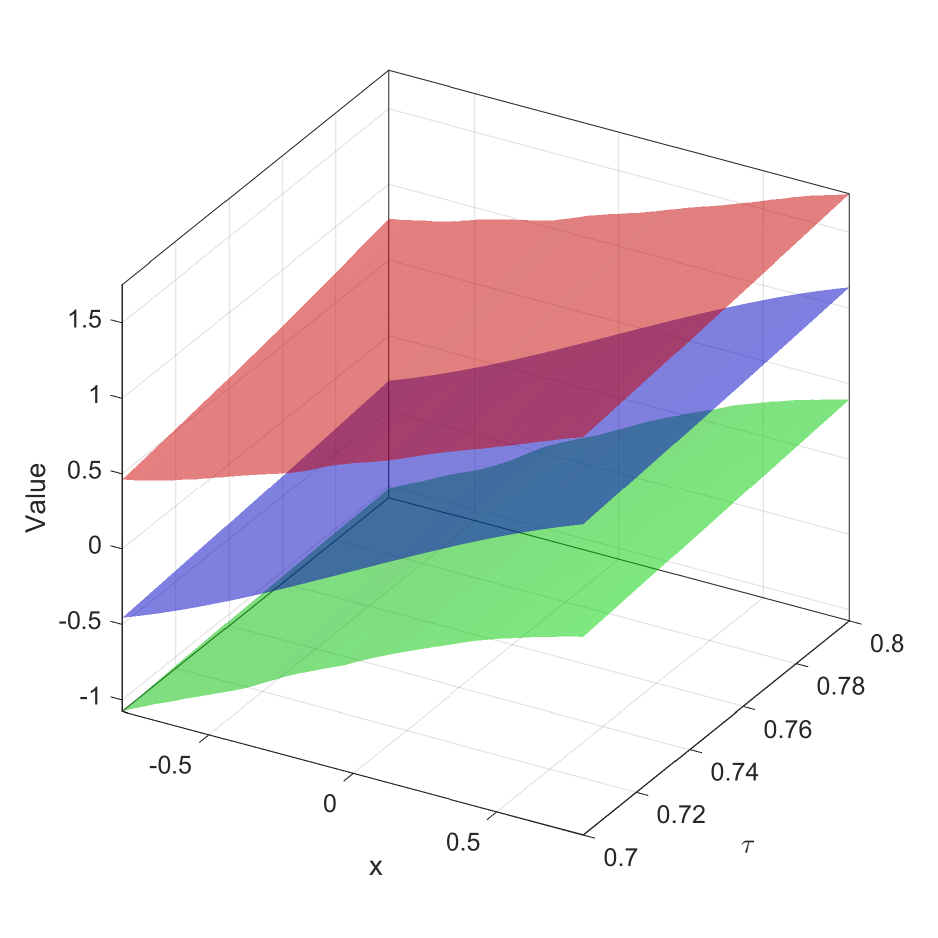}

    \end{minipage}
    
    \vspace{0.2cm}
    
    \begin{minipage}{0.48\textwidth}
        \centering
        \includegraphics[width=\linewidth]{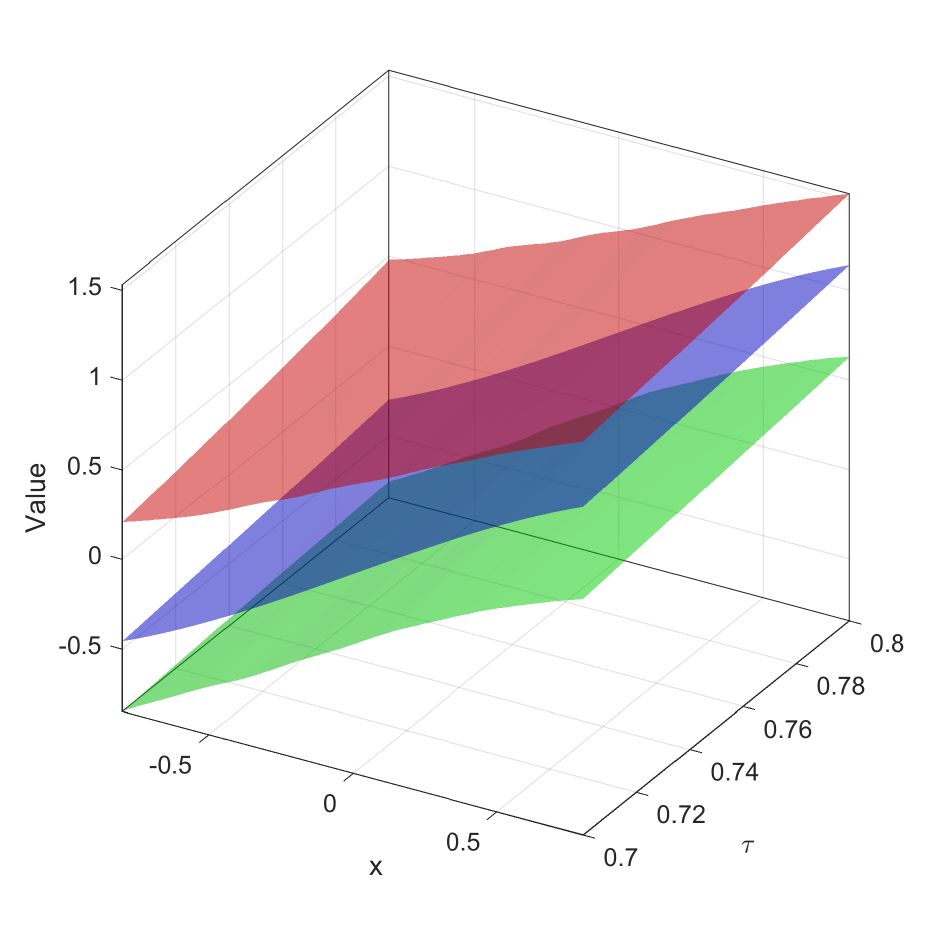}

    \end{minipage}%
    \hspace{0.1cm}
    \begin{minipage}{0.48\textwidth}
        \centering
        \includegraphics[width=\linewidth]{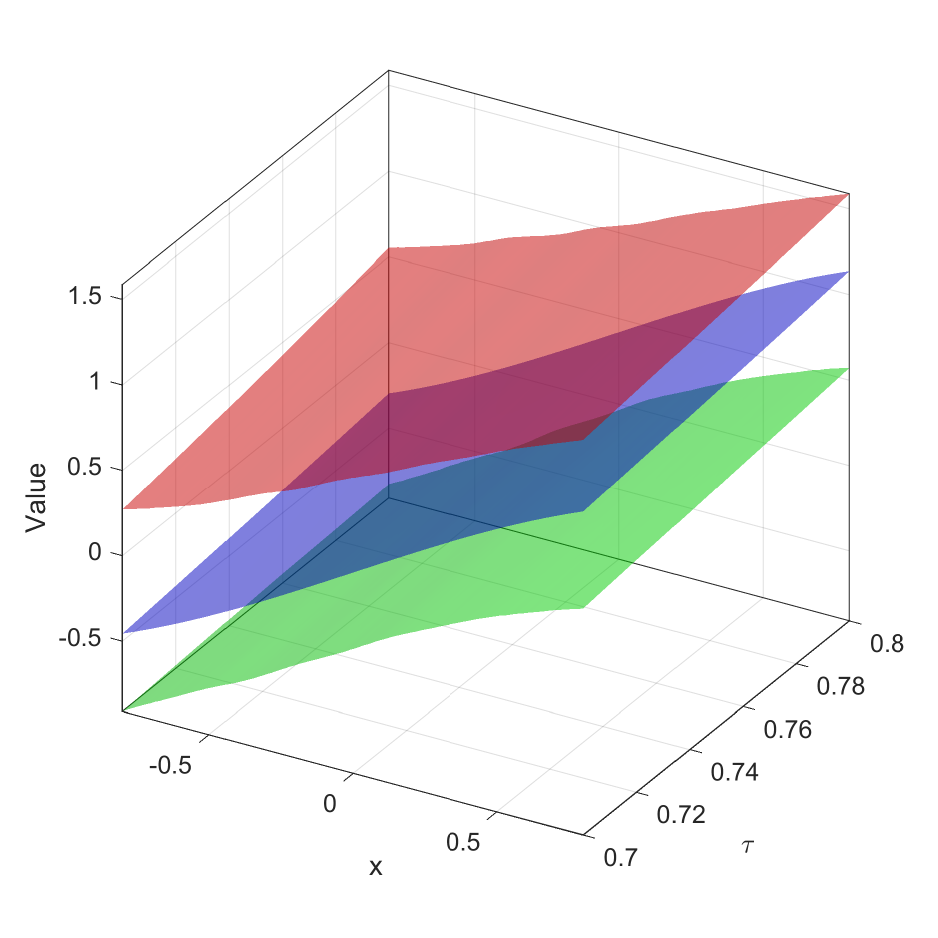}
    \end{minipage}
    
    \caption{The average of 1000 $(1-\alpha)$ uniform confidence bands for DGP3 with Gaussian error for $\tau\in[0.7,0.8]$ and $x\in[-0.8,0.8]$, for $\alpha = 10\%$ (left) and $\alpha = 5\%$ (right), with $n=250$ (Row 1) and $n=500$ (Row 2). The red and green surfaces represent the upper and lower bound of the  confidence band, and the blue surface is the real  quantile regression function.}
\end{figure}

\subsection{Singleton Quantile Level}\label{appendix:singleton}
    In this section, we present more simulation results for singleton $\mathcal{T}=\{0.25\}, \{0.5\}$ and $\{0.75\}$.

\begin{table}[htbp]
	\caption{Empirical Coverage probability, Mean of Sizes (MS) of 1000 uniform confidence bands for singleton quantile of DGPs~1, 2 and 3.}
	\label{tab:single}
	\resizebox{\textwidth}{!}{%
		\begin{tabular}{ccccllcclllcclll}
			\hline
			\multicolumn{1}{|c}{}                   & \multicolumn{1}{c|}{}                      & cp                    & N   & \multicolumn{1}{c}{empcp} & \multicolumn{1}{c|}{MS} & cp                    & N   & \multicolumn{1}{c}{empcp} & \multicolumn{1}{c|}{MS} & cp                    & N   & \multicolumn{1}{c}{empcp} & \multicolumn{1}{c|}{MS} \\ \hline
			\multicolumn{1}{|c}{DGP}                & \multicolumn{1}{c|}{ME}                    & \multicolumn{4}{c|}{$\tau=0.25$}                                                & \multicolumn{4}{c|}{$\tau=0.5$}                                                 & \multicolumn{4}{c|}{$\tau=0.75$}                                                \\ \hline
			\multicolumn{1}{|c}{\multirow{4}{*}{1}} & \multicolumn{1}{c|}{\multirow{4}{*}{Lap}}  & \multirow{2}{*}{0.9}  & 250 & 0.976                     & \multicolumn{1}{l|}{1.804} & \multirow{2}{*}{0.9}  & 250 & 0.961                     & \multicolumn{1}{l|}{1.590} & \multirow{2}{*}{0.9}  & 250 & 0.966                     & \multicolumn{1}{l|}{1.753} \\
			\multicolumn{1}{|c}{}                   & \multicolumn{1}{c|}{}                      &                       & 500 & 0.963                     & \multicolumn{1}{l|}{1.322} &                       & 500 & 0.962                     & \multicolumn{1}{l|}{1.181} &                       & 500 & 0.967                     & \multicolumn{1}{l|}{1.351} \\ \cline{3-14} 
			\multicolumn{1}{|c}{}                   & \multicolumn{1}{c|}{}                      & \multirow{2}{*}{0.95} & 250 & 0.988                     & \multicolumn{1}{l|}{2.010} & \multirow{2}{*}{0.95} & 250 & 0.978                     & \multicolumn{1}{l|}{1.776} & \multirow{2}{*}{0.95} & 250 & 0.982                     & \multicolumn{1}{l|}{1.947} \\
			\multicolumn{1}{|c}{}                   & \multicolumn{1}{c|}{}                      &                       & 500 & 0.982                     & \multicolumn{1}{l|}{1.458} &                       & 500 & 0.979                     & \multicolumn{1}{l|}{1.304} &                       & 500 & 0.980                     & \multicolumn{1}{l|}{1.486} \\ \hline
			\multicolumn{1}{|c}{\multirow{4}{*}{1}} & \multicolumn{1}{c|}{\multirow{4}{*}{Norm}} & \multirow{2}{*}{0.9}  & 250 & 0.955                     & \multicolumn{1}{l|}{1.686} & \multirow{2}{*}{0.9}  & 250 & 0.949                     & \multicolumn{1}{l|}{1.480} & \multirow{2}{*}{0.9}  & 250 & 0.953                     & \multicolumn{1}{l|}{1.639} \\
			\multicolumn{1}{|c}{}                   & \multicolumn{1}{c|}{}                      &                       & 500 & 0.945                     & \multicolumn{1}{l|}{1.217} &                       & 500 & 0.937                     & \multicolumn{1}{l|}{1.072} &                       & 500 & 0.945                     & \multicolumn{1}{l|}{1.220} \\ \cline{3-14} 
			\multicolumn{1}{|c}{}                   & \multicolumn{1}{c|}{}                      & \multirow{2}{*}{0.95} & 250 & 0.977                     & \multicolumn{1}{l|}{1.876} & \multirow{2}{*}{0.95} & 250 & 0.970                     & \multicolumn{1}{l|}{1.655} & \multirow{2}{*}{0.95} & 250 & 0.977                     & \multicolumn{1}{l|}{1.825} \\
			\multicolumn{1}{|c}{}                   & \multicolumn{1}{c|}{}                      &                       & 500 & 0.969                     & \multicolumn{1}{l|}{1.342} &                       & 500 & 0.964                     & \multicolumn{1}{l|}{1.185} &                       & 500 & 0.968                     & \multicolumn{1}{l|}{1.344} \\ \hline
			\multicolumn{1}{|c}{\multirow{4}{*}{2}} & \multicolumn{1}{c|}{\multirow{4}{*}{Lap}}  & \multirow{2}{*}{0.9}  & 250 & 0.907                     & \multicolumn{1}{l|}{1.472} & \multirow{2}{*}{0.9}  & 250 & 0.950                     & \multicolumn{1}{l|}{1.525} & \multirow{2}{*}{0.9}  & 250 & 0.977                     & \multicolumn{1}{l|}{1.982} \\
			\multicolumn{1}{|c}{}                   & \multicolumn{1}{c|}{}                      &                       & 500 & 0.916                     & \multicolumn{1}{l|}{1.148} &                       & 500 & 0.958                     & \multicolumn{1}{l|}{1.139} &                       & 500 & 0.971                     & \multicolumn{1}{l|}{1.428} \\ \cline{3-14} 
			\multicolumn{1}{|c}{}                   & \multicolumn{1}{c|}{}                      & \multirow{2}{*}{0.95} & 250 & 0.952                     & \multicolumn{1}{l|}{1.634} & \multirow{2}{*}{0.95} & 250 & 0.977                     & \multicolumn{1}{l|}{1.711} & \multirow{2}{*}{0.95} & 250 & 0.990                     & \multicolumn{1}{l|}{2.220} \\
			\multicolumn{1}{|c}{}                   & \multicolumn{1}{c|}{}                      &                       & 500 & 0.954                     & \multicolumn{1}{l|}{1.261} &                       & 500 & 0.981                     & \multicolumn{1}{l|}{1.252} &                       & 500 & 0.991                     & \multicolumn{1}{l|}{1.572} \\ \hline
			\multicolumn{1}{|c}{\multirow{4}{*}{2}} & \multicolumn{1}{c|}{\multirow{4}{*}{Norm}} & \multirow{2}{*}{0.9}  & 250 & 0.898                     & \multicolumn{1}{l|}{1.358} & \multirow{2}{*}{0.9}  & 250 & 0.924                     & \multicolumn{1}{l|}{1.371} & \multirow{2}{*}{0.9}  & 250 & 0.972                     & \multicolumn{1}{l|}{1.876} \\
			\multicolumn{1}{|c}{}                   & \multicolumn{1}{c|}{}                      &                       & 500 & 0.867                     & \multicolumn{1}{l|}{1.025} &                       & 500 & 0.911                     & \multicolumn{1}{l|}{1.040} &                       & 500 & 0.952                     & \multicolumn{1}{l|}{1.309} \\ \cline{3-14} 
			\multicolumn{1}{|c}{}                   & \multicolumn{1}{c|}{}                      & \multirow{2}{*}{0.95} & 250 & 0.932                     & \multicolumn{1}{l|}{1.503} & \multirow{2}{*}{0.95} & 250 & 0.956                     & \multicolumn{1}{l|}{1.532} & \multirow{2}{*}{0.95} & 250 & 0.982                     & \multicolumn{1}{l|}{2.122} \\
			\multicolumn{1}{|c}{}                   & \multicolumn{1}{c|}{}                      &                       & 500 & 0.919                     & \multicolumn{1}{l|}{1.129} &                       & 500 & 0.954                     & \multicolumn{1}{l|}{1.144} &                       & 500 & 0.972                     & \multicolumn{1}{l|}{1.444} \\ \hline
			\multicolumn{1}{|c}{\multirow{4}{*}{3}} & \multicolumn{1}{c|}{\multirow{4}{*}{Lap}}  & \multirow{2}{*}{0.9}  & 250 & 0.969                     & \multicolumn{1}{l|}{1.428} & \multirow{2}{*}{0.9}  & 250 & 0.975                     & \multicolumn{1}{l|}{1.223} & \multirow{2}{*}{0.9}  & 250 & 0.954                     & \multicolumn{1}{l|}{1.370} \\
			\multicolumn{1}{|c}{}                   & \multicolumn{1}{c|}{}                      &                       & 500 & 0.963                     & \multicolumn{1}{l|}{1.029} &                       & 500 & 0.963                     & \multicolumn{1}{l|}{0.841} &                       & 500 & 0.967                     & \multicolumn{1}{l|}{1.035} \\ \cline{3-14} 
			\multicolumn{1}{|c}{}                   & \multicolumn{1}{c|}{}                      & \multirow{2}{*}{0.95} & 250 & 0.986                     & \multicolumn{1}{l|}{1.626} & \multirow{2}{*}{0.95} & 250 & 0.981                     & \multicolumn{1}{l|}{1.405} & \multirow{2}{*}{0.95} & 250 & 0.982                     & \multicolumn{1}{l|}{1.550} \\
			\multicolumn{1}{|c}{}                   & \multicolumn{1}{c|}{}                      &                       & 500 & 0.980                     & \multicolumn{1}{l|}{1.147} &                       & 500 & 0.983                     & \multicolumn{1}{l|}{0.941} &                       & 500 & 0.985                     & \multicolumn{1}{l|}{1.159} \\ \hline
			\multicolumn{1}{|c}{\multirow{4}{*}{3}} & \multicolumn{1}{c|}{\multirow{4}{*}{Norm}} & \multirow{2}{*}{0.9}  & 250 & 0.926                     & \multicolumn{1}{l|}{1.319} & \multirow{2}{*}{0.9}  & 250 & 0.940                     & \multicolumn{1}{l|}{1.153} & \multirow{2}{*}{0.9}  & 250 & 0.919                     & \multicolumn{1}{l|}{1.240} \\
			\multicolumn{1}{|c}{}                   & \multicolumn{1}{c|}{}                      &                       & 500 & 0.867                     & \multicolumn{1}{l|}{0.943} &                       & 500 & 0.882                     & \multicolumn{1}{l|}{0.791} &                       & 500 & 0.878                     & \multicolumn{1}{l|}{0.951} \\ \cline{3-14} 
			\multicolumn{1}{|c}{}                   & \multicolumn{1}{c|}{}                      & \multirow{2}{*}{0.95} & 250 & 0.957                     & \multicolumn{1}{l|}{1.503} & \multirow{2}{*}{0.95} & 250 & 0.967                     & \multicolumn{1}{l|}{1.324} & \multirow{2}{*}{0.95} & 250 & 0.954                     & \multicolumn{1}{l|}{1.406} \\
			\multicolumn{1}{|c}{}                   & \multicolumn{1}{c|}{}                      &                       & 500 & 0.920                     & \multicolumn{1}{l|}{1.055} &                       & 500 & 0.935                     & \multicolumn{1}{l|}{0.887} &                       & 500 & 0.918                     & \multicolumn{1}{l|}{1.067} \\ \hline
			&                                            &                       &     & \multicolumn{1}{c}{}      & \multicolumn{1}{c}{}      &                       &     & \multicolumn{1}{c}{}      & \multicolumn{1}{c}{}      &                       &     & \multicolumn{1}{c}{}      & \multicolumn{1}{c}{}     
		\end{tabular}%
	}
\end{table}
Table \ref{tab:single} shows that the uniform bands for a fixed quantile level also perform stable and well. As $n$ increases, the coverage probabilities go towards the corresponding nominal coverage probabilities.
\FloatBarrier
    
\subsection{Sensitivity analysis of uniform inference to parameter choices}\label{appendix:sensitivity}

The bandwidth selection method described in Section \ref{Section:bandchoose} {requires} two parameters $\rho$ and $L$. To {analyze} the sensitivity {of the uniform inference results} with respect to $\rho$ and $L$, in the following tables \ref{tab:k20rho13}-\ref{tab:k30rho17},  we present more simulation results for {$\rho\in\{2.5,3,3.5\}$} and $L\in\{20,30\}$. 

In general, the uniform bands, for all the DGPs, quantiles and sample sizes, perform well and {are stable across} {$(L,\rho)\in\{20,30\}\times\{2.5,3,3.5\}.$} As evident from Tables \ref{tab:k20rho13}-\ref{tab:k30rho17}, increasing $\rho$ leads to a higher level of undersmoothing, resulting in wider confidence bands and higher empirical coverage probability. Conversely, increasing $L$ reduces the degree of undersmoothing, yielding narrower confidence bands and slightly lower empirical coverage probability. The results suggest that our method is relatively robust to moderate changes in $\rho$ and $L$, with $\rho$ having a more significant effect on the confidence band properties than $L$. Researchers applying this method should be aware of these trade-offs when selecting $\rho$ and $L$ for their specific applications. 

{A fully data-driven approach to choose $L$ and $\rho$ is challenging and beyond the scope of the paper, but it can be} an interesting future topic.

\begin{table}[htbp]
	\caption{Empirical Coverage probability, Mean of Sizes (MS) of confidence bands for DGP1, 2 and 3 when $L=20$ and $\rho=2.5$.}
	\label{tab:k20rho13}
	\resizebox{\textwidth}{!}{%
\begin{tabular}{ccccllcclllcclll}
			\hline
			\multicolumn{1}{|c}{}                   & \multicolumn{1}{c|}{}                      & cp                    & N   & \multicolumn{1}{c}{empcp} & \multicolumn{1}{c|}{MS} & cp                    & N   & \multicolumn{1}{c}{empcp} & \multicolumn{1}{c|}{MS} & cp                    & N   & \multicolumn{1}{c}{empcp} & \multicolumn{1}{c|}{MS} \\ \hline
			\multicolumn{1}{|c}{DGP}                & \multicolumn{1}{c|}{ME}                    & \multicolumn{4}{c|}{$\tau\in[0.2,0.3]$}                                                & \multicolumn{4}{c|}{$\tau\in[0.45,0.55]$}                                                 & \multicolumn{4}{c|}{$\tau\in[0.7,0.8]$}                                                \\ \hline
			\multicolumn{1}{|c}{\multirow{4}{*}{1}} & \multicolumn{1}{c|}{\multirow{4}{*}{Lap}}  & \multirow{2}{*}{0.9}  & 250 & 0.959                     & \multicolumn{1}{l|}{1.58} & \multirow{2}{*}{0.9}  & 250 & 0.946                     & \multicolumn{1}{l|}{1.38} & \multirow{2}{*}{0.9}  & 250 & 0.944                     & \multicolumn{1}{l|}{1.51} \\
			\multicolumn{1}{|c}{}                   & \multicolumn{1}{c|}{}                      &                       & 500 & 0.945                     & \multicolumn{1}{l|}{1.17} &                       & 500 & 0.947                     & \multicolumn{1}{l|}{1.05} &                       & 500 & 0.950                     & \multicolumn{1}{l|}{1.18} \\ \cline{3-14} 
			\multicolumn{1}{|c}{}                   & \multicolumn{1}{c|}{}                      & \multirow{2}{*}{0.95} & 250 & 0.977                     & \multicolumn{1}{l|}{1.76} & \multirow{2}{*}{0.95} & 250 & 0.963                     & \multicolumn{1}{l|}{1.53} & \multirow{2}{*}{0.95} & 250 & 0.967                     & \multicolumn{1}{l|}{1.67} \\
\hline
		\end{tabular}%
	}
\end{table}
\begin{table}[htbp]
   \caption{Empirical Coverage probability, Mean of Sizes (MS) of confidence bands for DGP1, 2 and 3 when $L=20$ and $\rho=3$.}
   \label{tab:k20rho15}
   \resizebox{\textwidth}{!}{%
   	\begin{tabular}{ccccllcclllcclll}
   		\hline
   		\multicolumn{1}{|c}{}                   & \multicolumn{1}{c|}{}                      & cp                    & N   & \multicolumn{1}{c}{empcp} & \multicolumn{1}{c|}{MS} & cp                    & N   & \multicolumn{1}{c}{empcp} & \multicolumn{1}{c|}{MS} & cp                    & N   & \multicolumn{1}{c}{empcp} & \multicolumn{1}{c|}{MS} \\ \hline
   		\multicolumn{1}{|c}{DGP}                & \multicolumn{1}{c|}{ME}                    & \multicolumn{4}{c|}{$\tau\in[0.2,0.3]$}                                                & \multicolumn{4}{c|}{$\tau\in[0.45,0.55]$}                                                 & \multicolumn{4}{c|}{$\tau\in[0.7,0.8]$}                                                \\ \hline
   		\multicolumn{1}{|c}{\multirow{4}{*}{1}} & \multicolumn{1}{c|}{\multirow{4}{*}{Lap}}  & \multirow{2}{*}{0.9}  & 250 & 0.976                     & \multicolumn{1}{l|}{1.80} & \multirow{2}{*}{0.9}  & 250 & 0.961                     & \multicolumn{1}{l|}{1.59} & \multirow{2}{*}{0.9}  & 250 & 0.966                     & \multicolumn{1}{l|}{1.75} \\
   		\multicolumn{1}{|c}{}                   & \multicolumn{1}{c|}{}                      &                       & 500 & 0.963                     & \multicolumn{1}{l|}{1.32} &                       & 500 & 0.962                     & \multicolumn{1}{l|}{1.18} &                       & 500 & 0.967                     & \multicolumn{1}{l|}{1.35} \\ \cline{3-14} 
   		\multicolumn{1}{|c}{}                   & \multicolumn{1}{c|}{}                      & \multirow{2}{*}{0.95} & 250 & 0.988                     & \multicolumn{1}{l|}{2.01} & \multirow{2}{*}{0.95} & 250 & 0.978                     & \multicolumn{1}{l|}{1.78} & \multirow{2}{*}{0.95} & 250 & 0.982                     & \multicolumn{1}{l|}{1.95} \\
\hline
   	\end{tabular}%
   }
\end{table}
\begin{table}[htbp]
   \caption{Empirical Coverage probability, Mean of Sizes (MS) of confidence bands for DGP1, 2 and 3 when $L=20$ and $\rho=3.5$.}
   \label{tab:k20rho17}
   \resizebox{\textwidth}{!}{%
   	\begin{tabular}{ccccllcclllcclll}
   		\hline
   		\multicolumn{1}{|c}{}                   & \multicolumn{1}{c|}{}                      & cp                    & N   & \multicolumn{1}{c}{empcp} & \multicolumn{1}{c|}{MS} & cp                    & N   & \multicolumn{1}{c}{empcp} & \multicolumn{1}{c|}{MS} & cp                    & N   & \multicolumn{1}{c}{empcp} & \multicolumn{1}{c|}{MS} \\ \hline
   		\multicolumn{1}{|c}{DGP}                & \multicolumn{1}{c|}{ME}                    & \multicolumn{4}{c|}{$\tau\in[0.2,0.3]$}                                                & \multicolumn{4}{c|}{$\tau\in[0.45,0.55]$}                                                 & \multicolumn{4}{c|}{$\tau\in[0.7,0.8]$}                                                \\ \hline
   		\multicolumn{1}{|c}{\multirow{4}{*}{1}} & \multicolumn{1}{c|}{\multirow{4}{*}{Lap}}  & \multirow{2}{*}{0.9}  & 250 & {0.985}                     & \multicolumn{1}{l|}{{2.06}} & \multirow{2}{*}{0.9}  & 250 & 0.970                     & \multicolumn{1}{l|}{1.80} & \multirow{2}{*}{0.9}  & 250 & 0.976                     & \multicolumn{1}{l|}{1.99} \\
   		\multicolumn{1}{|c}{}                   & \multicolumn{1}{c|}{}                      &                       & 500 & 0.982                     & \multicolumn{1}{l|}{1.47} &                       & 500 & 0.976                     & \multicolumn{1}{l|}{1.33} &                       & 500 & 0.972                     & \multicolumn{1}{l|}{1.51} \\ \cline{3-14} 
   		\multicolumn{1}{|c}{}                   & \multicolumn{1}{c|}{}                      & \multirow{2}{*}{0.95} & 250 & 0.995                     & \multicolumn{1}{l|}{2.30} & \multirow{2}{*}{0.95} & 250 & 0.985                     & \multicolumn{1}{l|}{2.02} & \multirow{2}{*}{0.95} & 250 & 0.988                     & \multicolumn{1}{l|}{2.22} \\
\hline
   	\end{tabular}%
   }
\end{table}
\begin{table}[htbp]
   \caption{Empirical Coverage probability, Mean of Sizes (MS) of confidence bands for DGP1, 2 and 3 when $L=30$ and $\rho=2.5$.}
   \label{tab:k30rho13}
   \resizebox{\textwidth}{!}{%
   	\begin{tabular}{ccccllcclllcclll}
   		\hline
   		\multicolumn{1}{|c}{}                   & \multicolumn{1}{c|}{}                      & cp                    & N   & \multicolumn{1}{c}{empcp} & \multicolumn{1}{c|}{MS} & cp                    & N   & \multicolumn{1}{c}{empcp} & \multicolumn{1}{c|}{MS} & cp                    & N   & \multicolumn{1}{c}{empcp} & \multicolumn{1}{c|}{MS} \\ \hline
   		\multicolumn{1}{|c}{DGP}                & \multicolumn{1}{c|}{ME}                    & \multicolumn{4}{c|}{$\tau\in[0.2,0.3]$}                                                & \multicolumn{4}{c|}{$\tau\in[0.45,0.55]$}                                                 & \multicolumn{4}{c|}{$\tau\in[0.7,0.8]$}                                                \\ \hline
   		\multicolumn{1}{|c}{\multirow{4}{*}{1}} & \multicolumn{1}{c|}{\multirow{4}{*}{Lap}}  & \multirow{2}{*}{0.9}  & 250 & 0.947                     & \multicolumn{1}{l|}{1.47} & \multirow{2}{*}{0.9}  & 250 & 0.932                     & \multicolumn{1}{l|}{1.30} & \multirow{2}{*}{0.9}  & 250 & 0.936                     & \multicolumn{1}{l|}{1.44} \\
   		\multicolumn{1}{|c}{}                   & \multicolumn{1}{c|}{}                      &                       & 500 & 0.934                     & \multicolumn{1}{l|}{1.11} &                       & 500 & 0.927                     & \multicolumn{1}{l|}{0.98} &                       & 500 & 0.933                     & \multicolumn{1}{l|}{1.11} \\ \cline{3-14} 
   		\multicolumn{1}{|c}{}                   & \multicolumn{1}{c|}{}                      & \multirow{2}{*}{0.95} & 250 & 0.976                     & \multicolumn{1}{l|}{1.63} & \multirow{2}{*}{0.95} & 250 & 0.960                     & \multicolumn{1}{l|}{1.44} & \multirow{2}{*}{0.95} & 250 & 0.968                     & \multicolumn{1}{l|}{1.60} \\
\hline
   	\end{tabular}%
   }
\end{table}
\begin{table}[htbp]
   \caption{Empirical Coverage probability, Mean of Sizes (MS) of confidence bands for DGP1, 2 and 3 when $L=30$ and $\rho=3$.}
   \label{tab:k30rho15}
   \resizebox{\textwidth}{!}{%
   	\begin{tabular}{ccccllcclllcclll}
   		\hline
   		\multicolumn{1}{|c}{}                   & \multicolumn{1}{c|}{}                      & cp                    & N   & \multicolumn{1}{c}{empcp} & \multicolumn{1}{c|}{MS} & cp                    & N   & \multicolumn{1}{c}{empcp} & \multicolumn{1}{c|}{MS} & cp                    & N   & \multicolumn{1}{c}{empcp} & \multicolumn{1}{c|}{MS} \\ \hline
   		\multicolumn{1}{|c}{DGP}                & \multicolumn{1}{c|}{ME}                    & \multicolumn{4}{c|}{$\tau\in[0.2,0.3]$}                                                & \multicolumn{4}{c|}{$\tau\in[0.45,0.55]$}                                                 & \multicolumn{4}{c|}{$\tau\in[0.7,0.8]$}                                                \\ \hline
   		\multicolumn{1}{|c}{\multirow{4}{*}{1}} & \multicolumn{1}{c|}{\multirow{4}{*}{Lap}}  & \multirow{2}{*}{0.9}  & 250 & 0.971                     & \multicolumn{1}{l|}{1.70} & \multirow{2}{*}{0.9}  & 250 & 0.953                     & \multicolumn{1}{l|}{1.50} & \multirow{2}{*}{0.9}  & 250 & 0.964                     & \multicolumn{1}{l|}{1.67} \\
   		\multicolumn{1}{|c}{}                   & \multicolumn{1}{c|}{}                      &                       & 500 & 0.959                     & \multicolumn{1}{l|}{1.25} &                       & 500 & 0.958                     & \multicolumn{1}{l|}{1.13} &                       & 500 & 0.955                     & \multicolumn{1}{l|}{1.27} \\ \cline{3-14} 
   		\multicolumn{1}{|c}{}                   & \multicolumn{1}{c|}{}                      & \multirow{2}{*}{0.95} & 250 & 0.986                     & \multicolumn{1}{l|}{1.90} & \multirow{2}{*}{0.95} & 250 & 0.974                     & \multicolumn{1}{l|}{1.67} & \multirow{2}{*}{0.95} & 250 & 0.979                     & \multicolumn{1}{l|}{1.85} \\
\hline
   	\end{tabular}%
   }
\end{table}

\begin{table}[htbp]
   \caption{Empirical Coverage probability, Mean of Sizes (MS) of confidence bands for DGP1, 2 and 3 when $L=30$ and $\rho=3.5$.}
   \label{tab:k30rho17}
   \resizebox{\textwidth}{!}{%
   	\begin{tabular}{ccccllcclllcclll}
   		\hline
   		\multicolumn{1}{|c}{}                   & \multicolumn{1}{c|}{}                      & cp                    & N   & \multicolumn{1}{c}{empcp} & \multicolumn{1}{c|}{MS} & cp                    & N   & \multicolumn{1}{c}{empcp} & \multicolumn{1}{c|}{MS} & cp                    & N   & \multicolumn{1}{c}{empcp} & \multicolumn{1}{c|}{MS} \\ \hline
   		\multicolumn{1}{|c}{DGP}                & \multicolumn{1}{c|}{ME}                    & \multicolumn{4}{c|}{$\tau\in[0.2,0.3]$}                                                & \multicolumn{4}{c|}{$\tau\in[0.45,0.55]$}                                                 & \multicolumn{4}{c|}{$\tau\in[0.7,0.8]$}                                                \\ \hline
   		\multicolumn{1}{|c}{\multirow{4}{*}{1}} & \multicolumn{1}{c|}{\multirow{4}{*}{Lap}}  & \multirow{2}{*}{0.9}  & 250 & {0.983}                     & \multicolumn{1}{l|}{{1.96}} & \multirow{2}{*}{0.9}  & 250 & 0.968                     & \multicolumn{1}{l|}{1.70} & \multirow{2}{*}{0.9}  & 250 & 0.969                     & \multicolumn{1}{l|}{1.89} \\
   		\multicolumn{1}{|c}{}                   & \multicolumn{1}{c|}{}                      &                       & 500 & 0.973                     & \multicolumn{1}{l|}{1.41} &                       & 500 & 0.971                     & \multicolumn{1}{l|}{1.26} &                       & 500 & 0.967                     & \multicolumn{1}{l|}{1.42} \\ \cline{3-14} 
   		\multicolumn{1}{|c}{}                   & \multicolumn{1}{c|}{}                      & \multirow{2}{*}{0.95} & 250 & 0.991                     & \multicolumn{1}{l|}{2.19} & \multirow{2}{*}{0.95} & 250 & 0.981                     & \multicolumn{1}{l|}{1.91} & \multirow{2}{*}{0.95} & 250 & 0.983                     & \multicolumn{1}{l|}{2.10} \\
\hline
   	\end{tabular}%
   }
\end{table}
\FloatBarrier

\newpage
	\section{Proof}\label{appendix:proof}
	\subsection{Notations}\label{section:notations}
	In this section, we list the notations which appear in the proof below.
	\begin{align*}
		&\widehat{M}_n(\theta;x,\tau) :=\frac{1}{n}\sum_{i=1}^n\psi_{\tau}(Y_i-\theta)\widehat{K}_{U,h}(x-W_i)\\
		&M_n(\theta;x,\tau) := \frac{1}{n}\sum_{i=1}^n\psi_{\tau}(Y_i-\theta){K}_{U,h}(x-W_i),\\
		&M(\theta;x,\tau) := \mathbb{E}\left\{\psi_{\tau}(Y-\theta)|X=x\right\}f_X(x),\\
		&B_n(\theta;x,\tau) :=h^s \cdot\kappa_{s1} \cdot\partial_x^s\varepsilon_\tau f_X(x;\theta), \\
		&V_n(\theta;x,\tau) := \frac{1}{n}\sum_{i=1}^n\psi_{\tau}(Y_i-\theta){K}_{U,h}(x-W_i) - \mathbb{E}\left\{ \psi_{\tau}(Y-\theta){K}_{U,h}(x-W)\right\}\\
		&\hspace{4.6em}=M_n(\theta;x,\tau) - \mathbb{E}\left\{M_n(\theta;x,\tau)\right\},\\
		&\phi_{YW}(t;\tau,\theta)=\mathbb{E}\left\{\psi_{\tau}(Y-\theta)e^{itW}\right\},\\
		&\phi_{YX}(t;\tau,\theta) = \mathbb{E}\left\{\psi_{\tau}(Y-\theta)e^{itX}\right\},\\
		&\widehat{\phi_{YW}}(t;\tau,\theta) = \frac{1}{m}\sum_{j=1}^m \psi_{\tau}(Y_j-\theta)e^{itW_j}.
	\end{align*}
	\subsection{Assumptions}
    For convenience, we list all the assumptions that appear in the main text:
    
    \subsection{Auxiliary Lemmas}
	\begin{lemma}
		(Bound for deconvolution kernel) \label{lemma:bound}
		\begin{itemize}
			\item[(i)] Suppose that $U$ is an ordinary smooth error satisfying \eqref{def:os} of order $\beta$, under Assumption \ref{Assumption:Kernel}, \ref{Assumption:ME} and OS,
			\begin{align*}
				&\sup_{x\in\mathbb{R}}|K_{U,h}(x)| = O\left(h^{-\beta-1}\right),\\
				&\int K_{U,h}^2(w) dw \asymp h^{-2\beta-1}.
			\end{align*}
			\item[(ii)] Suppose that $U$ is a supersmooth error satisfying \eqref{def:ss} of order $(\beta,\beta_0)$, under Assumption \ref{Assumption:Kernel}, \ref{Assumption:ME} and SS,
			\begin{align*}
				&\sup_{x\in\mathbb{R}}|K_{U,h}(x)| = O\left\{ e_U(h)\right\},\\
				&\int K_{U,h}^2(w) dw = O\left\{h^{-\beta} e_U^2(h)\right\}.
			\end{align*}
			\item[(iii)] Suppose that $U$ is a supersmooth error satisfying \eqref{def:ss} of order $(\beta,\beta_0)$, under Assumption \ref{Assumption:Kernel}, \ref{Assumption:ME}, and SS,
			\begin{align*}
				|K_{U,h}(x)|\ge a\cdot H(x) h^{-1/2}e_L(h),
			\end{align*}
			for $x\in \mathcal{X}$, where $a>0$ and 
			\begin{align*}
				\begin{split}
					H(x)= \left \{
					\begin{array}{ll}
						|\cos x|,& \text{if } I_U(t)=o\{R_U(t)\},\\
						|\sin x|, & \text{if } R_U(t)=o\{I_U(t)\}.\\
					\end{array}
					\right.
				\end{split}
			\end{align*}
		\end{itemize}
	\end{lemma}
	\begin{proof} (i) The first conclusion follows by Lemma 3 in \cite{fan1993nonparametric}. For the second conclusion, by the dominated convergence theorem (DCT), we have
		\begin{align*}
			h^{\beta+1} K_{U,h}(hy)\xrightarrow{h\to 0} \frac{1}{2\pi}\int_{-\infty}^{\infty}\exp(-ity)\frac{t^\beta}{c}\phi_K(t)dt =: J(y).
		\end{align*}
		Then, 
		\begin{align*}
			\int K_{U,h}^2(w)dw = h\int K_{U,h}^2(wh)dw \rightarrow h^{-2\beta-1} \int J^2(y)dy = \frac{h^{-2\beta-1}}{2\pi c^2}\int t^{2\beta}\phi_K^2(t)dt,
		\end{align*}
		where the last equality holds by Parseval's identity. Hence, the desired conclusion follows.
		
		The part of (ii) and (iii) follows by the Lemma 3.1 in \cite{fan1992multivariate}.
	\end{proof}
	Consider the function class
	\begin{align}\label{def:functionclassFn}
		\mathcal{F}_n:=\{(w,y)\mapsto K_{U,h}(x-w)\psi_{\tau}(y-\theta):x\in \mathcal{X}, \theta\in\Theta,\tau\in\mathcal{T}\}.
	\end{align}
	\begin{lemma}\label{lemma:VCtype} (VC-type class)
		\begin{itemize}
			
			\item[(i)] Suppose that $U$ is an ordinary smooth error satisfying \eqref{def:os} of order $\beta$ and Assumption \ref{Assumption:Kernel}, \ref{Assumption:ME} are satisfied. Let $D$ be a positive constant such that $\sup_{x\in \mathbb{R}}|K_{U,h}(x)|\le Dh^{-\beta-1}$ . Then there exist constants $A, v \geq e$ independent of $n$ such that
			$$
			\sup _Q N\left(\mathcal{F}_n,\|\cdot\|_{Q, 2}, \delta Dh^{-\beta-1}\right) \leq(A / \delta)^v,\  0<\forall \delta \leq 1
			$$
			where $\sup _Q$ is taken over all finitely discrete distributions on $\mathbb{R}^2$.
			\item[(ii)]  Suppose that $U$ is a supersmooth error satisfying \eqref{def:ss} of order $(\beta,\beta_0)$ and Assumption \ref{Assumption:Kernel}, \ref{Assumption:ME} are satisfied. Let $D$ be a positive constant such that $\sup_{x\in \mathbb{R}}|K_{U,h}(x)|\le Dh^{-\beta/2}e_U(h)$. Then there exist constants $A, v \geq e$ independent of $n$ such that
			$$
			\sup _Q N\left(\mathcal{F}_n,\|\cdot\|_{Q, 2}, \delta Dh^{-\beta/2}e_U(h)\right) \leq(A / \delta)^v,\  0<\forall \delta \leq 1
			$$
			where $\sup _Q$ is taken over all finitely discrete distributions on $\mathbb{R}^2$.
		\end{itemize}
	\end{lemma}
	\begin{proof}
		(i) We first focus on the function class 
		\begin{align*}
			\mathcal{A}=\left\{(w,y)\mapsto \psi_{\tau}(y-\theta):\tau\in\mathcal{T},\theta\in\Theta\right\}.
		\end{align*}
		We want to verify that there exists some constant $A_0,v_0$ such that \begin{align}\label{eq:vctypequantileloss}
			\sup _Q N\left(
			\mathcal{A},\|\cdot\|_{Q, 2}, \delta \right) \leq(A_0 / \delta)^{v_0},\  0<\forall \delta \leq 1.
		\end{align}
		For every fixed $\tau$ and $\theta$, the subgraph of the function $y\mapsto\psi_{\tau}(y-\theta)$,
		\begin{align*}
			&\{(y,u) : u\le\psi_{\tau}(y-\theta)\} \\
			=& \{(y,u):y \ge\theta, \tau-1\le u\le\tau \}\cup \{(y,u): u<\tau-1\}\\
			=&\left(\{(y,u):y \ge\theta,  u\le\tau \}\cap\{(y,u):y \le\theta,  u\ge\tau-1 \} \right)\cup \{(y,u): u<\tau-1\}\\
			=&\left(\{(y,u):y \ge\theta\}\cap\{(y,u):  u\le\tau \}\cap\{(y,u):u\ge\tau-1 \} \right)\cup \{(y,u): u<\tau-1\}\\
			\in& (\mathcal{G}\cap\mathcal{G}\cap\mathcal{G} )\cup \mathcal{G}
		\end{align*}
		where $\mathcal{A}\cap\mathcal{B}: = \{A\cap B: A\in\mathcal{A},B\in\mathcal{B}\}$ and $\mathcal{A}\cup\mathcal{B}: = \{A\cup B: A\in\mathcal{A},B\in\mathcal{B}\}$ and $\mathcal{G}$ is the collection of the negativity set of the vector space spanned by three functions $(y,u)\mapsto y$, $(y,u)\mapsto u$ and $(y,u)\mapsto 1$, i.e.
		\begin{align*}
			\mathcal{G}=\{\{(y,u):a\cdot y + b\cdot u + c\le 0\}:a,b,c\in\mathbb{R}\}.
		\end{align*} 
		By   \citet[Lemma 2.6.15, Lemma 2.6.18 (iii)]{van1996weak},  $\mathcal{G}$ is a VC class of set. Then by  \citet[Lemma 2.6.17 (iii)]{van1996weak}, $(\mathcal{G}\cap\mathcal{G}\cap\mathcal{G} )\cup \mathcal{G}$ is also a VC class of set. Hence,  $\mathcal{A}$ is a VC-subgraph class. Then by Theorem 2.6.17 in \cite{van1996weak}, we know that \eqref{eq:vctypequantileloss} is satisfied.
		
		By Lemma 1 in \cite{kato2018uniform}, we know that there exists $A_1,v_1\ge e$ such that
		\begin{align*}
			\sup _Q N\left(\left\{w\mapsto K_{U,h}(x-w):x\in\mathbb{R}\right\},\|\cdot\|_{Q, 2}, \delta Dh^{-\beta-1}\right) \leq(A_1 / \delta)^{v_1},\  0<\forall \delta \leq 1.
		\end{align*}
		By Corollary A.1 in \cite{chernozhukov2014gaussian}, the desired result follows.
		
		(ii) We shall show that there exists $A,v\ge e$ such that
		\begin{align}\label{l3:cover}
			\sup _Q N\left(\left\{w\mapsto K_{U,h}(x-w):x\in\mathbb{R}\right\},\|\cdot\|_{Q, 2}, \delta Dh^{-\beta/2}e_U(h)\right) \leq(A / \delta)^v,\  0<\forall \delta \leq 1.
		\end{align}
		Then the conclusion follows by \eqref{eq:vctypequantileloss} and Corollary A.1 in  \cite{chernozhukov2014gaussian}. According to Lemmas 1 and 2 of \cite{kato2018uniform}, a sufficient condition of \eqref{l3:cover} is 
		\begin{align}\label{l3:besov}
			\Vert K_{U,h}\Vert _{\dot{B}_{2,1}^{1/2}} = O\left\{h^{-\beta/2}e_U(h)\right\}
		\end{align}
		where $\Vert K_{U,h}\Vert _{\dot{B}_{2,1}^{1/2}}$ is the Besov norm of $K_{U,h}$, defined by
		\begin{align*}
			\Vert K_{U,h}\Vert _{\dot{B}_{2,1}^{1/2}} := \int_{\mathbb{R}}\frac{1}{|u|^{3/2}}\left(\int_{\mathbb{R}} |K_{U,h}(x+u)-K_{U,h}(x)|^2dx\right)^{1/2}du.
		\end{align*}
		We have
		\begin{equation}\nonumber
			\begin{aligned}
				&\Vert  K_{U,h}\Vert _{\dot{B}_{2,1}^{1/2}}
				= \int_\mathbb{R} \frac{1}{|u^{3/2}|} \left(\int_{\mathbb{R}}  |K_{U,h}(x+u)-K_{U,h}(x)|^2dx\right)^{1/2} du \\
				=&\frac{1}{\pi}\int_{\mathbb{R}}\frac{1}{|u^{3/2}|}\left[\int_{\mathbb{R}}\{1-\cos(ut)\} \frac{|\phi_K(th)|^2}{|\phi_U(t)|^2} dt\right]^{1/2}du \\
				\le&\frac{1}{\pi}\left[ \sqrt{2}\int_{[-1,1]^c}\frac{1}{|u^{3/2}|}du\left(\int_{\mathbb{R}}\frac{|\phi_K(th)|^2}{|\phi_U(t)|^2} dt\right)^{1/2} +\frac{1}{\sqrt{2}}\int_{[-1,1]}\frac{1}{|u^{1/2}|}du\left(\int_{\mathbb{R}}\frac{t^2|\phi_K(th)|^2}{|\phi_U(t)|^2}dt\right)^{1/2} \right],
			\end{aligned}
		\end{equation}
		where the second equality comes from Parseval's identity and the first inequality follows from the fact $1-\cos(tu) \le \min \{2,(tu)^2/2\}$. By Parseval's identity and Lemma \ref{lemma:bound}, 
		\begin{align*}
			\int_{\mathbb{R}}\frac{|\phi_K(th)|^2}{|\phi_U(t)|^2} dt = \int K_{U,h}^2(x) dx = O\left\{h^{-\beta}e_U^2(h)\right\}.
		\end{align*}
		By slightly modifying the proof of Lemma 3.1 in \cite{fan1992multivariate}, we can conclude that
		\begin{align*}
			\int_{\mathbb{R}}\frac{t^2|\phi_K(t)|^2}{|\phi_U(t/h)|^2} dt = O\left\{h^{-\beta}e_U^2(h)\right\}.
		\end{align*}
		Hence, \eqref{l3:besov} is satisfied, then the desired result follows.
	\end{proof}
	\begin{lemma} (Convergence of the characteristic function)\label{lemma:concf}
		\begin{align*}
			\sup_{t}\mathbb{E}\sup_{\tau\in\mathcal{T},\theta\in\Theta} \left|\widehat{\phi_{YW}}(t)-\phi_{YW}(t)\right|^2 = O(m^{-1})
		\end{align*}
	\end{lemma}
	\begin{proof}
		Consider the function class
		\begin{align*}
			\mathcal{A}_t:=\left\{(w,y)\mapsto \psi_{\tau}(y-\theta)\cos(tw):\theta\in\Theta,\tau\in\mathcal{T}\right\}
		\end{align*}
		By \eqref{eq:vctypequantileloss}, we know that we know that there exists $A_1,v_1\ge e$ such that for any $t$,
		\begin{align*}
			\sup _Q N\left(\mathcal{A}_t,\|\cdot\|_{Q, 2}, \delta \right) \leq(A_1 / \delta)^{v_1},\  0<\forall \delta \leq 1.
		\end{align*}
		Then by Theorem 2.14.1 in \cite{van1996weak}, there exists some constant $C>0$ independent of $t$ such that
		\begin{align*}
			\mathbb{E}\sup_{\tau\in\mathcal{T},\theta\in\Theta} \left|\frac{1}{m}\sum_{j=1}^n\psi_{\tau}(Y_j-\theta)\cos(tW_j) -\mathbb{E}\left\{\psi_{\tau}(Y-\theta)\cos(tW)\right\}\right|^2 \le \frac{C}{m}
		\end{align*}
		By similar argument (larger $C$ if necessary), 
		\begin{align*}
			\mathbb{E}\sup_{\tau\in\mathcal{T},\theta\in\Theta} \left|\frac{1}{m}\sum_{j=1}^n\psi_{\tau}(Y_j-\theta)\sin(tW_j) -\mathbb{E}\left\{\psi_{\tau}(Y-\theta)\sin(tW)\right\}\right|^2 \le \frac{C}{m}
		\end{align*}
		Hence, we can conclude that
		\begin{align*}
			\sup_{t}\mathbb{E}\sup_{\tau\in\mathcal{T},\theta\in\Theta} \left|\widehat{\phi_{YW}}(t)-\phi_{YW}(t)\right|^2 \le \frac{2C}{m}
		\end{align*}
	\end{proof}
	
	\subsection{Key Lemmas}
	\begin{lemma}\label{lemma:bias}
		 Under Assumption \ref{Assumption:Continuity} and \ref{Assumption:ME}, 
		\begin{align*}
			\sup_{x\in \mathcal{X}, \theta\in\Theta,\tau\in\mathcal{T}}\left| \mathbb{E}\left\{M_n(\theta;x,\tau)\right\} - M(\theta;x,\tau)-B_n(\theta;x,\tau)\right| = o(h^s).
		\end{align*}
	\end{lemma}
	\begin{proof}
		By the definition of deconvolution kernel,
		\begin{align*}
			\mathbb{E}\left\{M_n(\theta;x,\tau)\right\}=&\bbE\left[ K_{U,h}\left(x-W\right)\psi_{\tau}(Y-\theta)\right]\\
			=&\bbE\left[ \bbE\left\{K_{U,h}\left(x-X-U\right)|X,Y\right\}\psi_{\tau}(Y-\theta)\right]\\
			=&\bbE\left[ \bbE\left\{K_{U,h}\left(x-X-U\right)|X\right\}\psi_{\tau}(Y-\theta)\right]\\
			=&\bbE \left[\left\{ \int \exp\left(-it\cdot\frac{x-X}{h}\right)\phi_K(t) \frac{\bbE\{\exp(itU/h)\}}{\phi_U(t/h)} dt\right\} \psi_{\tau}(Y-\theta) \right]\\
			=&\bbE\left[K_h\left({x-X}\right)\psi_{\tau}(Y-\theta)\right].
		\end{align*}
		Using the Taylor expansion argument, we have
		\begin{align*}
			&\bbE\left[K_h\left({x-X}\right)\psi_{\tau}(Y-\theta)\right]\\
			=&\bbE\left[K_h\left({x-X}\right)\mathbb{E}\left\{\psi_{\tau}(Y-\theta)|X\right\}\right]\\
			=& \int K_{h}(x-x_0)\mathbb{E}\left\{\psi_{\tau}(Y-\theta)|X=x_0\right\}f_X(x_0)dx_0\\
			=& \int K(x_0)\mathbb{E}\left\{\psi_{\tau}(Y-\theta)|X=x-hx_0\right\}f_X(x-hx_0)dx_0\\
			=&\int K(x_0)\{\varepsilon_\tau(x;\theta)f_X(x) + h^s\cdot \partial_x^s\varepsilon_\tau f_X(x;\theta)\cdot x_0^s + h^{s+1}\cdot\partial_x^{s+1} \varepsilon_\tau f_X(\xi,\theta)\cdot x_0^{s+1}\}dx_0\\
			=& M(\theta;x,\tau) + B_n(\theta;x,\tau) + h^{s+1}\int K(x_0)\cdot\partial_x^{s+1} \varepsilon_\tau f_X(\xi,\theta)\cdot x_0^{s+1}dx_0,
		\end{align*}
		where $\xi$ lies between $x$ and $x-hx_0$. 
		Then the conclusion follows from Assumption \ref{Assumption:Continuity}.
	\end{proof}

	\begin{lemma}\label{lemma:variance}\
		\begin{itemize}
			\item[(i)] Suppose that $U$ is an ordinary smooth error satisfying \eqref{def:os} of order $\beta$, under Assumption \ref{Assumption:boundedsupport}-\ref{Assumption:PartialBoundAway} and OS,
			\begin{align*}
				\sup_{x\in \mathcal{X},\theta\in \Theta,\tau\in \mathcal{T}}\left|V_n(\theta;x,\tau)\right| = O_P\left(\frac{\{\log(1/h)\}^{1/2}}{\sqrt{nh^{2\beta+1}}}\right)
			\end{align*}
			\item[(ii)]  Suppose that $U$ is a supersmooth error satisfying \eqref{def:ss} of order $(\beta,\beta_0)$, under Assumption \ref{Assumption:boundedsupport}-\ref{Assumption:PartialBoundAway} and SS
			\begin{align*}
				\sup_{x\in \mathcal{X},\theta\in \Theta,\tau\in \mathcal{T}}\left|V_n(\theta;x,\tau)\right| = O_P\left(\frac{\{\log(1/h)\}^{1/2}e_U(h)}{\sqrt{n}}\right)
			\end{align*}
		\end{itemize} 
	\end{lemma}
	\begin{proof} (i) We apply Corollary 5.1 in \cite{chernozhukov2014gaussian} on the function class $\mathcal{F}_n$ given in \eqref{def:functionclassFn}. Define the envelope function of $\mathcal{F}_n$ be $F = Dh^{-\beta-1}$ with $D$ sufficiently large such that $\sup_{x\in\mathbb{R}}|K_{U,h}(x)|\le Dh^{-\beta-1} $. By Lemma \ref{lemma:VCtype}(i), the condition of Corollary 5.1 in \cite{chernozhukov2014gaussian} is satisfied. To apply the Corollary 5.1 in \cite{chernozhukov2014gaussian}, we bound the involved quantities:
		\begin{itemize}
			\item (The element-wise second moment)
			\begin{align*}
				\sup_{f\in\mathcal{F}}\mathbb{P}f^2 = &\sup_{x\in \mathcal{X},\tau\in\mathcal{T},\theta\in\Theta}\mathbb{E}\left\{ K_{U,h}^2(x-W)\psi_{\tau}^2(Y-\theta)\right\}\\
				=&\sup_{x\in \mathcal{X},\tau\in\mathcal{T},\theta\in\Theta} \int K_{U,h}^2(x-w)\mathbb{E}\left\{\psi_{\tau}^2(Y-\theta)|W=w\right\}f_W(w)dw \\
				=&\sup_{x\in \mathcal{X},\tau\in\mathcal{T},\theta\in\Theta}h\cdot\int K_{U,h}^2(wh)\mathbb{E}\left\{\psi_{\tau}^2(Y-\theta)|W=x-wh\right\}f_W(x-wh)dw \\
				\le & C_W\cdot \int K_{U,h}^2(w) dw \asymp h^{-2\beta-1},
			\end{align*}
			where the last equality comes from Lemma \ref{lemma:bound} and $C_W>0$ is the constant such that $f_W(x)\le C_W$.
			\item (The second moment of envelope function) By the definition of $F$, we immediately have
			\begin{align*}
				\|F\|^2_{\mathbb{P},2} = \mathbb{E}\left\{ F^2\right\} = O(h^{-2\beta-2}).
			\end{align*}
			\item (The second moment of maximal of envelope function)
			\begin{align*}
				\|M\|_2^2 = \mathbb{E}\left\{\max_{1\le i\le n} F^2(W_i,Y_i)\right\} = O(h^{-2\beta-2}).
			\end{align*}
		\end{itemize}
		Hence, by Corollary 5.1 in \cite{chernozhukov2014gaussian}, using their notations, we have $\sigma^2\asymp h^{-2\beta-1}$, $\|F\|_{P,2} = O(h^{-\beta-1})$, $\|M\|_2=O(h^{-\beta-1})$ and constant $A$ and $v$, then
		\begin{align*}
			&\mathbb{E}\sup_{x\in \mathcal{X},\tau\in\mathcal{T},\theta\in \Theta}\left|\frac{1}{n}\sum_{i=1}^n K_{U,h}(x-W)\psi_{\tau}(Y_i-\theta) - \mathbb{E}\left\{K_{U,h}(x-W)\psi_{\tau}(Y-\theta)\right\}\right|\\
			&=O\left(\frac{1}{\sqrt{n}}\left[\frac{\{\log(1/h)\}^{1/2}}{h^{\beta+1/2}} + \frac{\log(1/h)}{n^{1/2}h^{\beta+1}}\right]\right) = O\left[\frac{\{\log(1/h)\}^{1/2}}{\sqrt{nh^{2\beta+1}}}\right].
		\end{align*}
		where the last equality holds due to Assumption OS. Then the results follow by Markov's inequality.
		
		(ii) By Lemma \ref{lemma:VCtype}(ii), the condition of Corollary 5.1 in \cite{chernozhukov2014gaussian} is satisfied with $F = Dh^{-\beta/2}e_U(h)$ for some constant $D>0$. We bound the quantities appeared in (i) when $U$ is supersmooth. \begin{itemize}
			\item (The element-wise second moment)
			\begin{align*}
				\sup_{f\in\mathcal{F}_n}\mathbb{P}f^2=&\sup_{x\in \mathcal{X},\theta\in\Theta,\tau\in\mathcal{T}}\mathbb{E}\left\{ K_{U,h}^2(x-W)\psi_{\tau}^2(Y-\theta)\right\}\\
				\lesssim& e_U^2(h)\cdot\mathbb{E}\left\{ \sup_{\theta\in\Theta,\tau\in\mathcal{T}}\psi_{\tau}^2(Y-\theta)\right\} = O\{e^2_U(h)\},
			\end{align*}
			where the inequality follows from Lemma \ref{lemma:bound}(ii).
			\item (The second moment of envelope function) By the definition of $F$, we immediately have
			\begin{align*}
				\|F\|_{\mathbb{P},2}^2 = \mathbb{E}\left\{ F^2\right\} \asymp h^{-\beta}e_U^2(h)
			\end{align*}
			\item (The second moment of maximal of Envelope function)
			\begin{align*}
				\|M\|_2^2:=\mathbb{E}\left\{\max_{1\le i\le n} F^2(W_i,Y_i)\right\} =O\{h^{-\beta}e_U^2(h)\}.
			\end{align*}
		\end{itemize}
		Hence, by Corollary 5.1 in \cite{chernozhukov2014gaussian}, using their notations with $\sigma^2\asymp  e_U^2(h)$, $\|F\|_{P,2} \asymp h^{-\beta/2}e_U(h)$, $\|M\|_2=O\{h^{-\beta/2}e_U(h)\}$ and constant $A$ and $v$,
		\begin{align*}
			&\mathbb{E}\sup_{x\in \mathcal{X},\tau\in\mathcal{T},\theta\in \Theta}\left|\frac{1}{n}\sum_{i=1}^n K_{U,h}(x-W)\psi_{\tau}(Y_i-\theta) - \mathbb{E}\left\{K_{U,h}(x-W)\psi_{\tau}(Y-\theta)\right\}\right|\\
			&=O\left(\frac{1}{\sqrt{n}}\left[\{\log(1/h)\}^{1/2}e_U(h) + \frac{\log(1/h)h^{-\beta/2}e_U(h)}{n^{1/2}}\right]\right) = O\left[\frac{\{\log(1/h)\}^{1/2}e_U(h)}{\sqrt{n}}\right]
		\end{align*}
		where the last equality holds due to Assumption SS. Then the results follow by Markov's inequality.
	\end{proof}

	\begin{lemma} (Effect of estimating $U$) \label{lemma:effectU}
		\begin{itemize}
			\item[(i)]  Suppose that $U$ is an ordinary smooth error satisfying \eqref{def:os} of order $\beta$, under Assumption \ref{Assumption:boundedsupport}-\ref{Assumption:PartialBoundAway} and OS, we have
			\begin{align*}
				\sup_{x\in\mathcal{X},\tau\in\mathcal{T},\theta\in\Theta}\left|\widehat{M}_n(\theta;x,\tau) - M_n(\theta;x,\tau)\right| = O_P\left(m^{-1/2}n^{-1/2}h^{-2\beta-1}+m^{-1/2}h^{-\beta}\right).
			\end{align*}
			\item[(ii)] Suppose that $U$ is a supersmooth error satisfying \eqref{def:ss} of order $(\beta,\beta_0)$, under Assumption \ref{Assumption:boundedsupport}-\ref{Assumption:PartialBoundAway} and SS, we have
			\begin{align*}
				\sup_{x\in\mathcal{X},\tau\in\mathcal{T},\theta\in\Theta}\left|\widehat{M}_n(\theta;x,\tau) - M_n(\theta;x,\tau)\right| = &O_P\left\{m^{-1/2}n^{-1/2}h^{2\beta_0-1}\exp(2h^{-\beta}/\gamma)\right.\\
				&\qquad\qquad\left.+m^{-1/2}h^{\beta_0}\exp(h^{-\beta}/\gamma)\right\}.
			\end{align*}
		\end{itemize}
	\end{lemma}
	\begin{proof} With the notations in Section \ref{section:notations}, we first consider the ordinary smooth case.
		\begin{align*}
			&\left|\widehat{M}_n(\theta;x,\tau)-M_n(\theta;x,\tau)\right|\\
			=&\left|\frac{1}{n}\sum_{j=1}^n\psi_{\tau}(Y_j-\theta)\left\{\widehat{K}_{U,h}(x-W_j) - K_{U,h}(x-W_j)\right\}\right|\\
			=&\left|\frac{1}{2\pi nh}\sum_{j=1}^n\psi_{\tau}(Y_j-\theta)\int_{-\infty}^{\infty} \exp\{-it(x-W_j)/h\}\phi_K(t)\left\{ \frac{1}{\widehat{\phi}_U(t/h)} -\frac{1}{\phi_U(t/h)} \right\}dt\right| \\
			=&\left|\frac{1}{2\pi h}\int_{-\infty}^{\infty} \exp(-itx/h)\widehat{\phi_{YW}}(t/h;\tau,\theta)\phi_K(t)\left\{ \frac{1}{\widehat{\phi}_U(t/h)} -\frac{1}{\phi_U(t/h)} \right\}dt\right|\\
			\le&\left|\frac{1}{2\pi h}\int_{-\infty}^{\infty} \exp(-itx/h)\left\{\widehat{\phi_{YW}} (t/h;\tau,\theta) -{\phi_{YW}} (t/h;\tau,\theta)\right\}\phi_K(t)\left\{ \frac{1}{\widehat{\phi}_U(t/h)} -\frac{1}{\phi_U(t/h)} \right\}dt\right| \\
			+&\left|\frac{1}{2\pi h}\int_{-\infty}^{\infty} \exp(-itx/h){\phi_{YW}} (t/h;\tau,\theta)\phi_K(t)\left\{ \frac{1}{\widehat{\phi}_U(t/h)} -\frac{1}{\phi_U(t/h)} \right\}dt\right|\\
			=&\left|\frac{1}{2\pi h}\int_{-\infty}^{\infty} \exp(-itx/h)\left\{\frac{\widehat{\phi_{YW}} (t/h;\tau,\theta) - {\phi_{YW}} (t/h;\tau,\theta) }{\phi_{U}(t/h)} \right\}\phi_K(t)\left\{ \frac{\phi_U(t/h)}{\widehat{\phi}_U(t/h)} -1 \right\}dt\right| \\
			+&\left|\frac{1}{2\pi h}\int_{-\infty}^{\infty} \exp(-itx/h){\phi_{YX}} (t/h;\tau,\theta)\phi_K(t)\left\{ \frac{\phi_U(t/h)}{\widehat{\phi}_U(t/h)} -1 \right\}dt\right|\\
			\lesssim &  \frac{1}{h}\left[\int_{-1}^1  \left\{\frac{\widehat{\phi_{YW}} (t/h;\tau,\theta) - {\phi_{YW}} (t/h;\tau,\theta) }{\phi_{U}(t/h)} \right\}^2 dt \right]^{1/2}\cdot \left[\int_{-1}^1  \left\{ \frac{\phi_U(t/h)}{\widehat{\phi}_U(t/h)} -1 \right\}^2 dt\right]^{1/2} \\
			+&  \frac{1}{h}\left[\int_{-1}^1  \left|\phi_{YX}(t/h;\tau,\theta)\right| dt\right]^{1/2} \cdot\left[\int_{-1}^1  \left|\phi_{YX}(t/h;\tau,\theta)\right|\left\{ \frac{\phi_U(t/h)}{\widehat{\phi}_U(t/h)} -1 \right\}^2 dt\right]^{1/2} 
		\end{align*}
		where the last inequality follows from Cauchy-Schwartz inequality. 
		
		Next, we bound those quantities. We know that
		\begin{align*}
			\int_{-1}^1  \left\{ \frac{\phi_U(t/h)}{\widehat{\phi}_U(t/h)} -1 \right\}^2 dt &=  O_P\left(h^{-2\beta}\right)\cdot\int_{-1}^{1} \left\{\phi_U(t/h)-\widehat{\phi}_U(t/h)\right\}^2 dt= O_P\left(m^{-1}h^{-2\beta}\right),
		\end{align*}
		where the last equality follows from the fact that by Lemma \ref{lemma:concf},
		\begin{align*}
			\mathbb{E}\left[\int_{-1}^{1} \left\{\phi_U(t/h)-\widehat{\phi}_U(t/h)\right\}^2 dt\right] = \int_{-1}^{1}\mathbb{E}\left\{\phi_U(t/h)-\widehat{\phi}_U(t/h)\right\}^2dt = O\left(m^{-1}\right).
		\end{align*}
		For the remaining terms,
		\begin{align*}
			&\mathbb{E}\sup_{\theta\in\Theta,\tau\in\mathcal{T}}\int_{-1}^1  \left\{\frac{\widehat{\phi_{YW}} (t/h;\tau,\theta) - {\phi_{YW}} (t/h;\tau,\theta) }{\phi_{U}(t/h)} \right\}^2 dt \\
			\lesssim& h^{-2\beta} \mathbb{E} \sup_{\theta\in\Theta,\tau\in\mathcal{T}}\int_{-1}^1 \left\{\widehat{\phi_{YW}} (t/h;\tau,\theta) - {\phi_{YW}} (t/h;\tau,\theta)\right\}^2 dt \lesssim n^{-1}h^{-2\beta},
		\end{align*}
		and by Assumption \ref{Assumption:boundedsupport}(iii),
		\begin{align*}
			\sup_{\theta\in\Theta,\tau\in\mathcal{T}}\int_{-1}^1  \left|\phi_{YX}(t/h;\tau,\theta)\right| dt \le h\cdot \int_{-h^{-1}}^{h^{-1}} \sup_{\theta\in\Theta,\tau\in\mathcal{T}}\left|\phi_{YX}(t;\tau,\theta)\right|dt \le h\cdot \int_{-\infty}^{\infty} |\phi_{X}(t)|dt\lesssim h.
		\end{align*}
		and by Assumption \ref{Assumption:boundedsupport}(iii),
		\begin{align*}
			\int_{-1}^1  \left|\phi_{YX}(t/h;\tau,\theta)\right|\left\{ \frac{\phi_U(t/h)}{\widehat{\phi}_U(t/h)} -1 \right\}^2 dt  =& h\cdot\int_{-1/h}^{1/h}  \left|\phi_{YX}(t;\tau,\theta)\right|\left\{ \frac{\phi_U(t)}{\widehat{\phi}_U(t)} -1 \right\}^2 dt\\
			=& O_P(h^{-2\beta+1})\int_{-1/h}^{1/h}  \left|\phi_{YX}(t;\tau,\theta)\right| \left\{ \phi_U(t) - \widehat{\phi}_U(t)\right\}^2 dt\\
			=& O_P(m^{-1}h^{-2\beta+1})
		\end{align*}
		Hence, the result follows.
		
		For supersmooth measurement error, the argument is similar but using the fact   $|\phi_U(t)|\ge d_0|t|^{\beta_0}\exp(-|t|^{\beta}/\gamma)$.
		
	\end{proof}
	
	Combining Lemma \ref{lemma:bias}, \ref{lemma:variance} and \ref{lemma:effectU}, we immediately have the following Lemma.
	\begin{lemma}\label{lemma:biasandvariance}
		(i) Suppose that $U$ is an ordinary smooth error satisfying \eqref{def:os} of order $\beta$, under Assumption \ref{Assumption:boundedsupport}-\ref{Assumption:PartialBoundAway} and OS, we have
		\begin{align*}
			&\sup_{x\in \mathcal{X},\tau\in\mathcal{T},\theta\in\Theta} \left| \widehat{M}_n(\theta;x,\tau) - M(\theta;x,\tau) - B_n(\theta;x,\tau)-V_n(\theta;x,\tau)\right| = o\left(h^s \right)+O_P\left( R_{U,OS}\right),\\
			&\sup_{x\in \mathcal{X},\tau\in\mathcal{T},\theta\in\Theta} \left| \widehat{M}_n(\theta;x,\tau) - M(\theta;x,\tau)\right| = O_P\left(h^s + \frac{\{\log(1/h)\}^{1/2}}{\sqrt{nh^{2\beta+1}}}\right).
		\end{align*}
		(ii) Suppose that $U$ is a supersmooth error satisfying \eqref{def:ss} of order $(\beta,\beta_0,\beta_1)$, under Assumption \ref{Assumption:boundedsupport}-\ref{Assumption:PartialBoundAway} and SS, we have
		\begin{align*}
			&\sup_{x\in \mathcal{X},\tau\in\mathcal{T},\theta\in\Theta} \left| \widehat{M}_n(\theta;x,\tau) - M(\theta;x,\tau)- B_n(\theta;x,\tau)-V_n(\theta;x,\tau)\right|  = o\left(h^s\right) + O_P\left( R_{U,SS}\right),\\
			&\sup_{x\in \mathcal{X},\tau\in\mathcal{T},\theta\in\Theta} \left| \widehat{M}_n(\theta;x,\tau) - M(\theta;x,\tau)\right|  = O_P\left(h^s + \frac{\{\log (1/h)\}^{1/2}e_U(h)}{\sqrt{n}}\right).
		\end{align*}
	\end{lemma}
	
	Finally, we have a lemma of the lower bound of the variance.
	\begin{lemma} (Lower bound of the variance)\label{lemma:variancelowerboundos}\ 
		\begin{itemize}
			\item [(i)] Supposed $U$ is an ordinary smooth error of order $\beta$ satisfying \eqref{def:os} and Assumption \ref{Assumption:Continuity}, \ref{Assumption:Kernel} and \ref{Assumption:lowerbound} hold. Then $\inf_{x\in \mathcal{X},\tau\in\mathcal{T}}\sigma^2_n(x;\tau) \ge Cn^{-1}h^{-2\beta-1}$ for some constant $C>0$.
			\item [(ii)] 	Supposed $U$ is an supersmooth error of order $(\beta,\beta_0)$ satisfying \eqref{def:ss} and Assumption \ref{Assumption:Continuity}, \ref{Assumption:Kernel} and \ref{Assumption:lowerbound} hold. Then  $\inf_{x\in \mathcal{X},\tau\in\mathcal{T}}\sigma^2_n(x;\tau) \ge Cn^{-1}e_L^2(h)$ for some constant $C>0$.
		\end{itemize}
	\end{lemma}
	\begin{proof}
		(i) 
		From the proof of Lemma \ref{lemma:bias}, we have, for $(x,\tau)\in \mathcal{X}\times\mathcal{T}$ uniformly, 
		\begin{equation}
			\begin{aligned}\label{lmproof:bias}
				\mathbb{E}\left\{ K_{U,h}(x-W)\psi_{\tau}(Y-\theta_{\tau}(x))\right\} = O(h^{s}),
			\end{aligned}
		\end{equation}
		where the conclusion follows from Assumption \ref{Assumption:Continuity} and \ref{Assumption:Kernel} and the fact $M(x;\theta_{\tau}(x),\tau)\equiv 0$.
		\begin{align}
			&\mathbb{E}\left[ K_{U,h}^2(x-W)\psi_{\tau}^2\left\{Y-\theta_{\tau}(x)\right\}\right]\notag\\
			=&\int K_{U,h}^2(x-w)\mathbb{E}\left[\psi_{\tau}^2\{Y-\theta_{\tau}(x)\}|W=w\right]f_W(w)dw \notag\\
			=& h\int K_{U,h}^2(wh)\mathbb{E}\left[\psi_{\tau}^2\{Y-\theta_{\tau}(x)\}|W=x-hw\right]f_W(x-hw)dw\notag \\
			=& h\mathbb{E}\left\{\psi_{\tau}^2(Y-\theta_{\tau}(x))|W=x\right\}f_W(x) \int K_{U,h}^2(wh) dw\label{lm:term1} \\
			+& h \int K_{U,h}^2(wh) \Big[\mathbb{E}\left\{\psi_{\tau}^2(Y-\theta_{\tau}(x))|W=x-hw\right\}f_W(x-hw)\notag\\
			 &\qquad\qquad\qquad\qquad- \mathbb{E}\left\{\psi_{\tau}^2(Y-\theta_{\tau}(x))|W=x\right\}f_W(x) \Big]\label{lm:term2} dw.
		\end{align}
		
		For term \eqref{lm:term1}, by the dominated convergence theorem, we have
		\begin{align*}
			h^{\beta+1} K_{U,h}(hy)\rightarrow \frac{1}{2\pi}\int_{-\infty}^{\infty}\exp(-ity)\frac{t^\beta}{c}\phi_K(t)dt =: J(y).
		\end{align*}
		And by Lemma 3 in \cite{fan1993nonparametric}, we know that $|h^{\beta+1}K_{U,h}(xh)|\le C\min\{1,x^{-2}\}$ for some constant $C>0$. Hence, 
		\begin{align*}
			h^{2\beta+2}\int K_{U,h}^2(wh)dw \rightarrow \int J^2(w) dw < \infty.
		\end{align*}
		Then $\eqref{lm:term1}\asymp h^{-2\beta-1}.$
		
		For term \eqref{lm:term2}, by Assumption \ref{Assumption:lowerbound}, we know that for any $a>0$, there exists $b>0$ such that for any $|w_1-w_2|\le b$,
		$$\sup_{x\in\mathcal{X},\tau\in\mathcal{T}}\left|\mathbb{E}\left[\psi_{\tau}^2\{Y-\theta_{\tau}(x)\}|W=w_1\right]f_W(w_1) - \mathbb{E}\left[\psi_{\tau}^2\{Y-\theta_{\tau}(x)\}|W=w_2\right]f_W(w_2)\right|\le a. $$ Hence for any $a>0$,
		\begin{align*}
			\sup_{x\in \mathcal{X},\tau\in\mathcal{T}}|\eqref{lm:term2}| \le& ah\int_{|w|\le b/h} K_{U,h}^2(wh)dw + 2\|f_W(w)\|_{\infty}h\int_{|w|>b/h}  K_{U,h}^2(wh)dw\\
			\le & ah \int K_{U,h}^2(wh)dw + 2C\|f_W(w)\|_{\infty}h^{-2\beta-1}\int_{|w|> b/h} w^{-2}dw\\
			\le & D\left(ah^{-2\beta-1}+h^{-2\beta}/b\right),
		\end{align*}
		where $D>0$ is a constant. Then we have, uniformly over $(x,\tau)\in \mathcal{X}\times\mathcal{T}$, $\eqref{lm:term2} = o(h^{-2\beta-1})$.
		
		Combining all the results, by Assumption \ref{Assumption:lowerbound}, 
		\begin{align*}
			&\inf_{x\in \mathcal{X},\tau\in\mathcal{T}}\sigma_n^2(x)\\
			 &= \inf_{x\in \mathcal{X},\tau\in\mathcal{T}}n^{-1}\left( \mathbb{E}[K_{U,h}^2(x-W)\psi_{\tau}^2\{Y-\theta_{\tau}(x)\}] - \mathbb{E}[K_{U,h}(x-W)\psi_{\tau}\{Y-\theta_{\tau}(x)\}]^2\right)\\
			&\ge An^{-1}h^{-2\beta-1}\{1-o(1)\}
		\end{align*}
		for some constant $A>0$.
		
		(ii)
		Same as \eqref{lmproof:bias}, for some constant $c>0$, we have
		\begin{align*}
			\sup_{x\in \mathcal{X},\tau\in\mathcal{T}}\left|	\mathbb{E}\left\{ K_{U,h}(x-W)\psi_{\tau}(Y-\theta_{\tau}(x))\right\}\right| = O(h^s).
		\end{align*}
		For the second moment, by Lemma \ref{lemma:bound}(iii),
		\begin{align*}
			&\mathbb{E}\left\{ K_{U,h}^2(x-W)\psi_{\tau}^2(Y-\theta_{\tau}(x))\right\}\\
			=&\int K_{U,h}^2(x-w)\mathbb{E}\left[\psi_{\tau}^2\{Y-\theta_{\tau}(x)\}|W=w\right]f_W(w)dw\\
			\ge &\frac{a^2e_L^2(h)}{h} \int_{-1}^1 \cos^2\left(\frac{x-w}{h}\right)\mathbb{E}\left[\psi_{\tau}^2\{Y-\theta_{\tau}(x)\}|W=w\right]f_W(w)dw\\
			=& a^2 e_L^2(h) \int_{-1}^1 \cos^2(w)\mathbb{E}\left[\psi_{\tau}^2\{Y-\theta_{\tau}(x)\}|W=x-hw\right]f_W(x-hw) dw\\
			= & a^2 e_L^2(h) \mathbb{E}\left[\psi_{\tau}^2\{Y-\theta_{\tau}(x)\}|W=x\right]f_W(x)\int_{-1}^1 \cos^2(w)dw + O\{he_L^2(h)\},
		\end{align*}
		where the last equality follows from Assumption \ref{Assumption:lowerbound}.
		Hence, the desired result follows.
	\end{proof}
	
	\subsection{Proof of Theorem \ref{thm:criteron_consistency}}
	We first focus on the Bahadur expansion of $\widehat{\theta}_{\tau}(x)-\theta_{\tau}(x)$.
	
	\textbf{Step 1: Uniform Convergence Rate of $\widehat{\theta}_{\tau}(x)$}: We first show that
	\begin{align}\label{p1:unirate}
		\sup_{x\in \mathcal{X},\tau\in\mathcal{T}}\left| \widehat{\theta}_{\tau}(x) - \theta_{\tau}(x)\right| = O_P\left(\alpha_n \right),
	\end{align}
	where $\alpha_n:=h^s + \{nh^{2\beta+1}/\log (1/h)\}^{-1/2} + R_{U,OS}$. 
	By Lemma \ref{lemma:biasandvariance}, we know that 
	\begin{align}\label{p1:uniform}
		\widehat{M}_n(\theta;x,\tau) = M(\theta;x,\tau) + O_P(\alpha_n) \text{ uniformly in } (x,\tau,\theta)\in \mathcal{X}\times\mathcal{T}\times\Theta
	\end{align}
	By the definition of $\widehat{\theta}_{\tau}(x)$, we know that $\left|\widehat{M}_n\left\{\widehat{\theta}_{\tau}(x);x,\tau\right\}\right| \le \left|\widehat{M}_n\left\{\theta_{\tau}(x);x,\tau\right\}\right|$. By \eqref{p1:uniform}, we have, uniformly in $(x,\tau)\in \mathcal{X}\times\mathcal{T}$
	\begin{align*}
		\left|\widehat{M}_n\left\{\widehat{\theta}_{\tau}(x);x,\tau\right\}\right| \le \left|\widehat{M}_n\left\{\theta_{\tau}(x);x,\tau\right\}\right| = |M\left\{\theta_{\tau}(x);x,\tau\right\}| + O_P(\alpha_n).
	\end{align*}
	Since $M\left\{\theta_{\tau}(x);x,\tau\right\}=0$, we have
	\begin{equation}\label{p1:middlestep}
		\begin{aligned}
			\left|M\left\{\widehat{\theta}_{\tau}(x);x,\tau\right\}\right|  &= \left|M\left\{\widehat{\theta}_{\tau}(x);x,\tau\right\}\right| - \left|M\left\{{\theta_{\tau}}(x);x,\tau\right\}\right|\\
			&\le \left|M\left\{\widehat{\theta}_{\tau}(x);x,\tau\right\}\right| - \left|\widehat{M}_n\left\{\widehat{\theta}_{\tau}(x);x,\tau\right\}\right| + O_P(\alpha_n)\\
			&\le \left|M\left\{\widehat{\theta}_{\tau}(x);x,\tau\right\}-\widehat{M}_n\left\{\widehat{\theta}_{\tau}(x);x,\tau\right\}\right| + O_P(\alpha_n)\\
			& = O_P(\alpha_n) \text{ uniformly in } (x,\tau)\in \mathcal{X}\times\mathcal{T} \ (\text{by }\eqref{p1:uniform})
		\end{aligned}
	\end{equation}
	By Assumption \ref{Assumption:PartialBoundAway},
	\begin{align}\label{p1:taylorboundaway}
		\sup_{x\in \mathcal{X},\tau\in\mathcal{T}}\left| M\left\{\widehat{\theta}_{\tau}(x);x,\tau\right\}\right| =& \sup_{x\in \mathcal{X},\tau\in\mathcal{T}}\left| M\left\{\widehat{\theta}_{\tau}(x);x,\tau\right\} - M\left\{\theta_{\tau}(x);x,\tau\right\}\right|\notag\\
		\ge& c \cdot\sup_{x\in \mathcal{X},\tau\in\mathcal{T}}\left| \widehat{\theta}_{\tau}(x) - \theta_{\tau}(x)\right|,
	\end{align}
	where $c>0$ is a constant such that
	\begin{align*}
		\partial_{\theta}M(\theta;x,\tau) =& \partial_{\theta}\left\{\int_{-\infty}^{\theta} (\tau-1)f_{Y|X}(y|x)dy + \int_{\theta}^{\infty}\tau f_{Y|X}(y|x)dy \right\} f_X(x)\\
		=& \left\{(\tau-1)f_{Y|X}(\theta|x)dy - \tau f_{Y|X}(\theta|x) \right\}f_X(x) = - f_{X,Y}(x,\theta) \le -c.
	\end{align*}
	By Assumption \ref{Assumption:boundedsupport}(ii), we know that such $c$ exists.
	By \eqref{p1:middlestep} and \eqref{p1:taylorboundaway}, we proved \eqref{p1:unirate}.
	
	\textbf{Step 2: Uniform Linear Expansion of $\widehat{\theta}_{\tau}(x)$:} We approximate $\widehat{M}_{n}(\theta;x,\tau)$ by a linear function of $\theta$ around $\theta_{\tau}(x)$ defined by
	$$L_{n}(\theta;x,\tau):= \widehat{M}_n\left\{\theta_{\tau}(x);x,\tau\right\} + \partial_{\theta}M\left\{\theta_{\tau}(x);x,\tau\right\}\cdot\{\theta-\theta_{\tau}(x)\}.$$
	Denote
	\begin{align}\label{p1s2:deftilde}
		\widetilde{\theta}_{\tau}(x) = \theta_{\tau}(x) -  \frac{\widehat{M}_n\left\{\theta_{\tau}(x);x,\tau\right\}}{\partial_{\theta}M\left\{\theta_{\tau}(x);x,\tau\right\}}
	\end{align}
	Indeed, $L_n\left\{\widetilde{\theta}_{\tau}(x);x,\tau\right\}\equiv 0$. 
	In the following, we will show that
	\begin{align}\label{p1s2:first}
		&\sup_{x\in \mathcal{X},\tau\in\mathcal{T}}\left|\widehat{M}_{n}\left\{\widehat{\theta}_{\tau}(x);x,\tau\right\}-L_n\left\{\widehat{\theta}_{\tau}(x);x,\tau\right\}\right| = o(h^s)+ O_P(R_{U,OS}) + O_P(\alpha_n^{3/2}),
	\end{align}
	and
	\begin{align}\label{p1s2:second}
		&\sup_{x\in \mathcal{X},\tau\in\mathcal{T}}\left|\widehat{M}_{n}\left\{\widetilde{\theta}_{\tau}(x);x,\tau\right\}-L_n\left\{\widetilde{\theta}_{\tau}(x);x,\tau\right\}\right| = o(h^s) + O_P(R_{U,OS})+ o_P(\alpha_n^{3/2}).
	\end{align}
	To show \eqref{p1s2:first}, note that
	\begin{align}
		&\sup_{x\in \mathcal{X},\tau\in\mathcal{T}}\left|\widehat{M}_{n}\left\{\widehat{\theta}_{\tau}(x);x,\tau\right\}-L_n\left\{\widehat{\theta}_{\tau}(x);x,\tau\right\}\right|\notag \\
		\le&\sup_{x\in \mathcal{X},\tau\in\mathcal{T}}\left|M\left\{\widehat{\theta}_{\tau}(x);x,\tau\right\}-\partial_{\theta}M\left\{\theta_{\tau}(x);x,\tau\right\}\left\{\widehat{\theta}_{\tau}(x)-\theta_{\tau}(x)\right\}\right| \label{term:convergence6}\\
		+ &\sup_{x\in \mathcal{X},\tau\in\mathcal{T}}\left|\widehat{M}_n\left\{\widehat{\theta}_{\tau}(x);x,\tau\right\} - M\left\{\widehat{\theta}_{\tau}(x);x,\tau\right\} -\left[\widehat{M}_n\left\{\theta_{\tau}(x);x,\tau\right\} - M\left\{\theta_{\tau}(x);x,\tau\right\}\right]\right|. \label{term:convergence7}
	\end{align}

	For term \eqref{term:convergence6}, by Taylor expansion argument, Assumption \ref{Assumption:PartialBoundAway} and the result from step 1, we have
	$$\eqref{term:convergence6} = O_P\left\{\sup_{x\in\mathcal{X},\tau\in\mathcal{T}}\left|\widehat{\theta}_{\tau}(x)-\theta_{\tau}(x)\right|\right\}^2 = O_P(\alpha_n^2).$$	
	
	For term \eqref{term:convergence7}, by the result in Step 1, i.e., equation \eqref{p1:unirate}, we can find $\delta_n = O(\alpha_n)$ such that $\mathbb{P}(\sup_{x\in \mathcal{X},\tau\in\mathcal{T}}|\widehat{\theta}_{\tau}(x) - \theta_{\tau}(x)|>\delta_n)\rightarrow 0$. Then by Lemma \ref{lemma:biasandvariance}, with probability approaching one, we have
	\begin{align}
		\eqref{term:convergence7} &\le \sup_{x\in \mathcal{X},\tau\in\mathcal{T}, |\theta_1-\theta_2|\le \delta_n} \left|B_n(\theta_1;x,\tau)-B_n(\theta_2;x,\tau)\right| \label{p1s2:var}\\
		& + \sup_{x\in \mathcal{X},\tau\in\mathcal{T}, |\theta_1-\theta_2|\le \delta_n} \left|V_n(\theta_1;x,\tau)-V_n(\theta_2;x,\tau)\right|  \label{p1s2:bias}\\
		& + o(h^s) + O_P(R_{U,OS}).\notag
	\end{align}
	For the term \eqref{p1s2:var}, by the Assumption \ref{Assumption:Continuity} and mean value theorem, we have
	\begin{align*}
		\sup_{x\in \mathcal{X},\tau\in\mathcal{T}, |\theta_1-\theta_2|\le \delta_n} \left|B_n(\theta_1;x,\tau)-B_n(\theta_2;x,\tau)\right| = O_P\left(\delta_nh^s\right) = O_P(\alpha_n^2).
	\end{align*}
	For the term \eqref{p1s2:bias}, we use the Corollary 5.1 in 
	\cite{chernozhukov2014gaussian} to bound it. Define the function class 
	\begin{align*}
		\mathcal{F}_{\delta_n} = \left\{(w,y)\mapsto K_{U,h}(x - w)\{\psi_{\tau}(y-\theta_1) - \psi_{\tau}(y-\theta_2)\}:x\in \mathcal{X},|\theta_1-\theta_2|\le\delta_n,\tau\in\mathcal{T}\right\}
	\end{align*}
	In the proof of Lemma \ref{lemma:VCtype}, we know that there exists constant $A$ and $v$ for the function class
	\begin{align*}
		\{w \mapsto K_{U,h}(x-w):x\in \mathcal{X}\}
	\end{align*}
	such that
	\begin{align}\label{p1s2:KVCtype}
		N\left(\{w \mapsto K_{U,h}(x-w):x\in \mathcal{X}\},\|\cdot\|_{Q,2},\delta Dh^{-\beta-1}\right)\le (A/\delta)^{v}, 0<\forall\delta\le 1,
	\end{align}
	where $D>0$ is defined in Lemma \ref{lemma:VCtype} (sufficiently large such that $\sup_{x\in \mathbb{R}}h^{\beta+1}|K_{U,h}(x)|\le D$).
	We know that
	\begin{align}\label{p1s2:FminusF}
		\{y\mapsto\psi_{\tau}(y-\theta_1) - \psi_{\tau}(y-\theta_2):|\theta_1-\theta_2|\le\delta_n,\tau\in\mathcal{T}\} \subset \mathcal{G} - \mathcal{G} 
	\end{align}
	where $\mathcal{G}-\mathcal{G}:=\left\{g_1-g_2:g_1,g_2\in\mathcal{G}\right\}$ and $\mathcal{G}:=\{y\mapsto\psi_{\tau}(y-\theta):\theta\in\Theta,\tau\in\mathcal{T}\}$. By Lemma A.6 in \cite{chernozhukov2014gaussian}, there exists $A$ and $v$ such that
	\begin{align}\label{p1s2:FminusFVCtype}
		N\left(\mathcal{G}-\mathcal{G},\|\cdot\|_{Q,2},\delta\right)\le (A/\delta)^{v}, 0<\forall\delta\le 1.
	\end{align}
	In the view of Corollary A.1 in \cite{chernozhukov2014gaussian}, combined with \eqref{p1s2:KVCtype} and \eqref{p1s2:FminusF}, we know that there exists constant $A$ and $v$ such that
	\begin{align*}
		N(\mathcal{F}_{\delta_n},\|\cdot\|_{Q,2},\delta\|F_{\delta_n}\|_{Q,2}) \le (A/\delta)^{v},\ 0<\forall\delta\le 1,
	\end{align*}
	where $F_{\delta_n} = 2Dh^{-\beta-1}$. Hence, the conditions of Corollary 5.1 in \cite{chernozhukov2014gaussian} on the function class $\mathcal{F}_{\delta_n}$ is satisfied. To apply it, we first bound several involved quantities:
	\begin{itemize}
		\item (The element-wise second moment)
		We know that
		\begin{align*}
			&\sup_{f\in\mathcal{F}_{\delta_n}}\mathbb{P}f^2\\
			&:=\sup_{x\in \mathcal{X}, |\theta_1-\theta_2|\le \delta_n}\mathbb{E}\left[K_{U,h}^2(x-W)\{\psi_{\tau}(Y-\theta_1)-\psi_{\tau}(Y-\theta_2)\}^2\right]\\
			&= \sup_{x\in \mathcal{X}, |\theta_1-\theta_2|\le \delta_n}\int K_{U,h}^2(x-w)\mathbb{E}\left[\left\{\psi_{\tau}(Y-\theta_1) - \psi_{\tau}(Y-\theta_2)\right\}^2 | W=w\right]f_W(w) dw \\
			&\le \sup_{w\in\mathbb{R}}\sup_{|\theta_1 - \theta_2|\le \delta_n} \mathbb{E}\left[\left\{\psi_{\tau}(Y-\theta_1) - \psi_{\tau}(Y-\theta_2)\right\}^2 | W=w\right]f_W(w) \cdot\int K_{U,h}^2(x-w)dw\\
			&= O\left(\delta_n h^{-2\beta-1}\right).
		\end{align*}
		where the last equality follows from Lemma \ref{lemma:bound}(i), Assumption \ref{Assumption:PartialBoundAway} and mean value theorem.
		\item (The second moment of envelope function) By the definition of $F_{\delta_n}$, we immediately have
		\begin{align*}
			\|F\|_{\mathbb{P},2}^2:=\mathbb{E}\left\{ F_{\delta_n}^2\right\} = O(h^{-2\beta-2}).
		\end{align*}
		\item (The second moment of maximal of envelope function)
		\begin{align*}
			\|M\|_2^2:=\mathbb{E}\left\{\max_{1\le i\le n} F_{\delta_n}^2(W_i,Y_i)\right\} = O( h^{-2\beta-2}).
		\end{align*}
	\end{itemize}
	
	Then, by Corollary 5.1 in \cite{chernozhukov2014gaussian}, we know that
	\begin{align*}
		&\mathbb{E}\left\{ \sup_{x\in \mathcal{X},\tau\in\mathcal{T}, |\theta_1-\theta_2|\le \delta_n}\left|V_n(\theta_1;x,\tau)-V_n(\theta_2;x,\tau)\right|\right\}\\
		=& O\left(\frac{1}{\sqrt{n}}\left\{ \frac{\delta_n^{1/2}\{\log(h^{-1})\}^{1/2} }{h^{\beta+1/2}} + \frac{\log(h^{-1})}{\sqrt{n}h^{\beta+1}} \right\}\right)= O(\alpha_n^{3/2})
	\end{align*}
	where the equalities follow from the Assumption OS.
	
	Hence, combining the result of \eqref{term:convergence6} to \eqref{term:convergence7}, the result of \eqref{p1s2:first} follows. Note that in the argument above, we only use the result of $\sup_{x\in \mathcal{X},\tau\in\mathcal{T}}|\widehat{\theta}_{\tau}(x) - \theta_{\tau}(x)| = O_P(\alpha_n).$ By the definition of $\widetilde{\theta}_{\tau}(x)$ in \eqref{p1s2:deftilde} and $\sup_{x\in \mathcal{X},\tau\in\mathcal{T}}| \widehat{M}_n\left\{\theta_{\tau}(x);x,\tau\right\}-M\left\{\theta_{\tau}(x);x,\tau\right\}| = O_P(\alpha_n)$, we can also conclude that $\sup_{x\in \mathcal{X},\tau\in\mathcal{T}}|\widetilde{\theta}_{\tau}(x) - \theta_{\tau}(x)| = O_P(\alpha_n).$ Then \eqref{p1s2:second} can be verified through similar arguments.

	Then by \eqref{p1s2:first} and \eqref{p1s2:second}, we have
	\begin{align}
		&\sup_{x\in \mathcal{X},\tau\in\mathcal{T}}\left|L_n\left\{\widehat{\theta}_{\tau}(x);x,\tau\right\}\right|\\
		\le& \sup_{x\in \mathcal{X}}|\widehat{M}_{n}\left\{\widehat{\theta}_{\tau}(x);x,\tau\right\}| + o_P(h^s)+ O_P(R_{U,OS}+\alpha_n^{3/2})\notag \\
		\le& \sup_{x\in \mathcal{X},\tau\in\mathcal{T}}\left|\widehat{M}_{n}\left\{\widetilde{\theta}_{\tau}(x);x,\tau\right\}\right| +o_P(h^s)+ O_P(R_{U,OS}+\alpha_n^{3/2})\notag\\
		\le& \sup_{x\in \mathcal{X},\tau\in\mathcal{T}} \left|L_n\left\{\widetilde{\theta}_{\tau}(x);x,\tau\right\}\right| +o_P(h^s)+ O_P(R_{U,OS}+\alpha_n^{3/2})\notag\\
		=& o_P(h^s)+O_P(R_{U,OS}+\alpha_n^{3/2}),\label{p1s2:Ln}
	\end{align}
	where the second inequality follows from $\widehat{\theta}_{\tau}(x)$ is the minimizer of $|\widehat{M}_n(\theta;x,\tau)|$. 
	On the other hand, we have
	\begin{align*}
		L_n(\theta;x,\tau) =& \widehat{M}_n\left\{\theta_{\tau}(x);x,\tau\right\} + \partial_{\theta}M\left\{\theta_{\tau}(x);x,\tau\right\}\cdot\{\theta-\widetilde{\theta}_{\tau}(x)\} \\
		+& \partial_{\theta}M\left\{\theta_{\tau}(x);x,\tau\right\}\cdot\{\widetilde{\theta}_{\tau}(x)-\theta_{\tau}(x)\}\\ =&\partial_{\theta}M\left\{\theta_{\tau}(x);x,\tau\right\}\left\{\theta-\widetilde{\theta}_{\tau}(x)\right\}.
	\end{align*}
	
	Hence, \eqref{p1s2:Ln} implies
	$$\sup_{x\in \mathcal{X},\tau\in\mathcal{T}}\left|\partial_{\theta}M\left\{\theta_{\tau}(x);x,\tau\right\}\left\{\widehat{\theta}_{\tau}(x)-\widetilde{\theta}_{\tau}(x)\right\}\right| = o_P(h^s)+ O_P(R_{U,OS}+\alpha_n^{3/2}).$$
	By Assumption \ref{Assumption:boundedsupport}(iii) and the fact that $\partial_{\theta}M\left\{\theta_{\tau}(x);x,\tau\right\} = f_{X,Y}(x,\theta)$, we obtain that
	$$\sup_{x\in \mathcal{X},\tau\in\mathcal{T}}\left|\widehat{\theta}_{\tau}(x) - \widetilde{\theta}_{\tau}(x)\right| =o_P(h^s)+  O_P(R_{U,OS}+\alpha_n^{3/2}).$$
	Recall that the definition of $\widetilde{\theta}_{\tau}(x)$ in \eqref{p1s2:deftilde} and Lemma \ref{lemma:biasandvariance}(i), the first result of the Theorem follows.
	
	For the conclusion of $\widehat{\theta}_{\tau}^{(b)}(x)$, noting that the random variable $\chi_i^{(b)}$ for $i=1,2,\dots,n$ is independent of $(W_i,Y_i)$ and have mean zero. By analogy to the proof of the Bahadur expansion of $\widehat{\theta}_{\tau}(x)$, we have the following result,
	\begin{align*}
		&\widehat{\theta}_{\tau}^{(b)}(x)-\theta_{\tau}(x)\notag\\
		= &-\frac{1}{f_{X,Y}\{x,\theta_{\tau}(x)\}}B_n\{\theta_{\tau}(x);x,\tau\}\notag\\
		&- \frac{1}{f_{X,Y}\{x,\theta_{\tau}(x)\}}\Bigg(\frac{1}{n}\sum_{i=1}^n \chi_i^{(b)}K_{U,h}(x-W_i)\psi_{\tau}\{Y_i-\theta_{\tau}(x)\} \\
		&\qquad\qquad\qquad\qquad\qquad\qquad- \mathbb{E}\left[\chi_i^{(b)}K_{U,h}(x-W_i)\psi_{\tau}\{Y_i-\theta_{\tau}(x)\}\right]\Bigg)\notag\\
		& + o_P\left(h^s\right) + O_P\left(\left\{ \frac{\log(1/h)}{{nh^{2\beta+1}}}\right\}^{3/4} \right) + O_P\left(R_{U,OS}\right).
	\end{align*}
	Then by the Bahadur expansion of $\theta_{\tau}(x)$, we have
	\begin{align*}
		\widehat{\theta}_{\tau}^{(b)}(x) - \widehat{\theta}_{\tau}(x) =& \widehat{\theta}_{\tau}^{(b)}(x) - \theta_{\tau}(x) - \left\{\widehat{\theta}_{\tau}(x) - \theta_{\tau}(x)\right\}\\
		=&-\frac{1}{f_{X,Y}\{x,\theta_{\tau}(x)\}}V_n^{\chi}\{\theta_{\tau}(x);x,\tau\}\\
		& + o_P\left(h^s\right)+ O_P\left(\left\{ \frac{\log(1/h)}{{nh^{2\beta+1}}}\right\}^{3/4} \right) + O_P\left(R_{U,OS}\right).
	\end{align*}
	\subsection{Proof of Theorem \ref{thm:criteron_consistency_supersmooth}}
	\textbf{Step 1: Uniform Convergence Rate of $\widehat{\theta}_{\tau}(x)$:}
	Denote $\beta_n = h^s+n^{-1/2}e_U(h)\{\log(1/h)\}^{1/2}$. By the similar argument in step 1 of Theorem \ref{thm:criteron_consistency} combined with $\sup_{x\in \mathcal{X},\theta\in\Theta}|\widehat{M}_n(x)-M(x)| = O_P(\beta_n)$, from Lemma \ref{lemma:biasandvariance}(ii) we have
	\begin{align}\label{p2:unirate}
		\sup_{x\in \mathcal{X},\tau\in\mathcal{T}}\left| \widehat{\theta}_{\tau}(x) - \theta_{\tau}(x)\right| = O_P\left(\beta_n \right).
	\end{align}
	
	\textbf{Step 2: Uniform Linear Expansion of $\widehat{\theta}(x)$:}
	Using the same definition of $L_n$ and $\widetilde{\theta}_{\tau}(x)$ in Step 2 in the proof of Theorem \ref{thm:criteron_consistency}, in view of the proof of Theorem \ref{thm:criteron_consistency}, we only need to show \eqref{p1s2:first} and $\eqref{p1s2:second}$ with $\alpha_n$ replaced by $\beta_n$. That is, we only need to show \eqref{term:convergence6} and \eqref{term:convergence7} is $O_P\left(\beta_n^{3/2}\right) + O_P(R_{U,SS}) + O_P(h^s).$
	
	For \eqref{term:convergence6}, by Taylor expansion, we know that it is $O_P(\beta_n^2).$ The only difference is \eqref{term:convergence7}. We use the Corollary 5.1 in \cite{chernozhukov2014gaussian} to bound it. Define the function class 
	\begin{align*}
		\mathcal{F}_{\delta_n} = \left\{(w,y)\mapsto K_{U,h}(x - w)\{\psi_{\tau}(y-\theta_1) - \psi_{\tau}(y-\theta_2)\}:x\in \mathcal{X},|\theta_1-\theta_2|\le\delta_n\right\}.
	\end{align*}
	By Lemma \ref{lemma:VCtype}, we know that for the function class, there exists $A$ and $v$ such that
	\begin{align*}
		N\left(\{w \mapsto K_{U,h}(x-w):x\in \mathcal{X}\},\|\cdot\|_{Q,2},\delta Dh^{-\beta/2}e_U(h)\right)\le (A/\delta)^{v}, 0<\forall\delta\le 1.
	\end{align*}
	Then by  Corollary A.1 in \cite{chernozhukov2014gaussian}, we know that there exists $A$ and $v$ such that
	\begin{align*}
		N(\mathcal{F}_{\delta_n},\|\cdot\|_{Q,2},F_{\delta_n}) \le (A/\delta)^{v},\ 0<\forall\delta\le 1
	\end{align*}
	where $F_{\delta_n}$ is defined as $F_{\delta_n} = 2Dh^{-\beta/2}e_U(h)$. Then the conditions of Corollary 5.1 in \cite{chernozhukov2014gaussian} are satisfied. To apply it, we calculate several involved quantities:
	\begin{itemize}
		\item (The element-wise second moment)
		We know that
		\begin{align*}
			\sup_{f\in\mathcal{F}_{\delta_n}}\mathbb{P}f^2:=&\sup_{x\in \mathcal{X}, |\theta_1-\theta_2|\le \delta_n}\mathbb{E}\left\{K_{U,h}^2(x-W)\{\psi_{\tau}(Y-\theta_1)-\psi_{\tau}(Y-\theta_2)\}^2\right\}\\
			\le & e_U^2(h)\sup_{|\theta_1-\theta_2|\le \delta_n}\mathbb{E}\left\{\{\psi_{\tau}(Y-\theta_1)-\psi_{\tau}(Y-\theta_2)\}^2\right\}
			= O\left\{\delta_n e_U^2(h)\right\},
		\end{align*}
		where the inequality follows from Lemma \ref{lemma:bound}.
		\item (The second moment of envelope function) By the definition of $F_{\delta_n}$, we immediately have
		\begin{align*}
			\|F\|_{P,2}^2 := \mathbb{E}\left\{ F_{\delta_n}^2\right\} = O\left\{h^{-\beta}e_U^2(h)\right\}
		\end{align*}
		\item (The second moment of maximal of envelope function)
		\begin{align*}
			\|M\|_2^2:=\mathbb{E}\left\{\max_{1\le i\le n} F_{\delta_n}^2(W_i,Y_i)\right\} &\le O\left\{ h^{-\beta}e_U^2(h)\right\}.
		\end{align*}
	\end{itemize}
	Then, by Corollary 5.1 in \cite{chernozhukov2014gaussian}, we know that
	\begin{align*}
		&\mathbb{E}\left\{ \sup_{x\in \mathcal{X}, |\theta_1-\theta_2|\le \delta_n}\left|V_n(\theta_1;x,\tau)-V_n(\theta_2;x,\tau)\right|\right\}\\
		=& O\left(\frac{1}{\sqrt{n}}\left\{ \delta_n^{1/2}e_U(h)\{\log(h^{-1})\}^{1/2} + \frac{e_U(h)\log(h^{-1})}{\sqrt{n}h^{\beta/2}} \right\}\right) = O(\beta_n^{3/2}).
	\end{align*}
	where the last equality by Assumption SS. Hence, the desired result follows.
	
	\subsection{Proof of Theorem \ref{thm:validity}}
	Before we start to proof the main theorem, we present an useful lemma. The proof of the following lemma is given in Section \ref{proof:lemma}.
	\begin{lemma}\label{thm:varianceconvergence}
		Under Assumptions \ref{Assumption:boundedsupport}-\ref{Assumption:VarianceContinuity}, if (a) $U$ is an ordinary smooth error satisfying \eqref{def:os} of order $\beta$ and Assumptions OS, OS$'$ hold  or (b) $U$ is a supersmooth error satisfying \eqref{def:ss} of order $(\beta,\beta_0)$ and Assumptions SS, SS$'$ hold, we have
		\begin{align*}
			\sup_{(x,\tau) \in \mathcal{X}\times\mathcal{T}}\left|\frac{\widehat{\sigma}_n^2(x,\tau)}{\sigma^2_n(x,\tau)}-1\right| = o_P\left\{ 1/\log (1/h)\right\}.
		\end{align*}
	\end{lemma}
	
	Denote the processes
	{
	\begin{align*}
		&\widehat{Z}_n(x,\tau) = \{\widehat{\theta}_{\tau}(x)-\theta_{\tau}(x)\}/\sigma_n(x,\tau),\\
		&Z_n(x,\tau) = -\frac{V_n\{\theta_{\tau}(x);x,\tau\}}{f_{X,Y}\{x,\theta_{\tau}(x)\}\sigma_n(x,\tau)},\\
		&\widehat{\widehat{Z}_n}(x,\tau) = \{\widehat{\theta}_{\tau}(x)-\theta_{\tau}(x)\}/\widehat{\sigma}_n(x),\\
		&\widehat{Z}_n^{\chi}(x) = \{\widehat{\theta}^{\chi}_{\tau}(x)-\widehat{\theta}_{\tau}(x)\}/\sigma_n(x,\tau), \\
		&Z_n^{\chi}(x,\tau) = - \frac{1}{n\sigma_n(x,\tau)f_{X,Y}(x,\theta_{\tau}(x)}\sum_{i=1}^n(\chi_i-1)\Bigg[ K_{U,h}(x-W_i)\psi_{\tau}\left\{Y_i-\theta_{\tau}(x)\right\}\\
		&\qquad\qquad\qquad\qquad\qquad\left.- \frac{1}{n}\sum_{j=1}^nK_{U,h}(x-W_j)\psi_{\tau}\left\{Y_j-\theta_{\tau}(x)\right\}\right]\\
		&\widehat{\widehat{Z}^{\chi}_n}(x,\tau) = \{\widehat{\theta}^{\chi}_{\tau}(x)-\widehat{\theta}_{\tau}(x)\}/\widehat{\sigma}_n(x)
	\end{align*}
}
\subsubsection{Proof of Theorem \ref{thm:validity}: Ordinary Smooth Case}
	\textbf{Preliminary results:}
	By Theorem \ref{thm:criteron_consistency}, Lemma \ref{lemma:variancelowerboundos}(i) and Assumption OS, OS', we have
	\begin{align}\label{eq:process}
		\left\|\widehat{Z}_n(x,\tau)  - Z_n(x,\tau)\right\|_{\mathcal{X}\times\mathcal{T}} &= O_P\left\{ \left[\left\{\frac{\log(1/h)}{{nh^{2\beta+1}}}\right\}^{3/4} + h^s + R_{U,OS}\right]\times \left(nh^{2\beta+1}\right)^{1/2}\right)\notag\\
   &= o_P\left(\{\log(1/h)\}^{-1/2}\right\}.
	\end{align}
	By analogy, we have
	\begin{align}\label{eq:bsprocess1}
		\left\|\widehat{Z}_n^\chi(x,\tau) + \frac{V_n^{\chi}\{\theta_{\tau}(x);x,\tau\}}{n\sigma_n(x,\tau)f_{X,Y}\{x,\theta_{\tau}(x)\}}\right\|_{x\in \mathcal{X},\tau\in\mathcal{T}}=o_P\left(\{\log (1/h)\}^{-1/2}\right).
	\end{align}
	By Lemma \ref{lemma:bias} and \ref{lemma:variance} and Assumption OS', we know that
	\begin{align}\label{eq:bsprocess2}
		\sup_{x\in \mathcal{X},\tau\in\mathcal{T}}\left|\frac{1}{n\sigma_n(x,\tau)}\sum_{i=1}^{n}(\chi_i-1)\frac{1}{n}\sum_{j=1}^n K_{U,h}(x-W_j)\psi_{\tau}\{Y_j-\theta_{\tau}(x)\}\right| = o_P\left(\{\log(1/h)\}^{-1/2}\right).
	\end{align}
	Hence, by the definition of $Z_n^{\chi}$ and combining \eqref{eq:bsprocess1} and \eqref{eq:bsprocess2}, we have
	\begin{align}\label{eq:bsprocess}
		\|\widehat{Z}_n^\chi - Z_n^\chi\|_{\mathcal{X}\times\mathcal{T}} = o_P\left(\left\{\log (1/h)\right\}^{-1/2}\right).
	\end{align}
	\textbf{Step 1:} We first show that there exists a tight Gaussian process $Z_n^G$ in $\ell^{\infty}(I\times \mathcal{T})$ with mean zero and same covariance function with $Z_n$ and such that 
	$$\sup_{t\in\mathbb{R}}\left|\mathbb{P}\left(\|\widehat{Z}_n(x,\tau)\|_{\mathcal{X}\times\mathcal{T}}\le t\right) - \mathbb{P}\left(\|{Z}_n^{G}(x,\tau)\|_{\mathcal{X}\times\mathcal{T}}\le t\right)\right|\rightarrow 0.$$
	
	To prove that, we verify the conditions of Theorem 2.1 in \cite{chernozhukov2016empirical} on the function class
	\begin{align*}
		\mathcal{F}_{n,2}:=\left\{(w,y)\mapsto \frac{1}{\sqrt{n}\sigma_n(x,\tau)f_{X,Y}\{x,\theta_{\tau}(x)\}} K_{U,h}(x-w)\psi_{\tau}\{y-\theta_{\tau}(x)\}:x\in \mathcal{X},\tau\in\mathcal{T}\right\}.
	\end{align*}
	By Lemma \ref{lemma:VCtype} and \ref{lemma:variancelowerboundos},
	\begin{align}\label{eq:Fn2}
		\sup_Q N(\mathcal{F}_{n,2},\|\cdot\|_{Q,2},2\delta Dh^{-1/2})\le (A/\delta)^v,\ 0<\forall \delta\le 1.
	\end{align}
	
	We know that by Lemma \ref{lemma:variancelowerboundos},  $\sup_{f\in\mathcal{F}_{n,2}}\mathbb{P}f^2 = 1 $, $\sup_{f\in\mathcal{F}_{n,2}}\mathbb{P}f^3 =  O(h^{3\beta+3/2}h^{-3\beta-2}) = O(h^{-1/2}) $ and $\sup_{f\in\mathcal{F}_{n,2}}\mathbb{P}f^3 =  O(h^{4\beta+2}h^{-4\beta-3}) = O(h^{-1}) $. By applying Theorem 2.1 in \cite{chernozhukov2016empirical} with $B(f)\equiv 0$, $b = O(h^{-1/2})$, $\sigma^2=O(1)$, $\gamma = O(1/\log n)$ and arbitrary $q\ge 4$, there exists a tight gaussian process $G_n$ on $\ell^{\infty}(\mathcal{F}_{n,2})$,
	$$
	\left|\|Z_n\|_{\mathcal{X}\times\mathcal{T}} - \|G_n\|_{\mathcal{F}_{n,2}}  \right| = O_P\left(\frac{(\log n)^{1+1/q}}{n^{1/2-1/q}h^{1/2}} + \frac{\log n}{(nh)^{1/6}}\right) = o_P\left(\{\log(1/h)\}^{-1/2}\right).
	$$
	Write $Z_n^G(x,\tau) = G_n(f_{n,x,\tau})$ for $f_{n,x,\tau}(w,y)\in\mathcal{F}_{n,2}$. Then $Z_n^G$ is a tight Gaussian process on $\ell^{\infty}(\mathcal{X}\times\mathcal{T})$ with mean zero and same covariance function with $Z_n$, and $\|Z_n^G\|_{\mathcal{X}\times\mathcal{T}}$ has the same distribution with $\|G_n\|_{\mathcal{F}_{n,2}}$. Then we know that
	\begin{align*}
		\left|\|\widehat{Z}_n\|_{\mathcal{X}\times\mathcal{T}} - \|Z_n^G\|_{\mathcal{X}\times\mathcal{T}} \right| \le& \left|\|\widehat{Z}_n\|_{\mathcal{X}\times\mathcal{T}} - \|Z_n\|_{\mathcal{X}\times\mathcal{T}}\right| + \left|\|Z_n\|_{\mathcal{X}\times\mathcal{T}}-\|Z_n^G\|_{\mathcal{X}\times\mathcal{T}}\right|\\
		\le& \|\widehat{Z}_n - Z_n\|_{\mathcal{X}\times\mathcal{T}} + o_P\left(\{\log(1/h)\}^{-1/2}\right) = o_P\left(\{\log(1/h)\}^{-1/2}\right),
	\end{align*}
	where the last equation holds because of \eqref{eq:process}.
	Hence, there exists sequence $\Delta_n\rightarrow 0$ such that
	\begin{align*}
		\mathbb{P}\left(\left|\|\widehat{Z}_n\|_{\mathcal{X}\times\mathcal{T}} - \|Z_n^G\|_{\mathcal{X}\times\mathcal{T}}\right|> \Delta_n \cdot \{\log(1/h)\}^{-1/2}\right) \rightarrow 0,
	\end{align*}
	which implies
	\begin{align*}
		\mathbb{P}\left(\|\widehat{Z}_n\|_{\mathcal{X}\times\mathcal{T}}\le t\right) \le \mathbb{P}\left(\|Z_n^G\|_{\mathcal{X}\times\mathcal{T}}\le t+\Delta_n\cdot \{\log(1/h)\}^{-1/2}\right) + o(1)
	\end{align*}
	uniformly over $t\in\mathbb{R}$. By Corollary 2.1 in \cite{chernozhukov2014anti}, we have
	\begin{align*}
		\mathbb{P}\left(\|Z_n^G\|_{\mathcal{X}\times\mathcal{T}}\le t+\Delta_n\cdot \{\log(1/h)\}^{-1/2}\right)\le \mathbb{P}\left(\|Z_n^G\|_{\mathcal{X}\times\mathcal{T}}\le t\right) + \frac{4\Delta_n}{\sqrt{\log(1/h)}}\{1+\mathbb{E}\|Z_n^G\|_{\mathcal{X}\times\mathcal{T}}\}.
	\end{align*}
	By Dudley's entropy bound (Corollary 2.2.8 in \cite{van1996weak}) and \eqref{eq:Fn2}, we have
	\begin{align}
		\mathbb{E}\|Z_n^G\|_{\mathcal{X}\times\mathcal{T}} = O\left(\int_{0}^1\sqrt{1+\log(1/\epsilon h^{1/2} )}d\epsilon\right) = O\left(\{\log(1/h)\}^{1/2}\right).
	\end{align}
	Hence, uniformly over $t\in\mathbb{R}$, 
	\begin{align*}
		\mathbb{P}\left(\|\widehat{Z}_n\|_{\mathcal{X}\times\mathcal{T}}\le t\right)\le \mathbb{P}\left(\|Z_n^G\|_{\mathcal{X}\times\mathcal{T}}\le t\right) + o(1).
	\end{align*}
	For the other side, note that
	\begin{align*}
		\mathbb{P}\left(\|\widehat{Z}_n\|_{\mathcal{X}\times\mathcal{T}}\ge t\right) \le \mathbb{P}\left(\|Z_n^G\|_{\mathcal{X}\times\mathcal{T}}\ge t-\Delta_n (\log(1/h))^{-1/2}\right) + o(1).
	\end{align*}
	By similar argument, we have
	\begin{align*}
		\mathbb{P}\left(\|\widehat{Z}_n\|_{\mathcal{X}\times\mathcal{T}}\ge t\right) \le \mathbb{P}\left(\|Z_n^G\|_{\mathcal{X}\times\mathcal{T}}\ge t\right) + o(1)
	\end{align*}
	uniformly over $t\in\mathbb{R}.$ Hence the desired result follows.
	
	\textbf{Step 2:} Next we show that, for the $Z_n^G$ defined in Step 1, 
	\begin{align}
		\sup_{t\in\mathbb{R}}\left|\mathbb{P}\left(\|\widehat{\widehat{Z}_n^{\chi}}\|_{\mathcal{X}\times\mathcal{T}}\le t | \mathcal{D}_n\right)-\mathbb{P}\left(\|Z_n^G\|_{\mathcal{X}\times\mathcal{T}}\le t\right)\right| = o_P(1).
	\end{align}

	By applying Theorem 2.2 in \cite{chernozhukov2014gaussian}  with $B\equiv 0$, $b = O(h^{-1/2})$, $\sigma^2=O(1)$, $\gamma = O(1/\log n)$ and arbitrary $q\ge 4$, we know that there exists random variable $V_n^{\chi}$ with conditional distribution on $\mathcal{D}_n$ identical to the distribution of $\|Z_n^G\|_{\mathcal{X}\times\mathcal{T}}$, such that
	\begin{align*}
		\left|\|Z_n^{\chi}\|_{\mathcal{X}\times\mathcal{T}} -V_n^{\chi} \right|= O_P\left(\frac{(\log n)^{2+1/q}}{n^{1/2-1/q}h^{1/2}} + \frac{(\log n)^{7/4+1/q}}{(nh)^{1/4}}\right) = o_P\left(\{\log (1/h)\}^{-1/2}\right).
	\end{align*}
	Hence, there exists $\Delta_n\rightarrow 0$ such that
	\begin{align*}
		\mathbb{P}\left(\left|\|Z_n^{\chi}\|_{\mathcal{X}\times\mathcal{T}}-V_n^{\chi}\right|>\Delta_n\cdot(\log (1/h))^{-1/2} |\mathcal{D}_n\right) = o_P(1).
	\end{align*}
	Then, 
	\begin{align*}
		\mathbb{P}\left(\|Z_n^{\chi}\|_{\mathcal{X}\times\mathcal{T}} \le t | \mathcal{D}_n\right) \le& \mathbb{P}\left( V_n^{\chi} \le t+\Delta_n\cdot\{\log (1/h)\}^{-1/2} |\mathcal{D}_n\right)+o_P(1)\\ 
		=& \mathbb{P}\left(\|Z_n^G\|_{\mathcal{X}\times\mathcal{T}}\le t+\Delta_n\cdot\{\log (1/h)\}^{-1/2}\right) + o_P(1)\\
		=& \mathbb{P}\left(\|Z_n^G\|_{\mathcal{X}\times\mathcal{T}} \le t \right) + o_P(1)
	\end{align*}
	uniformly over $t\in \mathbb{R}$, where the last equality follows from the anti-concentration inequality (Corollary 2.1 in \cite{chernozhukov2014anti}) and the fact $\mathbb{E}\|Z_n^G\|_{\mathcal{X}\times\mathcal{T}}= O\left(\{\log(1/h)\}^{1/2}\right)$.

	And we have
	\begin{align*}
		\|\widehat{\widehat{Z}^\chi_n} - \widehat{Z}_n^{\chi}\|_{\mathcal{X}\times\mathcal{T}} \le  \|\widehat{Z}_n^\chi\|_{\mathcal{X}\times\mathcal{T}}\|\widehat{\sigma}(\cdot)/\sigma(\cdot)-1\|_{\mathcal{X}\times\mathcal{T}}
	\end{align*}
	By  Lemmas \ref{lemma:bias},\ref{lemma:variance} and \ref{lemma:biasandvariance}, we know that $\|Z_n^{\chi}\|_{\mathcal{X}\times\mathcal{T}} = o_P\left(\{\log(1/h)\}^{1/2}\right)$, then by \eqref{eq:bsprocess}, we have $\|\widehat{Z}_n^{\chi}\|_{\mathcal{X}\times\mathcal{T}} = o_P\left(\{\log(1/h)\}^{1/2}\right).$ We have $\|\widehat{\sigma}(\cdot)/\sigma(\cdot)-1\|_{\mathcal{X}\times\mathcal{T}} = o_P(\{\log (1/h)\}^{-1})$, hence, by \eqref{eq:bsprocess},
	\begin{align*}
		\|\widehat{\widehat{Z}^\chi_n} - Z_n^{\chi}\|_{\mathcal{X}\times\mathcal{T}} \le&  \|\widehat{\widehat{Z}^\chi_n} - \widehat{Z}_n^{\chi}\|_{\mathcal{X}\times\mathcal{T}} + \|\widehat{Z}_n^{\chi} - Z_n^{\chi}\|_{\mathcal{X}\times\mathcal{T}}\\
		=& o_P\left(\{\log(1/h)\}^{-1/2}\right) + o_P\left(\{\log(1/h)\}^{-1/2}\right) = o_P\left(\{\log(1/h)\}^{-1/2}\right).
	\end{align*}
	By Corollary 2.1 in \cite{chernozhukov2014anti}, we can conclude that
	\begin{align*}
		\mathbb{P}\left(\|\widehat{\widehat{Z}_n^{\chi}}\|_{\mathcal{X}\times\mathcal{T}} \le t | \mathcal{D}_n\right)\le 	\mathbb{P}\left(\|Z_n^{\chi}\|_{\mathcal{X}\times\mathcal{T}} \le t | \mathcal{D}_n\right) \le \mathbb{P}\left(\|Z_n^G\|_{\mathcal{X}\times\mathcal{T}}\le t\right) + o_P(1),
	\end{align*}
	uniformly over $t\in\mathbb{R}$. Likewise, we have $	\mathbb{P}\left(\|\widehat{\widehat{Z}_n^{\chi}}\|_{\mathcal{X}\times\mathcal{T}} \le t | \mathcal{D}_n\right) \ge \mathbb{P}\left(\|Z_n^G\|_{\mathcal{X}\times\mathcal{T}}\le t\right) - o_P(1)$ uniformly over $t\in\mathbb{R}$.
	
	\textbf{Step 3:} Finally we prove the conclusion of the theorem. Note that 
	\begin{align*}
		\|\widehat{\widehat{Z}_n} - \widehat{Z}_n\|_{\mathcal{X}\times\mathcal{T}}\le\|\widehat{Z}_n\|_{\mathcal{X}\times\mathcal{T}}\cdot\|\widehat{\sigma}(\cdot)/\sigma(\cdot)-1\|_{\mathcal{X}\times\mathcal{T}} = o_P\left(\{\log(1/h)\}^{1/2}\right).
	\end{align*}
	Then by the similar argument in Step 1 and Step 2, we have
	\begin{align*}
		\sup_{t\in\mathbb{R}}\left|\mathbb{P}\left(\|\widehat{\widehat{Z}_n}\|_{\mathcal{X}\times\mathcal{T}}\le t\right)-\mathbb{P}\left(\|Z_n^G\|_{\mathcal{X}\times\mathcal{T}}\le t\right)\right| = o(1).
	\end{align*}
	By the result of Step 2, there exists a sequence $\Delta_n\rightarrow 0$ such that with probability greater than $1-\Delta_n$,
	\begin{align*}
		\sup_{t\in\mathbb{R}}\left|\mathbb{P}\left(\|\widehat{\widehat{Z}_n^{\chi}}\|_{\mathcal{X}\times\mathcal{T}}\le t|\mathcal{D}_n\right)-\mathbb{P}\left(\|Z_n^G\|_{\mathcal{X}\times\mathcal{T}}\le t\right)\right|\le \Delta_n.
	\end{align*}
	Let $A_n$ be the event such that the above equation holds. Taking $\Delta_n$ smaller if necessary, we have
	\begin{align*}
		\sup_{t\in\mathbb{R}}\left|\mathbb{P}\left(\|\widehat{\widehat{Z}_n}\|_{\mathcal{X}\times\mathcal{T}}\le t\right)-\mathbb{P}\left(\|Z_n^G\|_{\mathcal{X}\times\mathcal{T}}\le t\right)\right|\le \Delta_n.
	\end{align*}
	Denote $c_n^{G}(1-q)$ be the $(1-q)$-quantile of $\|Z_n^G\|_{\mathcal{X}\times\mathcal{T}}.$ On the event $A_n$, we have
	\begin{align*}
		\mathbb{P}\left(\|\widehat{\widehat{Z}_n^{\chi}}\|_{\mathcal{X}\times\mathcal{T}}\le c_n^G(1-q+\Delta_n)|\mathcal{D}_n\right) \ge \mathbb{P}\left(\|Z_n^G\|_{\mathcal{X}\times\mathcal{T}}\le c_n^G(1-q+\Delta_n)\right) - \Delta_n = 1-q.
	\end{align*}
	Then on the event $A_n$, $\widehat{c}_n(1-q)\le c_n^G(1-q+\Delta_n)$, which holds with probability greater than $1-\Delta_n$. Hence,
	\begin{align*}
		\mathbb{P}\left(\|\widehat{\widehat{Z}_n}\|_{\mathcal{X}\times\mathcal{T}}\le \widehat{c}_n(1-q)\right)\le& \mathbb{P}\left(\|\widehat{\widehat{Z}_n}\|_{\mathcal{X}\times\mathcal{T}}\le c_n^G(1-q+\Delta_n)\right) + \Delta_n\\
		\le &\mathbb{P}\left(\|Z_n^G\|_{\mathcal{X}\times\mathcal{T}}\le c_n^G(1-q+\Delta_n)\right) + 2\Delta_n\\
		=& 1-q+3\Delta_n.
	\end{align*}
	By similar argument, the other side $\mathbb{P}\left(\|\widehat{\widehat{Z}_n}\|_{\mathcal{X}\times\mathcal{T}}\le \widehat{c}_n(1-q)\right)\ge 1-q-3\Delta_n$ also holds. Hence, the desired result follows.
	\subsubsection{Proof of Theorem \ref{thm:validity}: Supersmooth Case}
	The proof is quite similar to the ordinary smooth case of Theorem \ref{thm:validity} except for the use of the gaussian approximation. Consider the same $\mathcal{F}_{n,2}$ defined in the proof of Theorem \ref{thm:validity}, by Lemma \ref{lemma:VCtype} and \ref{lemma:variancelowerboundos}, we know that for supersmooth error,
	\begin{align*}
		\sup_Q N(\mathcal{F}_{n,2},\|\cdot\|_{Q,2},2Dh^{-\beta/2}e_U(h)/e_L(h))\le (A/\delta)^v,\ 0<\forall \delta\le 1.
	\end{align*}
	Note that $e_U(h)/e_L(h) = h^{-1/2}\{\log(1/h)\}^{\ell_k}$. Hence,
	By applying Theorem 2.1 in \cite{chernozhukov2016empirical} with $B\equiv 0$, $b = O(h^{-(\beta+1)/2}\{\log(1/h)\}^{\ell_k})$, $\sigma^2=O(1)$, $\gamma = O(1/\log n)$ and arbitrary $q\ge 4$, there exists a tight gaussian process $G_n$ on $\ell^{\infty}(\mathcal{F}_{n,2})$,
	\begin{align*}
		\left|\|Z_n\|_{\mathcal{X}\times\mathcal{T}} - \|G_n\|_{\mathcal{F}_{n,2}}  \right| =& O_P\left(\frac{(\log n)^{1+1/q} (\log (1/h))^{\ell_k}}{n^{1/2-1/q}h^{(\beta+1)/2}} + \frac{\log n \{\log (1/h)\}^{\ell_k/3}}{n^{1/6}h^{(\beta+1)/6}}\right)\\
		=& o_P\left\{(\log(1/h))^{-1/2}\right\}
	\end{align*}
	By Theorem 2.2 in \cite{chernozhukov2016empirical} with $B\equiv 0$, $b = O(h^{-(\beta+1)/2}\{\log(1/h)\}^{\ell_k})$, $\sigma^2=O(1)$, $\gamma = O(1/\log n)$ and arbitrary $q\ge 4$, we know that there exists random variable $V_n^{\chi}$ which conditional distribution on $\mathcal{D}_n$ is identical to the distribution of $\|Z_n^G\|_{\mathcal{X}\times\mathcal{T}}$
	\begin{align*}
		\left|\|Z_n^{\chi}\|_{\mathcal{X}\times\mathcal{T}} -V_n^{\chi} \right|=& O_P\left(\frac{(\log n)^{2+1/q}(\log (1/h))^{\ell_k}}{n^{1/2-1/q}h^{(\beta+1)/2}} + \frac{(\log n)^{7/4+1/q}\{\log (1/h)\}^{1/2}}{n^{1/4}h^{(\beta+1)/4}}\right)\\
		=& o_P\left\{(\log (1/h))^{-1/2}\right\}.
	\end{align*}
	Hence, the desired result follows.
	\subsubsection{Proof of Lemma \ref{thm:varianceconvergence}}\label{proof:lemma}
	\textbf{Ordinary Smooth Case:}
	Note that
	\begin{align}
		&\widehat{\sigma}_n^2(x;\tau) - \sigma^2_n(x;\tau)\notag\\
		=& \frac{1}{\left\{n \widehat{f}_{X,Y}(x,\widehat{\theta}_{\tau}(x))\right\}^2} \sum_{i=1}^n \left\{\widehat{K}_{U,h}^2(x-W_i) - K_{U,h}^2(x-W_i)\right\}\psi_{\tau}^2\{Y-\widehat{\theta}_{\tau}(x)\}\notag\\
		+& \frac{1}{\left\{n \widehat{f}_{X,Y}(x,\widehat{\theta}_{\tau}(x))\right\}^2} \sum_{i=1}^n K_{U,h}^2(x-W_i)\psi_{\tau}^2(Y-\widehat{\theta}_{\tau}(x)) - \sigma^2_n(x;\tau)\notag\\
		=&\frac{1}{\left\{n \widehat{f}_{X,Y}(x,\widehat{\theta}_{\tau}(x))\right\}^2} \sum_{i=1}^n \left\{\widehat{K}_{U,h}^2(x-W_i) - K_{U,h}^2(x-W_i)\right\}\psi_{\tau}^2\{Y-\widehat{\theta}_{\tau}(x)\}\label{varproof:term0}\\
		+&\frac{1}{\left\{n \widehat{f}_{X,Y}(x,\widehat{\theta}_{\tau}(x))\right\}^2} \sum_{i=1}^n\left[ K_{U,h}^2(x-W_i)\psi_{\tau}^2(Y-{\theta_{\tau}}(x)) - \mathbb{E}\left\{K_{U,h}^2(x-W_i)\psi_{\tau}^2(Y-{\theta_{\tau}}(x)) \right\}\right]\label{varproof:term1}\\
		+&\frac{1}{n{f}_{X,Y}^2(x,{\theta}_{\tau}(x))}\mathbb{E}\left\{K_{U,h}^2(x-W_i)\psi_{\tau}^2(Y-{\theta_{\tau}}(x)) \right\}-\sigma^2_n(x;\tau)\label{varproof:termbias}\\
		+&\frac{1}{\left\{n \widehat{f}_{X,Y}(x,\widehat{\theta}_{\tau}(x))\right\}^2} \sum_{i=1}^n K_{U,h}^2(x-W_i)\left\{\psi_{\tau}^2(Y-\widehat{\theta}_{\tau}(x)) - \psi_{\tau}^2(Y-\theta_{\tau}(x))\right\}\label{varproof:term2}\\
		+& \left\{\frac{1}{n\widehat{f}_{X,Y}^2(x,\widehat{\theta}_{\tau}(x))}- \frac{1}{n{f}_{X,Y}^2(x,{\theta}_{\tau}(x))}\right\}\mathbb{E}\left\{K_{U,h}^2(x-W_i)\psi_{\tau}^2(Y-{\theta_{\tau}}(x)) \right\} \label{varproof:term3}.
	\end{align}
	
	Before we give the bound of \eqref{varproof:term0} to \eqref{varproof:term3}, we first focus on $\widehat{f}_{X,Y}$. Let $\widetilde{f}_{X,Y}(x,y) = n^{-1}\sum_{i=1}^n K_{h_Y}(y-Y_i)K_{U,h_W}(x-W_i)$.  We decompose $\widehat{f}_{X,Y}(x,y)-\widetilde{f}_{X,Y}(x,y)$, in the similar way as in the proof of Lemma \ref{lemma:effectU},
	{\small
	\begin{align*}
		&\widehat{f}_{X,Y}(x,y) -\widetilde{f}_{X,Y}(x,y)\\
		&= \frac{1}{(2\pi)^2h_Yh_W}\iint \exp\left(-\frac{it_1x}{h_W}-\frac{it_2y}{h_Y}\right)\widehat{\varPsi}\left(\frac{t_1}{h_{W}},\frac{t_2}{h_{Y}}\right)\phi_K(t_1)\phi_K(t_2)\left\{\frac{1}{\widehat{\phi}_U(t_1/h)}-\frac{1}{\phi_U(t_1/h)}\right\}dt_1dt_2
	\end{align*}
	}
	with $\widehat{\varPsi}(t_1,t_2) = n^{-1}\sum_{j=1}^n\exp(it_1W_j + it_2Y_j)$. Then, as in the proof of Lemma \ref{lemma:effectU}, we can obtain
	\begin{align}\label{eq:star1}
		\sup_{(x,y) \in \mathcal{X}\times\Theta}\left|\widehat{f}_{X,Y}(x,y)-\widetilde{f}_{X,Y}(x,y)\right| = O_P\left(m^{-1/2}n^{-1/2}h_W^{-2\beta-1}h_Y^{-1}+m^{-1/2}h_W^{-\beta}\right).
	\end{align}
	Using the arguments as in the proof of Lemmas \ref{lemma:bias}, \ref{lemma:variance} and \ref{lemma:biasandvariance}, we have
	\begin{align}\label{eq:star2}
		\sup_{x\in\mathcal{X},\theta\in\Theta}\left|\widetilde{f}_{X,Y}(x,\theta) - f_{X,Y}(x,\theta)\right| = O(h_Y^s + h_{W}^s) + O_P\left(\sqrt{\frac{\log\{1/(h_Y\land h_W)\}}{nh_Yh_{W}^{2\beta+1}}}\right).
	\end{align}
	Hence, by Assumption \ref{Assumption:VarianceContinuity}, we know that $\widehat{f}_{X,Y}(x,\theta) - {f}_{X,Y}(x,\theta) = o_P(1)$ uniformly over $(x,\theta)\in\mathcal{X}\times\Theta.$ Then by Assumption \ref{Assumption:boundedsupport}(ii), we know that 
	\begin{align}\label{eq:den}
		\sup_{(x,\tau)\in\mathcal{X}\times\mathcal{T}} \left\{\widehat{f}_{X,Y}(x,\widehat{\theta}_{\tau}(x))\right\}^{-1} = O_P(1).
	\end{align}

	For term \eqref{varproof:term0}, we know that
	\begin{align*}
		&\sup_{x\in\mathcal{R}}\left| \widehat{K}_{U,h}(x) - K_{U,h}(x) \right|\\
		&\le  \frac{1}{2\pi h}\int |\phi_K(t)|\left|\frac{1}{\phi_U(t/h)}- \frac{1}{\widehat{\phi}_U(t/h)}\right|dt\\
		&\lesssim O_P\left(h^{-2\beta-1}\right)\int |\phi_K(t)|\left|{\phi_U(t/h)}- {\widehat{\phi}_U(t/h)}\right|dt \lesssim O_P\left(m^{-1/2}h^{-2\beta-1}\right).
	\end{align*}
	Hence, from Lemma \ref{lemma:bound}
	\begin{align*}
		\sup_{x\in\mathcal{R}}\left| \widehat{K}_{U,h}^2(x) - K_{U,h}^2(x) \right| = \sup_{x\in\mathcal{R}}\left| \widehat{K}_{U,h}(x) - K_{U,h}(x) \right|\left| \widehat{K}_{U,h}(x) + K_{U,h}(x) \right| = O_P\left(m^{-1/2}h^{-3\beta-2}\right).
	\end{align*}
	Then by \eqref{eq:den} and Assumption OS$'$(iii), uniformly in $(x,\tau)\in\mathcal{X}\times\mathcal{T}$,
	\begin{align*}
		\eqref{varproof:term0} = O_P\left(n^{-1}m^{-1/2}h^{-3\beta-2}\right).
	\end{align*}
	
	For the term \eqref{varproof:term1}, we use Corollary 5.1 in \cite{chernozhukov2014gaussian} to bound it. Define the function class
	$$\mathcal{F}_n=\left\{(w,y)\mapsto K_{U,h}^2(x-w)\psi_{\tau}^2(y-\theta_{\tau}(x)):x\in \mathcal{X},\tau\in\mathcal{T}\right\}.$$
	Let the envelope function $F:= Dh^{-2\beta-2}$ with constant $D>0$ such that $\sup_{x\in\mathbb{R}}\left|K_{U,h}^2(x)\right| \le Dh^{-2\beta-2}$. By Lemma \ref{lemma:VCtype} and Corollary A.1 in \cite{chernozhukov2014gaussian}, we know that there exists constants $A$ and $v$ such that
	\begin{align*}
		N(\mathcal{F}_n,\|\cdot\|_{Q,2},\|F\|_{Q,2})\le (A/\delta)^v,\ 0<\forall \delta\le 1.
	\end{align*}
	Then the conditions of Corollary 5.1 in \cite{chernozhukov2014gaussian} are satisfied. To apply it, we calculate several involved quantities:
	\begin{itemize}
		\item (The element-wise second moment)
		We know that by Lemma \ref{lemma:bound}
		\begin{align*}
			\sup_{f\in\mathcal{F}_n}\mathbb{P}f^2 = &\sup_{x\in \mathcal{X},\tau\in\mathcal{T}}\mathbb{E}\left\{K_{U,h}^4(x-W)\psi_{\tau}^4(Y-\theta)\right\}\\
			\lesssim & \sup_{x\in \mathcal{X}}\mathbb{E}\left\{K_{U,h}^4(x-W)\right\} \lesssim h^{-2\beta-2}\mathbb{E}\left\{K_{U,h}^2(x-W)\right\}\lesssim h^{-4\beta-3}.
		\end{align*}
		\item (The second moment of envelope function) By the definition of $F$, we immediately have
		\begin{align*}
			\mathbb{E}\left\{ F^2\right\} = O(h^{-4\beta-4})
		\end{align*}
		\item (The second moment of maximal of envelope function)
		\begin{align*}
			\|M\|_2^2 = \mathbb{E}\left\{\max_{1\le i\le n} F^2(W_i,Y_i)\right\} &\le O(h^{-4\beta-4} )
		\end{align*}
	\end{itemize}
	Then, by Corollary 5.1 in \cite{chernozhukov2014gaussian}, we know that
	\begin{align*}
		&\mathbb{E}\left\{ \sup_{(x,\tau) \in \mathcal{X}\times\mathcal{T}}\left|\frac{1}{n^2 } \sum_{i=1}^n\left[ K_{U,h}^2(x-W_i)\psi_{\tau}^2(Y-{\theta_{\tau}}(x)) - \mathbb{E}\left\{K_{U,h}^2(x-W_i)\psi_{\tau}^2(Y-{\theta_{\tau}}(x)) \right\}\right]\right|\right\}\\
		=& O\left(\frac{1}{n^{3/2}}\left\{ \{h^{-4\beta-3}\log(1/h)\}^{1/2}  + \frac{\log(h^{-1})}{\sqrt{n}h^{4\beta+4}} \right\}\right) = O\left(\frac{\{\log(1/h)\}^{1/2}}{n^{3/2}h^{2\beta+3/2}}\right).
	\end{align*}
	Then by \eqref{eq:den}, uniformly over $(x,\tau)\in\mathcal{X}\times\mathcal{T},$
	\begin{align*}
		\eqref{varproof:term1} = O_P\left(\frac{\{\log(1/h)\}^{1/2}}{n^{3/2}h^{2\beta+3/2}}\right).
	\end{align*}
	
	For term \eqref{varproof:termbias}, recall that the definition of $\sigma_n^2(x,\tau)= [f_{X,Y}\{x,\theta_\tau(x)\}]^{-2} \Var [V_n\{\theta_{\tau}(x);x,\tau\}]$. We know that
	\begin{align*}
		&\Var [V_n\{\theta_{\tau}(x);x,\tau\}]\\
		 &= \frac{1}{n}\Var [K_{U,h}(x-W)\psi_{\tau}\{Y-\theta_{\tau}(x)\}]\\
		&=\frac{1}{n} \left\{\mathbb{E}\left[K_{U,h}^2(x-W)\psi_{\tau}^2\{Y-\theta_{\tau}(x)\}\right] - \left(\mathbb{E}\left[K_{U,h}(x-W)\psi_{\tau}\{Y-\theta_{\tau}(x)\} \right]\right)^2\right\},
	\end{align*}
	Then,
	\begin{align*}
		\eqref{varproof:termbias} = \frac{1}{nf_{X,Y}^2\{x,\theta_{\tau}(x)\}}\left(\mathbb{E}\left[K_{U,h}(x-W)\psi_{\tau}\{Y-\theta_{\tau}(x)\} \right]\right)^2 = O(n^{-1}h^{2s}),
	\end{align*}
	where the last equality holds by Assumption \ref{Assumption:boundedsupport}(ii) and Lemma \ref{lemma:bias}.

	For the term \eqref{varproof:term2}, we know from Theorem \ref{thm:criteron_consistency} and Assumption OS$'$ that
	\begin{align*}
		\sup_{x\in \mathcal{X},\tau\in\mathcal{T}} \left|\widehat{\theta}_{\tau}(x)-\theta_{\tau}(x)\right| = O_P\left(\sqrt{\frac{\log (1/h)}{{nh^{2\beta+1}}}}\right).
	\end{align*}
	Hence, there exists constant $C$ such that
	\begin{align*}
		\mathbb{P}\left(\sup_{x\in \mathcal{X}}\left|\widehat{\theta}_{\tau}(x)-\theta_{\tau}(x)\right| \le C\sqrt{\frac{\log (1/h)}{{nh^{2\beta+1}}}} \right)\rightarrow 1.
	\end{align*}
	Then, denote $\delta_n = C\sqrt{\frac{\log (1/h)}{{nh^{2\beta+1}}}}$, we have, with probability approaching one, 
	\begin{align}
		&\left\{\widehat{f}_{X,Y}(x,\widehat{\theta}_{\tau}(x))\right\}^2\cdot\eqref{varproof:term2}\notag\\
		&\le  \sup_{x\in \mathcal{X},\tau\in\mathcal{T}}\left|\frac{1}{n^2}\sum_{i=1}^n K_{U,h}^2(x-W_i) \left\{\psi_{\tau}^2(Y-\widehat{\theta}(x)) -\psi_{\tau}^2(Y-\theta_{\tau}(x))\right\} \right|\notag\\
		&\le \sup_{x\in \mathcal{X},\tau\in\mathcal{T}, |\theta_1-\theta_2|\le C\delta_n} \left|\frac{1}{n^2}\sum_{i=1}^n K_{U,h}^2(x-W_i)\left\{\psi_{\tau}^2(Y-{\theta}_1) -\psi_{\tau}^2(Y-\theta_2)\right\}\right|\notag\\
		&\le \sup_{x\in \mathcal{X},\tau\in\mathcal{T}, |\theta_1-\theta_2|\le C\delta_n}\left|\frac{1}{n}\mathbb{E}\left[K_{U,h}^2(x-W)\left\{\psi_{\tau}^2(Y-{\theta}_1) -\psi_{\tau}^2(Y-\theta_2)\right\}\right]\right|\label{p4:eq1}\\
		&+\sup_{x\in \mathcal{X},\tau\in\mathcal{T},|\theta_1-\theta_2|\le C\delta_n} \left|\frac{1}{n^2}\sum_{i=1}^n \left(K_{U,h}^2(x-W_i)\left\{\psi_{\tau}^2(Y-{\theta}_1) -\psi_{\tau}^2(Y-\theta_2)\right\}\right.\right.\notag\\
		&\qquad\qquad\qquad\qquad\qquad-\left.\left.\mathbb{E}\left[K_{U,h}^2(x-W)\left\{\psi_{\tau}^2(Y-{\theta}_1)\label{p4:eq2} -\psi_{\tau}^2(Y-\theta_2)\right\}\right]\right)\right|
	\end{align}
	
	For \eqref{p4:eq1}, using change of variable and Lemma \ref{lemma:bound}, we have
	{\small
	\begin{align*}
		&\eqref{p4:eq1}\\
		 =&\frac{1}{n} \sup_{x\in \mathcal{X},\tau\in\mathcal{T},|\theta_1-\theta_2|\le C\delta_n} \left|\int K^2_{U,h}(x-w) \mathbb{E}\left\{\psi_{\tau}^2(Y-\theta_1)-\psi_{\tau}^2(Y-\theta_2)|W=w\right\}f_W(w)dw\right|\\
		=&\frac{h}{n}\sup_{x\in \mathcal{X},\tau\in\mathcal{T},|\theta_1-\theta_2|\le C\delta_n}\left|\int K^2_{U,h}(wh) \mathbb{E}\left\{\psi_{\tau}^2(Y-\theta_1)-\psi_{\tau}^2(Y-\theta_2)|W=x-wh\right\}f_W(x-wh)dw\right|\\
		\le&\frac{1}{n}\sup_{w\in\mathbb{R}}\sup_{\tau\in\mathcal{T},|\theta_1-\theta_2|\le C\delta_n} \left| \mathbb{E}\left\{\psi_{\tau}^2(Y-\theta_1)-\psi_{\tau}^2(Y-\theta_2)|W=w\right\}f_W(w) \right|\cdot \int K_{U,h}^2(w)dw\\
		=& O(n^{-1}h^{-2\beta-1}\delta_n).
	\end{align*}
	}
	
	For \eqref{p4:eq2}, consider the function class
	\begin{align*}
		\mathcal{F}_{\delta_n}:=\left\{(w,y)\mapsto K_{U,h}^2(x-w)\{\psi_{\tau}^2(Y-\theta_1) - \psi_{\tau}^2(Y-\theta_2)\}:x\in \mathcal{X},\tau\in\mathcal{T}, |\theta_1-\theta_2|\le \delta_n\right\}
	\end{align*}
	Denote ${F}_{\delta_n} = 2D^2h^{-2\beta-2}$ with constant $D>0$ such that $\sup_{x\in\mathbb{R}}|K_{U,h}(x)|\le Dh^{-\beta-1}$. By Lemma \ref{lemma:VCtype} and Corollary A.1 in \cite{chernozhukov2014gaussian} we know that there exists $A$ and $v$ such that
	\begin{align}\label{p4:squareKVC}
		N\left(	\mathcal{F}_{\delta_n},\|\cdot\|_{Q,2},D^2h^{-2\beta-2}\right) \le (A/\delta)^{v},\ 0< \forall \delta\le 1.
	\end{align}
	Then the conditions of Corollary 5.1 in \cite{chernozhukov2014gaussian} are satisfied. Next, we bound several involved quantities:
	\begin{itemize}
		\item (The element-wise second moment)
		We know that
		\begin{align*}
			&\sup_{x\in \mathcal{X}, |\theta_1-\theta_2|\le \delta_n}\mathbb{E}\left[K_{U,h}^4(x-W)\{\psi_{\tau}^2(Y-\theta_1)-\psi_{\tau}^2(Y-\theta_2)\}^2\right]\\
			\le& C h^{-4\beta-4}\sup_{|\theta_1-\theta_2|\le \delta_n}\mathbb{E}\left[\psi_{\tau}^2(Y-\theta_1)-\psi_{\tau}^2(Y-\theta_2)\right]^2
			=O\left(h^{-4\beta-4}\delta_n\right),
		\end{align*}
		where the last equality follows from Assumption \ref{Assumption:PartialBoundAway}.
		\item (The second moment of envelope function) By the definition of $F_{\delta_n}$, we immediately have
		\begin{align*}
			\mathbb{E}\left\{ F_{\delta_n}^2\right\} = O(h^{-4\beta-4})
		\end{align*}
		\item (The second moment of maximal of envelope function)
		\begin{align*}
			\mathbb{E}\left\{\max_{1\le i\le n} F_{\delta_n}^2(W_i,Y_i)\right\}= O(h^{-4\beta-4}).
		\end{align*}
	\end{itemize}
	Hence, by Corollary 5.1 in \cite{chernozhukov2014gaussian}, using their notations with $\sigma^2\asymp h^{-4\beta-4}\delta_n$, $\|F\|_{P,2} = O(h^{-2\beta-2})$, $\|M\|_2=O(h^{-2\beta-2})$ and constant $A$ and $v$, we have,
	\begin{align*}
		&\mathbb{E}\sup_{x\in \mathcal{X},|\theta_1-\theta_2|\le C\delta_n} \left|\frac{1}{\sqrt{n}}\sum_{i=1}^n \left(K_{U,h}^2(x-W_i)\left\{\psi_{\tau}^2(Y-{\theta}_1) -\psi_{\tau}^2(Y-\theta_2)\right\}\right.\right.\notag\\
		&\qquad\qquad\qquad\qquad\qquad-\left.\left.\mathbb{E}\left[K_{U,h}^2(x-W)\left\{\psi_{\tau}^2(Y-{\theta}_1) -\psi_{\tau}^2(Y-\theta_2)\right\}\right]\right)\right|\\
		&=O\left(\frac{\delta_n^{1/2}\{\log(1/h)\}^{1/2}}{h^{2\beta+2}} + \frac{\log(1/
			h)}{n^{1/2}h^{2\beta+2}}\right) = O\left(\frac{\delta_n^{1/2}\{\log(1/h)\}^{1/2}}{h^{2\beta+2}}\right)
	\end{align*}
	Then uniformly over $(x,\tau)\in\mathcal{X}\times\mathcal{T}$,
	\begin{align*}
		\eqref{p4:eq2} = O_P\left(\frac{\delta_n^{1/2}\{\log(1/h)\}^{1/2}}{n^{3/2}h^{2\beta+2}}\right).
	\end{align*}
	Combine the results of \eqref{p4:eq1} and \eqref{p4:eq2} and by \eqref{eq:den}, we know that
	\begin{align*}
		\eqref{varproof:term2} =&  O(n^{-1}h^{-2\beta-1}\delta_n) + O_P\left(\frac{\delta_n^{1/2}\{\log(1/h)\}^{1/2}}{n^{3/2}h^{2\beta+2}}\right)\\
		=&O_P\left(\frac{\{\log(1/h)\}^{1/2}}{n^{3/2}h^{3\beta+3/2}}\right) + O_P\left(\frac{\{\log(1/h)\}^{3/4}}{n^{7/4}h^{5\beta/2+9/4}}\right) = O_P\left(\frac{\{\log(1/h)\}^{1/2}}{n^{3/2}h^{3\beta+3/2}}\right).
	\end{align*}
	
	For term \eqref{varproof:term3}, from \eqref{eq:star1}, \eqref{eq:star2}, Theorem \ref{thm:criteron_consistency} and Assumption OS$'$, we know that
	\begin{align*}
		&\widehat{f}_{X,Y}(x,\widehat{\theta}_{\tau}(x)) - f_{X,Y}\{x,\theta_{\tau}(x)\}\\
		=& \widehat{f}_{X,Y}(x,\widehat{\theta}_{\tau}(x)) - f_{X,Y}(x,\widehat{\theta}_{\tau}(x))  + f_{X,Y}(x,\widehat{\theta}_{\tau}(x))  -  f_{X,Y}(x,{\theta}_{\tau}(x))\\
		=&O_P\left(m^{-1/2}n^{-1/2}h_W^{-2\beta-1}h_Y^{-1}+m^{-1/2}h_W^{-\beta}+h_Y^s + h_{W}^s+\sqrt{\frac{\log\{1/(h_Y\land h_W)\}}{nh_Yh_{W}^{2\beta+1}}} + \sqrt{\frac{\log (1/h)}{{nh^{2\beta+1}}}} \right)\\
		=&o_P\left\{\frac{1}{\log (1/h)}\right\}.
	\end{align*}
	And $\mathbb{E}\{K_{U,h}^2(x-W)\psi_{\tau}^2(Y-\theta_{\tau}(x))\}\lesssim h^{-2\beta-1}$, then uniformly over $(x,\tau)\in\mathcal{X}\times\mathcal{T}$,
	\begin{align*}
		\eqref{varproof:term3} =&o_P\left\{\frac{1}{nh^{2\beta+1}\log(1/h)} \right\}.
	\end{align*}
	Combine all the results,
	\begin{align*}
		\widehat{\sigma}_n^2(x;\tau) - \sigma_n^2(x;\tau) = o_P\left\{\frac{1}{nh^{2\beta+1}\log(1/h)}\right\}
	\end{align*}
	By Lemma \ref{lemma:variancelowerboundos}, we have our desired result.
	
	\textbf{Supersmooth Case:}
	We follow the decomposition in the ordinary smooth case and re-bound every term. 
	
	For the term \eqref{varproof:term0}, analogous to ordinary smooth case, we have
	\begin{align*}
		\eqref{varproof:term0} = O_P\left\{n^{-1}m^{-1/2}h^{2\beta_0-1}\exp(2h^{-\beta}/\gamma) e_U(h)\right\}
	\end{align*}
	
	For the term \eqref{varproof:term1}, we use Corollary 5.1 in \cite{chernozhukov2014gaussian} to bound it. Define the function class
	$$\mathcal{F}_n=\left\{(w,y)\mapsto K_{U,h}^2(x-w)\psi_{\tau}^2(y-\theta_{\tau}(x)):x\in \mathcal{X},\tau\in\mathcal{T}\right\}.$$
	Let the envelope function $F:= Dh^{-\beta}e_U^2(h)$ with constant $D>0$ such that $\sup_{x\in\mathbb{R}}\left|K_{U,h}^2(x)\right| \le Dh^{-\beta}e_U^2(h)$. By Lemma \ref{lemma:VCtype} and Corollary A.1 in \cite{chernozhukov2014gaussian}, we know that there exist constants $A$ and $v$ such that
	\begin{align*}
		N(\mathcal{F}_n,\|\cdot\|_{Q,2},\|F\|_{Q,2})\le (A/\delta)^v,\ 0<\forall \delta\le 1.
	\end{align*}
	Then the conditions of Corollary 5.1 in \cite{chernozhukov2014gaussian} are satisfied. To apply it, we calculate several involved quantities:
	\begin{itemize}
		\item (The element-wise second moment)
		We know that by Lemma \ref{lemma:bound}
		\begin{align*}
			\sup_{f\in\mathcal{F}_n}\mathbb{P}f^2 = &\sup_{x\in \mathcal{X},\tau\in\mathcal{T}}\mathbb{E}\left\{K_{U,h}^4(x-W)\psi_{\tau}^4(Y-\theta)\right\} \lesssim e_U^4(h).
		\end{align*}
		\item (The second moment of envelope function) By the definition of $F$, we immediately have
		\begin{align*}
			\mathbb{E}\left\{ F^2\right\} = O\{h^{-2\beta}e_U^4(h)\}
		\end{align*}
		\item (The second moment of maximal of envelope function)
		\begin{align*}
			\|M\|_2^2 = \mathbb{E}\left\{\max_{1\le i\le n} F^2(W_i,Y_i)\right\} &= O\{h^{-2\beta}e_U^4(h)\}
		\end{align*}
	\end{itemize}
	Then, by Corollary 5.1 in \cite{chernozhukov2014gaussian}, we know that
	\begin{align*}
		&\mathbb{E}\left\{ \sup_{(x,\tau) \in \mathcal{X}\times\mathcal{T}}\left|\frac{1}{n^2 } \sum_{i=1}^n\left[ K_{U,h}^2(x-W_i)\psi_{\tau}^2(Y-{\theta_{\tau}}(x)) - \mathbb{E}\left\{K_{U,h}^2(x-W_i)\psi_{\tau}^2(Y-{\theta_{\tau}}(x)) \right\}\right]\right|\right\}\\
		=& O\left(\frac{1}{n^{3/2}}\left\{ e_U^2(h)\{\log(1/h)\}^{1/2}  + \frac{h^{-\beta}e_U^2(h)\log(h^{-1})}{\sqrt{n}} \right\}\right) = O\left(\frac{e_U^2(h)\{\log(1/h)\}^{1/2}}{n^{3/2}} \right).
	\end{align*}
	For term \eqref{varproof:termbias}, analogous to the ordinary smooth case, 
	\begin{align*}
		\eqref{varproof:termbias} =  O(n^{-1}h^{2s}).
	\end{align*}
	For term \eqref{p4:eq1}, we have
	\begin{align*}
		\eqref{p4:eq1} = O_P\left(n^{-1}e_U^2(h)\sqrt{\frac{e_U^2(h)\log(1/h)}{n}}\right)
	\end{align*}
	For term \eqref{p4:eq2}, we have
	\begin{align*}
		\eqref{p4:eq2} = O_P\left(n^{-3/2}e_U^2(h)\{\log(1/h)\}^{1/2}\sqrt{\frac{e_U^2(h)\log(1/h)}{n}}\right)
	\end{align*}
	For term \eqref{varproof:term3}, by analogous to Lemma \ref{lemma:effectU}, we have
 \begin{align*}
\widehat{f}_{X,Y}(x,y) - f_{X,Y}(x,y) = O_P\left\{m^{-1/2}n^{-1/2}h_W^{2\beta_0-1}h_Y^{-1}\exp(2h_W^{-\beta}/\gamma)+m^{-1/2}h_W^{\beta_0}\exp(h_W^{-\beta}/\gamma)\right.\\
\left.+ h_W^s + h_Y^s + \sqrt{\frac{e_U^2(h_W)\log\{1/(h_Y\land h_W)\}}{nh_Y}}\right\}
 \end{align*}
Then by Assumption SS$'$, we have
	\begin{align*}
		\eqref{varproof:term3} = o_P\left(n^{-1}e_L^2(h)\{\log(1/h)\}^{-1}\right).
	\end{align*}
	Combine all the results,
	\begin{align*}
		\widehat{\sigma}_n^2(x;\tau) - \sigma_n^2(x;\tau) = o_P\left(n^{-1}e_L^2(h)\{\log(1/h)\}^{-1}\right).
	\end{align*}
	By Lemma \ref{lemma:variancelowerboundos}, we have our desired result.

\newpage
\section{Discussion on \cite{schennach2008quantile}}

This section aims to discuss \cite{schennach2008quantile} and highlight its differences compared to our work. Both papers consider the estimation of the $\tau^{th}$ conditional quantile function $\theta_{\tau}(\cdot)$, defined as a unique solution to the conditional moment $\mathbb{E}[\psi_{\tau}(Y-\theta)|X=x]=0$, in the presence of mismeasured covariates $W=X+U$. 

The main results of \cite{schennach2008quantile} rely on the instrumental variables (IV) approach of
identifying the unobservable quantity $\mathbb{E}[\psi_{\tau}(Y-\theta)|X=x]$ and the consistent estimation of $\theta_{\tau}(\cdot)$. We summarize the result of \cite{schennach2008quantile}:
\begin{itemize}
    \item \textbf{Identification}. Based on the instrumental variables (IV) condition, with an application of the Fourier and the inverse Fourier transform techniques,  \cite{schennach2008quantile} identified the unobservable quantity $\mathbb{E}[\psi_{\tau}(Y-\theta)|X=x]$ as an inverse Fourier transform of an estimable IV function. \cite{schennach2008quantile} assumes there is an IV $Z$ satisfying the following IV condition:
   \begin{align}\label{as:IV}
  \text{(IV Condition)}: \  Y\perp Z|X, \quad X^*\perp Z, \quad \mathbb{E}[U|Z,Y]=0,
   \end{align}
 where $X^*:=X-\mathbb{E}[X|Z]$ is the prediction error of the instrument.  
By the IV Condition, $\mathbb{E}[\psi_{\tau}(Y-\theta)|X=\cdot]$ satisfies the following convolution equation:
\begin{align}\label{eq:inteq}
  \mathbb{E}[\psi_{\tau}(Y-\theta)|Z^*=z^*]=\int  \mathbb{E}[\psi_{\tau}(Y-\theta)|X=z^*-x^*]f_{X^*}(-x^*)dx^*, 
\end{align}
where $Z^*:=\mathbb{E}[W|Z]$.  \cite{schennach2008quantile} also identified $\mathbb{E}[e^{-i\zeta X^*}]$, the Fourier transform of $f_{X^*}(-x^*)$, as a nonlinear functional of derivatives of conditional expectations given IV. With the identified $\mathbb{E}[e^{-i\zeta X^*}]$, by consecutively applying the Fourier transform and the inverse Fourier transform on  both sides of \eqref{eq:inteq}, the unobservable quantity $\mathbb{E}[\psi_{\tau}(Y-\theta)|X=x]$ can be identified as follows:
\begin{align*}
 \mathbb{E}[\psi_{\tau}(Y-\theta)|X=x]=\int F_{Z^*}(\zeta,\theta)e^{-i\zeta x}d\zeta,   
\end{align*}
where $F_{Z^*}$ is an estimable IV function.
\item \textbf{Estimation}. In estimation, they first construct a plug-in estimator for $F_{Z^*}$, denoted by $\widehat{F}_{Z^*}$, then estimate $\theta_{\tau}(\cdot)$ based on a global sieve optimization:
\begin{align}\label{IVestimator}
    \widehat{\theta}^{IV}_{\tau}(\cdot)= \arg\min_{\theta(\cdot)\in \mathcal{L}_n}\int \left\{\int \widehat{F}_{Z^*}(\zeta,\theta(x))e^{-i\zeta x}d\zeta\right\}^2dx
\end{align}
 where $\mathcal{L}_n$ is a specified sieve space increasing with the sample size $n$. They proved the uniform consistency result $\sup_{x\in\mathcal{X}}|\widehat{\theta}_{\tau}(x)-\theta_{\tau}(x)|=o_P(1)$ over a compact set $\mathcal{X}$ for any fixed quantile level $\tau\in (0,1)$. 
\end{itemize} 


Our paper provides a uniform inference approach for the conditional quantile function when the measurement errors are classical with unknown distribution, which is not covered by \cite{schennach2008quantile}; indeed,  \cite{schennach2008quantile} aims for identification and consistent estimation of the conditional quantile function with the aid of appropriate IVs, when the covariates subject to non-classical measurement errors, i.e. $U$ and $(X,Y)$ are allowed to be correlated. \cite{schennach2008quantile} neither studied the convergence rate of her estimator nor the valid uniform inference approach.

\textcolor{black}{Although our paper is  generally distinct from \cite{schennach2008quantile}, there still exists some connections in some special circumstances. As an anonymous referee pointed out, in the repeated measurement error scenario discussed in Section 2, the repeated observation can serve as a valid instrumental variable satisfying the IV Condition (D1) in \cite{schennach2008quantile}, provided that some additional conditions hold. Specifically, let $W^{(1)}=X+U^{(1)}$ and $W^{(2)}=X+U^{(2)}$ be two repeated observation subject to classic measurement errors, where $X\perp \{U^{(1)},U^{(2)}\}$. Let $\varepsilon:=Y-\mathbb{E}[Y|X]$, if we further assume that
\begin{align}\label{as:addition}
X \sim N\left(0, \sigma_x^2\right), \ U^{(2)} \sim N\left(0, \sigma_v^2\right),\  \varepsilon \perp (X, U^{(1)}, U^{(2)}) \ \text{and} \ \mathbb{E}[U^{(1)}] 
  = 0, 
\end{align}
 then $W^{(2)}$ is a valid instrument for $X$ that satisfying the IV Condition (D1). This can be done by verifying that
\begin{align*}
Y\perp W^{(2)}|X, \  \mathbb{E}[U^{(1)}|W^{(2)},Y] = 0\ \text{and} \ (X - \mathbb{E}[X|W^{(2)}])\perp  W^{(2)}.
\end{align*} 
 The first and second conditions are obviously satisfied. For the third condition, note that $$
\begin{aligned}
X - \mathbb{E}[X|W^{(2)}] & =W^{(2)}-U^{(2)}-\mathbb{E}[W^{(2)} - U^{(2)} \mid W^{(2)}] \\
& =-\left(U^{2} - \mathbb{E}[U^{(2)} \mid W^{(2)}]\right).
\end{aligned}
$$
Thus, $X - \mathbb{E}[X|W^{(2)}]$ and $W^{(2)}$ are always mean independent. Note that $(W^{(2)}, U^{(2)})$ are jointly normal, we have
$$
\mathbb{E}[U^{(2)} \mid W^{(2)}]=\underbrace{\mathbb{E}[U^{(2)}]}_0+\frac{\overbrace{\operatorname{Cov}(U^{(2)}, W^{(2)})}^{\sigma_v^2}}{\underbrace{\operatorname{Var}[W^{(2)}]}_{\sigma_x^2+\sigma_v^2}}\{W^{(2)}-\underbrace{\mathbb{E}[W^{(2)}]}_0\}=\frac{\sigma_v^2}{\sigma_x^2+\sigma_v^2} W^{(2)}
$$
which implies $U^{(2)}-\mathbb{E}[U^{(2)} \mid W^{(2)}]=U^{(2)}-\frac{\sigma_v^2}{\sigma_x^2+\sigma_v^2} W^{(2)}$. Also note that
$$
\operatorname{Cov}\left(U^{(2)}-\frac{\sigma_v^2}{\sigma_x^2+\sigma_v^2} W^{(2)}, W^{(2)}\right)=0
$$
which together with the fact $(U^{(2)}, W^{(2)})$ are jointly normal implies $X - \mathbb{E}[X|W^{(2)}]=-\{U^{(2)}-\mathbb{E}[U^{(2)} \mid W^{(2)}]\}$ and $W^{(2)}$ are fully independent. }

\end{document}